\documentclass[prc,twocolumn,superscriptaddress,
amsmath,amssymb,
aps]{revtex4-2}

\usepackage{array}[=2016-10-06] 

\usepackage{extarrows}
\usepackage{subcaption}

\usepackage{mathtools}

\usepackage{siunitx}
\usepackage{physics} % für Differntiale und ähnliches
\usepackage{slashed}
\usepackage{booktabs}
\usepackage{multirow}
\usepackage{csquotes}
\usepackage{graphicx}
\usepackage{subdepth} 
\usepackage{ragged2e}
\usepackage{hyperref}
\usepackage{xstring}
\usepackage[dvipsnames]{xcolor} % for color notation during creation and for the peer revies process

\begin{document}
\allowdisplaybreaks

\newcommand{\ii}{i}

\newcommand{\IS}{\mspace{2mu}} 

%%%
% kernel definitions
%%%
\newcommand{\kernelQ}[3][Q]{\ensuremath{K_{#1,#2}^{(#3)}}}\
\newcommand{\kernelR}[3][R]{\ensuremath{K_{#1,#2}^{(#3)}}}
\newcommand{\kernelL}[3][L]{\ensuremath{K_{#1,#2}^{(#3)}}}

\newcommand{\kernelQR}[4][Q-R]{\ensuremath{K_{\StrBefore{#1}{-}\StrBehind{#1}{-}^{#2},#3}^{(#4)}}}
\newcommand{\kernelLQ}[4][L-Q]{\ensuremath{K_{\StrBefore{#1}{-}^{#2}\StrBehind{#1}{-},#3}^{(#4)}}}
\newcommand{\kernelLQR}[5][L-Q-R]{\ensuremath{K_{\StrBefore{#1}{-}^{#2}\StrBetween[1,2]{#1}{-}{-}\StrBehind[2]{#1}{-}^{#3},#4}^{(#5)}}}

%%%
% momenta
%%%
\newcommand{\kl}[1][]{\vec{l}^{\,#1}}
\newcommand{\kx}{\vec{x}_n}
\newcommand{\xn}{x_n}
\newcommand{\ppf}[1][]{\ensuremath{\bar{p}^{#1}}}
\newcommand{\kf}[1][]{\ensuremath{\vec{\bar{p}}^{\,#1}}}
\newcommand{\ppi}[1][]{\ensuremath{p^{#1}}}
\newcommand{\ki}[1][]{\ensuremath{\vec{p}^{\,#1}}}
\newcommand{\kfi}[1][]{\ensuremath{\vec{q}^{\,#1}}}
\newcommand{\kq}[1][]{\ensuremath{\vec{q}^{\,#1}}}
\newcommand{\kQ}[1][]{\ensuremath{\vec{Q}^{\,#1}}}
\newcommand{\Kfi}[1][]{\ensuremath{\vec{Q}^{\,#1}}}
\newcommand{\qqf}[1][]{\ensuremath{\bar{q}^{#1}}}
\newcommand{\qqi}[1][]{\ensuremath{q^{#1}}}

%%%
%  Integrals
%%%
\newcommand{\IntdDPi}[2][d]{\ensuremath{\int\frac{\dd[#1]{#2}}{(2\,\pi)^{#1}}}}
\newcommand{\SumAll}[1][n]{\ensuremath{\sum\limits_{\vec{#1}\in\mathbb{Z}^3}}}
\newcommand{\IntFeyPa}[2][1]{\ensuremath{\int_0^{#1}\dd{#2}}}
\newcommand{\FPR}[1][]{z_R^{#1}\mspace{1mu}}
\newcommand{\FPL}[1][]{z_L^{#1}\mspace{1mu}}

%%%
% Commands over-complete basis elements
%%%
\newcommand{\genQBox}[5][Q]{\ensuremath{ J_{#1}^{(#2,#3,#4;#5)}[\kx]}} 
\newcommand{\genQRBox}[5][QR]{\ensuremath{ J_{#1}^{(#2,#3,#4;#5)}[\kx]}} 
\newcommand{\genLQBox}[5][LQ]{\ensuremath{ J_{#1}^{(#2,#3,#4;#5)}[\kx]}} 
\newcommand{\genLQRBox}[5][LQR]{\ensuremath{ J_{#1}^{(#2,#3,#4;\,#5)}[\kx]}} 

\newcommand{\aaf}{\bar{a}}
\newcommand{\bbf}{\bar{b}}
\newcommand{\ccf}{\bar{c}}
\newcommand{\eef}{\bar{e}}
\newcommand{\kkf}{\bar{k}}

\newcommand{\aai}{a}
\newcommand{\bbi}{b}
\newcommand{\cci}{c}
\newcommand{\eei}{e}
\newcommand{\hhi}{h}
\newcommand{\kki}{k}
\newcommand{\pll}{h}

%%%
% Commands basis elements
%%%
\newcommand{\basisQ}[4][Q]{\ensuremath{I_{#1}^{(#2,#3,#4)}}} 
\newcommand{\basisQR}[4][QR]{\ensuremath{I_{#1}^{(#2,#3,#4)}}} 
\newcommand{\basisLQ}[4][LQ]{\ensuremath{I_{#1}^{(#2,#3,#4)}}} 

\newcommand{\renQ}[4][Q]{\ensuremath{\bar{I}_{#1}^{(#2,#3,#4)}}} 
\newcommand{\renQR}[4][QR]{\ensuremath{\bar{I}_{#1}^{(#2,#3,#4)}}} 
\newcommand{\renLQ}[4][LQ]{\ensuremath{\bar{I}_{#1}^{(#2,#3,#4)}}}
\newcommand{\renLQR}[6][L-Q-R]{\ensuremath{\bar{I}_{\StrBefore{#1}{-}^{#2}\StrBetween[1,2]{#1}{-}{-}\StrBehind[2]{#1}{-}^{#3}}^{(#4,#5,#6)}}}

\newcommand{\renSQR}[2][QR]{\ensuremath{\bar{S}_{#1}^{(#2)}}} 
\newcommand{\renSLQ}[2][LQ]{\ensuremath{\bar{S}_{#1}^{(#2)}}}
\newcommand{\renSLQR}[4][L-Q-R]{\ensuremath{\bar{S}_{\StrBefore{#1}{-}^{#2}\StrBetween[1,2]{#1}{-}{-}\StrBehind[2]{#1}{-}^{#3}}^{(#4)}}}

%%%
% Alpha factors
%%%
\newcommand{\alpA}[3]{\ensuremath{\alpha^{A,#3}_{#1,#2}}} 
\newcommand{\alpP}[3]{\ensuremath{\alpha^{P,#3}_{#1,#2}}} 
\newcommand{\alpT}[3]{\ensuremath{\alpha^{T,#3}_{#1,#2}}} 
\newcommand{\betT}[3]{\ensuremath{\beta^{T,#3}_{#1,#2}}} 
\newcommand{\alpX}[3]{\ensuremath{\alpha^{X,#3}_{#1,#2}}} 
\newcommand{\alpPS}[3]{\ensuremath{\alpha^{PS,#3}_{#1,#2}}}

\setlength{\parindent}{0pt}	% Einrücken der ersten Zeile
\setlength{\parskip}{0pt}	

\title{Decomposition of the axial-vector current in a finite box}

\author{Felix Hermsen}
\affiliation{Van Swinderen Institute for Particle Physics and Gravity, University of Groningen, 9747 AG Groningen, The Netherlands}
\affiliation{GSI Helmholtzzentrum f\"ur Schwerionenforschung GmbH, \\Planckstra\ss e 1, 64291 Darmstadt, Germany}

\author{Matthias F. M. Lutz}
\affiliation{GSI Helmholtzzentrum f\"ur Schwerionenforschung GmbH, \\Planckstra\ss e 1, 64291 Darmstadt, Germany}
\author{Rob G. E. Timmermans}
\affiliation{Van Swinderen Institute for Particle Physics and Gravity, University of Groningen, 9747 AG Groningen, The Netherlands}

\begin{abstract}
We consider the matrix element of the axial-vector current between two nucleon states in a finite box. Starting from the chiral Lagrangian density with nucleon and $\Delta$-isobar degrees of freedom, we study the finite-volume effects at the one-loop level. We show that the standard decomposition into the axial-vector and pseudoscalar form factor is incomplete in a finite box.
We derive expressions for the complete set of in-box form factors at one loop, and demonstrate how to extract the full set from lattice correlation functions. We verify that the axial Ward identity holds in the chiral limit. We derive the one-loop expressions for the pseudoscalar form factor and verify that the in-box axial Ward identity away from the chiral limit is fulfilled also.
Selected numerical results are shown for two flavor-SU(2) lattice ensembles. Sizable finite-volume effects are observed, with an important role for the $\Delta$-isobar. We discuss the implications of our results for lattice studies of the axial-vector current. We conclude that full finite-box results are crucial for a precise determination of the form factors.
\end{abstract}

\maketitle
\section{Introduction}
The axial-vector form factor of the nucleon is an important testing ground for the progress achieved in the lattice approach to QCD \cite{Beane:2004rf,Khan:2006de,Bali:2014nma,Capitani:2017qpc,Alexandrou:2017hac,Bali:2018qus}. It is possible nowadays in lattice-QCD to simulate specific hadronic quantities at physical pion masses \cite{Alexandrou:2023qbg,Tsuji:2025quu}. However, excited-state contamination \cite{Gupta:2024qip} in the extraction of the asymptotic signal from Euclidean-time simulations is increasingly difficult to control when the pion mass decreases to its physical value. When ensembles with non-physical pion masses are used (e.g. Refs. \cite{Khan:2006de,Beane:2004rf,Bali:2014nma,Capitani:2017qpc,Alexandrou:2017hac,Bali:2018qus,Djukanovic:2024krw,Bali:2023sdi,Gupta:2018qil}), one needs to extrapolate not only to the physical pion mass, but also from the finite-box results to the infinite volume. Often these extrapolations are done with phenomenological expressions, assuming in particular that finite-volume effects are proportional to $m_\pi^2\,\exp(-L\,m_\pi)/\sqrt{L\,m_\pi}$, see for instance Refs. \cite{Djukanovic:2024krw,Bali:2023sdi,Bali:2022qja,Gupta:2018qil,Park:2025rxi}. It is known \cite{Greil:2011aa,Hall:2025ytt}, however, that a correct, controlled extrapolation is challenging, especially if the $\Delta$-isobar plays an important role in the hadronic observables under consideration \cite{Lutz:2014oxa,Lutz:2018cqo,Lutz:2020dfi,Lutz:2023xpi,Hermsen:2024eth}. \\ \par

It is expected that the extrapolation to the physical pion mass can be done reliably by using the results of the flavor-SU(2) chiral Lagrangian density with nucleon, $\Delta$-isobar, and pion degrees of freedom \cite{Jenkins:1990jv,Jenkins:1991es,Fuchs:2003vw,Procura:2006gq,Ledwig:2011cx,Yao:2017fym,Lutz:2020dfi,Hermsen:2024eth,Alvarado:2021ibw,Alvarado:2021ibw}. From a conceptual point of view an application of the flavor-SU(2) chiral Lagrangian density to lattice-QCD ensembles at fixed physical strange-quark mass, but varying values of the light-quark masses, appears most promising. So far, mostly Coordinated Lattice Simulations (CLS) ensembles are available, where typically the sum of up-, down-, and strange-quark masses are held constant. Recently, the Mainz group performed a chiral-SU(2) extrapolation to such data, however, the dependence of the low-energy constants (LECs) on the varying strange-quark mass was not taken into account explicitly \cite{Djukanovic:2024krw}. Moreover, the size and even the sign of the required chiral correction effects appears to contradict the expectation of conventional Chiral Perturbation Theory (ChPT) \cite{Djukanovic:2024krw,Harris:2019bih,Ottnad:2022axz,Capitani:2017qpc,Bali:2023sdi}. \\ \par

In lattice-QCD it has been assumed that the decomposition of the matrix element of the axial-vector current in form factors in a finite box is the same as in the infinite volume, with only small corrections neglected. Previous studies \cite{Bijnens:2014yya,Greil:2011aa}, for example for the pion decay constant, have demonstrated that the decomposition of form factors in a finite box may differ from the one in an infinite volume. We supported this conclusion in our earlier work \cite{Hermsen:2025vds}. In this article, we will show how a complete set of form factors can be included in a finite box, and we verify that the chiral Ward identity holds in the finite box in the chiral limit. \\ \par 

In Ref. \cite{Hermsen:2025vds}, we investigated the impact of two types of finite-volume effects: the ``implicit'' effects due to the values of the nucleon and $\Delta$-isobar masses in the finite box, and the ``explicit'' effects caused by computing the in-box loop integrals with the values of the nucleon and $\Delta$-isobar masses obtained in the infinite-volume limit. Here, we study the finite-box effects at one-loop level for two flavor-SU(2) ensembles, with the complete set of in-box form factors. Even for a relatively large box we observe sizable finite-volume effects, with an intricate interplay of implicit and explicit finite-box effects. We confirm our previous conclusion \cite{Hermsen:2025vds} that the full finite-box results are crucial for a precise determination of the axial-vector form factors. \\ \par

The organisation of our paper is as follows. In Section \ref{sec:MatrixELement} we discuss the matrix element of the axial-vector current and the complete set of form factors in a finite box, starting from the Lagrangian density with nucleon, $\Delta$-isobar, and pion degrees of freedom. Next, in Section \ref{sec:Loop} we discuss the one-loop contributions to the form factors of the axial-vector current. 
In Section \ref{sec:complete} we discuss the relevance of the complete set of in-box form factors and how to extract the full set from lattice correlation functions.
In Section \ref{sec:Numerics} we present numerical results for the finite-box effects for two lattice ensembles. We discuss the implicit versus the explicit finite-volume effects and the role of the $\Delta$-isobar. We summarize our results and conclusions in Section \ref{sec:Summary}. Several Appendices are devoted to various technical developments and results.
%\\ \par

%\clearpage
\begin{widetext}
\section{Matrix element of the axial-vector current in a finite box}\label{sec:MatrixELement}
In our previous work \cite{Hermsen:2025vds}, we introduced the decomposition of the axial-vector current between two nucleon states in a finite box of length $L$ and periodic boundary conditions:
\begin{equation}\begin{aligned}[b]
&\langle N(\bar{p})\vert\,A^{\mu}_i(0)\,\vert N(p)\rangle = \bar{u}(\bar{p})\,\sum_{\vec{n}\in\mathbb{Z}^3}\,F^\mu_A(\bar{p},p,\xn)\,\frac{\tau_i}{2}\,u(p)\,,\\
&F^\mu_A(\bar{p},p,\xn) = \Big(\gamma^\mu\,G_A(\bar{p},p,\xn) + \frac{q^\mu}{2\,M_N}\,G_P(\bar{p},p,\xn) + \frac{Q^{\mu}}{2\,M_N}\,G_T(\ppf,\ppi,\xn) 
+ 2\,M_N\,x_n^{\mu}\, G_X(\ppf,\ppi,\xn)\,\Big)\,\gamma_5\,,
\label{eqn:AxialMatrixElement}
\end{aligned}\end{equation}
where we used $q^{\mu} = \bar{p}^{\mu} - p^{\mu},\,Q^{\mu} = \bar{p}^{\mu} + p^{\mu},\,x_n = (0,\vec{x}_n)$ and $\vec{x}_n = L\,\vec{n}$. The sum is split into two parts: The terms with $\vec n \neq (0,0,0)$ are the finite-volume corrections, while the infinite-volume limit, $L \to \infty$, corresponds to the   $\vec{n}=(0,0,0)$ term. In this case, we recover our previous decomposition for the infinite-volume limit \cite{Hermsen:2024eth,Hermsen:2025vds,Schindler:2006it}. We note that the form factor $G_T$ is typically associated with the Dirac structure $\ii\,\sigma^{\mu\nu}\,q_{\nu}$ instead of $Q^{\mu}$ \cite{Weinberg:1958ut,Schindler:2006jq,Cabibbo:2003cu}. However, in the SU(2) case the two structures cannot be discriminated. 
%This follows from the relation $ 2\,\bar{u}(\ppf)\,\ii\,\sigma^{\mu\nu}\,q_\nu\,\gamma_5\,u(\ppi) = -\,\bar{u}(\ppf)\,Q^{\mu}\,\gamma_5\,u(\ppi)$. 
In Appendix \ref{app:decompDirac} we confirm that the set of Dirac structures in the decomposition in Eq. \eqref{eqn:AxialMatrixElement} is indeed minimal. \\\par 

Although at the physical point the form factors depend only on $q^2$, the finite box introduces a separate dependence on the three-momenta $\vec {\bar p}$ and $\vec p$.  Applying G-parity to the decomposition in Eq. \eqref{eqn:AxialMatrixElement} we find the following symmetry constraints for the form factors:
\begin{equation}\begin{aligned}[b]
&G_A(\bar{p},p, x_n) = +\,G_A(p,\bar{p}, x_n ) \,, \qquad
    G_P(\bar{p},p, x_n) = +\,G_P(p,\bar{p}, x_n ) \,,
    \qquad \text{ but } \quad
\\
&G_T(\bar{p},p, x_n) = -\,G_T(p,\bar{p} , x_n) \,, \qquad
    G_X(\bar{p},p , x_n) = -\,G_X(p,\bar{p}, x_n ) \,.
\label{eqn:SymRelFF}
\end{aligned}\end{equation}
These relations are derived in Appendix \ref{app:decompDirac} also. As we discussed in our previous work \cite{Hermsen:2024eth}, the axial Ward identity in the infinite volume dictates that $G_A$ and $G_P$ are not independent in the chiral limit. This relation was instrumental to establish consistent results for the form factors. With the additional form factor $G_{X}$, the axial Ward identity has to be updated in the finite box and we find in the chiral limit:
\begin{equation}
\lim_{m\to 0}\sum_{\vec{n}\in\mathbb{Z}^3}\bigg[G_A(\bar{p},p,x_n) + \frac{t}{4\,M_N^2}\,G_P(\bar{p},p,x_n) + \,(q\cdot x_n)\,G_{X}(\bar{p},p,x_n)\bigg] = 0\,.
\label{eqn:axialWard:chiralLimit}
\end{equation}

In Ref. \cite{Hermsen:2025vds} we derived the results for the axial-vector form factor $G_A$ in the finite box in a minimal set of basis integrals. The main goal of this work is to derive explicit expressions for the induced pseudoscalar form factors $G_P$, the induced pseudotensor form factor $G_T$, and for $G_X$. \\\par 

To extract the form factors from the amplitude $F^{\mu}_A(\bar{p}, p , x_n)$, it is convenient to express them as a Dirac trace in $d$ dimensions,
\begin{align}
&G_A(\bar{p},p, x_n) =
\frac{2 \, M_N^2}
{\big(d-2\big)\,\big(t- 4\,M_N^2\big)}\,\bigg\{ H + \frac{2\,M_N}{t}\,q_{\mu}\,H^{\mu}\bigg\}\,,\notag \\[1em]
%%%
%
%%%
&G_P(\bar{p},p, x_n)=
-\,\frac{8\,M_N^4}
{\big(d-2\big)   \, \big(t- 4\,M_N^2\big)\, t}\,\bigg\{H 
-\,\frac{4\,(d-1) \,M_N^2 -  (d-2) \, t}{2\,M_N\,t}\,q_{\mu}\,H^{\mu}\bigg\}
\notag \\ &\hspace{9em}
%%% endlicher Volumen Anteil
-\,\frac{4\,M_N^3\,(q\cdot\xn)}{t^2\,\big(4\,M_N^2\,(q \cdot\xn)^2 + 4\,t\,(\ppf\cdot\xn)\,(\ppi\cdot\xn) + (t-4\,M_N^2)\,t\,\xn^2\big)}
\notag \\ &\hspace{12em}
\times\,\Big\{4\,M_N^2\,(q \cdot x_n)\,q_{\mu} + 2\,t\,\big( (p\cdot x_n)\,\bar{p}_{\mu} + (\bar{p}\cdot x_n)\,p_{\mu} \big) + \big(t-4\,M_N^2\big)\,t\,x_{n,\mu}
\Big\}\,H^{\mu}\,,\notag \\[1em]
%%%
%
%%%
&G_T(\bar{p},p, x_n)=
-\,\frac{4\,M_N^3}{\big(t-4\,M_N^2\big)\,t}\,Q_{\mu}\,H^{\mu}
\notag \\ &\hspace{9em}
%%% endlicher Volumen Anteil
+\,\frac{4\,M_N^3\,(Q\cdot\xn)}{\big(t-4\,M_N^2\big)\,t\,\big(4\,M_N^2\,(q \cdot\xn)^2 + 4\,t\,(\ppf\cdot\xn)\,(\ppi\cdot\xn) + (t-4\,M_N^2)\,t\,\xn^2\big)}
\notag \\ &\hspace{12em}\times\,
\Big\{4\,M_N^2\,(q \cdot x_n)\,q_{\mu} + 2\,t\,\big( (p\cdot x_n)\,\bar{p}_{\mu} + (\bar{p}\cdot x_n)\,p_{\mu} \big) + \big(t-4\,M_N^2\big)\,t\,x_{n,\mu}
\Big\}\,H^{\mu}\,,\notag \\[1em]
%%%
%
%%%
&G_X(\bar{p},p, x_n)=
\frac{M_N}{t\,\big(4\,M_N^2\,(q \cdot\xn)^2 + 4\,t\,(\ppf\cdot\xn)\,(\ppi\cdot\xn) + (t-4\,M_N^2)\,t\,\xn^2\big)}
\notag \\ &\hspace{12em}\times\,
\Big\{4\,M_N^2\,(q \cdot x_n)\,q_{\mu} + 2\,t\,\big( (p\cdot x_n)\,\bar{p}_{\mu} + (\bar{p}\cdot x_n)\,p_{\mu} \big) + \big(t-4\,M_N^2\big)\,t\,x_{n,\mu}
\Big\}\,H^{\mu}\,,
\label{res:projection}
\end{align}
with $t = (\bar{p} - p)^2$ and the auxiliary quantities
\begin{align}
H =  \tr \Bigg[\gamma_{\mu} \, \gamma_5\,
\frac{\slashed{\bar{p}}+M_N}{2 \, M_N} \, F^{\mu}_A(\bar{p},p,x_n) \,\frac{\slashed{p}+M_N}{2 \, M_N}\Bigg]\,,\qquad 
%%%
H^{\mu} = \tr\Bigg[\gamma_5\,
\frac{\slashed{\bar{p}}+M_N}{2 \, M_N} \, F^{\mu}_A(\bar{p},p,x_n) \,\frac{\slashed{p}+M_N}{2 \, M_N}\Bigg]\,.
\label{def:H}
\end{align}
The use of projectors has the advantage that we do not need to consider tensor-type loop integrals. However, they come with a small complication because they introduce superficial singularities at $t=0$ and $x_n=0$. In the final results, the form factors are regular at $x_n=0$, as otherwise the infinite-volume limit cannot be approached. The kinematical region around $t=0$ is more intricate.\\\par 

We point out that due to G-parity the axial-vector matrix element at zero momentum transfer, i.e. the axial charge, is given by $G_A(\ppf,\ppi,\xn)\big\vert_{t=0}$ only. Therefore, the potential singularity at $t=0$ in the axial-vector form factor must be removed. In our previous work \cite{Hermsen:2025vds} we reduced $G_A(\bar{p},p,x_n)$ to our novel in-box basis, which meets this requirement. Consequently, potential singularities in the remaining form factors are not problematic, since we do not perform the limit $t\to 0$, as we have only discrete momenta. However, we showed previously \cite{Hermsen:2024eth} that in the infinite volume the limit $G_P(\ppf,\ppi)\big\vert_{t=0}$ exists. \\\par

We consider the chiral Lagrangian density with pion, nucleon, and $\Delta$-isobar  degrees of freedom, as developed in Refs. \cite{Bernard:1993bq,Bernard:1998gv,Fearing:1997dp,Schindler:2006it, Fuchs:2003qc, Chen:2012nx, Ando:2006xy,Hemmert:2003cb,Procura:2006gq,Yao:2017fym, Ellis:1997kc,Lutz:2020dfi}. Earlier work relied on heavy-baryon ChPT \cite{Bernard:1993bq,Bernard:1998gv,Fearing:1997dp}, while later the relativistic form of the chiral Lagrangian density was advocated \cite{Schindler:2006it,Fuchs:2003qc,Chen:2012nx,Ando:2006xy}. We will use the conventions of Refs. \cite{Lutz:2020dfi,Sauerwein:2021jxb}, in which a renormalization scheme based on the Passarino-Veltman reduction scheme \cite{Passarino:1978jh} was applied. \\ \par

We turn to the diagrams that enter the amplitude $F^{\mu}_A$. In Fig. \ref{chap:FFInf:figLoopContri} we show three different types of loop contributions, which characterize the amplitude up to chiral order four. First, we have one-loop corrections to the direct coupling of the axial-vector current to the nucleon line, which were determined in our previous work \cite{Lutz:2020dfi}. The second and third contributions originate from the axial-vector current coupling to a pion, which then couples to the nucleon line. While the baryonic Lagrangian density implies loop corrections to the pion-nucleon coupling, as displayed in the second picture of Fig. \ref{chap:FFInf:figLoopContri}, the mesonic Lagrangian density implies corrections to the axial-vector-current pion coupling, as displayed in the third picture of Fig. \ref{chap:FFInf:figLoopContri}. These two types of contributions were derived in our previous work, in the infinite volume only \cite{Hermsen:2024eth}. Altogether we have: 
\begin{equation}\begin{aligned}[b]
& F_A^\mu(\bar{p},p,x_n) = F_{\text{tree}}^{\mu}(\bar{p},p)\,\delta_{x_n,0} + F_{\text{loop}}^{\mu}(\bar{p},p,x_n)\,,\\
&\frac{1}{\sqrt{2}}\,F_{\text{tree}}^{\mu}(\bar{p},p) = 
\Big(  \frac{1}{\sqrt{2}} \,g_A\, \gamma^\nu \,\gamma_5 \Big)
\Big(Z_N\, g_{\nu}^{\;\mu} - \frac{q_\nu \,q^\mu}{q^2 -m_\pi^2}\,\big[ Z_N+f_\pi/f +\sqrt{Z_\pi} - 2\big] \Big)
\\  & \hspace{2.5em} + \frac{1}{\sqrt{2}}\,\big(
g_R \, q^2 +4 \, g^+_\chi \,( 2\,B_0\,m ) \big) \,\gamma^\mu \, \gamma_5 - \frac{2\,M}{\sqrt{2}}\,\big( 
4\,g_\chi ^++ g_\chi^- \big)\,\gamma_5\,\frac{q^\mu \,(2\,B_0\,m)}{q^2-m_\pi^2} - \frac{1}{\sqrt{2}}\,g_R\,\slashed{q}\,\gamma_5\,q^\mu\,,
\\
&\frac{1}{\sqrt{2}}\,F_{\text{loop}}^{\mu}(\bar{p},p,x_n) =  \frac{g_A}{f^2}\,  \Big\{J^\mu_{\pi}(\bar p, p,x_n) + J_{ N \pi }^\mu(\bar p, p,x_n)   +  J_{\pi N}^\mu(\bar p, p,x_n) \Big\}
+\, \frac{f_S}{3\,f^2}\,  \Big\{ J_{ \Delta \pi }^\mu(\bar p, p,x_n) + J_{\pi \Delta}^\mu (\bar p, p,x_n)  \Big\} 
\\  & \hspace{2.5em}
+ \frac{g_A^3}{4\,f^2}\,J^\mu_{N\pi N}(\bar p, p,x_n)  
+ \, \frac{2\,g_A\,f_S}{3\,f^2}\,\Big\{J^\mu_{\Delta \pi N }(\bar p, p,x_n) + J^\mu_{N\pi \Delta }(\bar p, p,x_n) \Big\}
+ \frac{5\,h_A\,f_S^2}{9\,f^2}\,J^\mu_{\Delta \pi \Delta}(\bar p, p,x_n) +{\mathcal O} \big( Q^4\big)\,.
\label{res-FA}
\end{aligned}\end{equation}

Here $f$ and $M$ are the chiral limits of the pion decay constant and the nucleon mass. Furthermore, the leading-order LECs $B_0, g_A, f_S, h_A$ and the subleading-order LECs $g_R,\,g_{\chi}^{+}$, and $g_{\chi}^{-}$ are involved.  We introduced the wave-function renormalization factors of the nucleon $Z_N$ and of the pion $Z_{\pi}$, the quark mass $m$, and the pion decay constant $f_\pi$. For explicit expressions for $Z_N$ we refer to Ref. \cite{Lutz:2020dfi} and for the mesonic quantities to Ref. \cite{Hermsen:2024eth} .\\\par

\begin{figure*}
    \centering
\begin{subfigure}{0.32\textwidth}
        \centering
        \includegraphics[height=0.17\textheight]{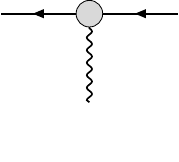}
    \end{subfigure}   
\begin{subfigure}{0.32\textwidth}
        \centering
        \includegraphics[height=0.17\textheight]{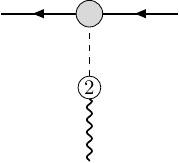}
    \end{subfigure}     
   \begin{subfigure}{0.32\textwidth}
    	\centering
    	 \includegraphics[height=0.17\textheight]{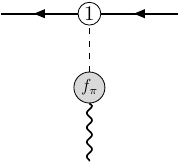}
    \end{subfigure} 
\caption{Different types of loop contributions that enter the matrix element of the axial-vector current between two nucleons. The  wavy, dotted, and solid  lines represent the axial current, pion, and nucleon. The chiral order of all vertices is indicated. The gray blobs indicate one-loop corrections. }
\label{chap:FFInf:figLoopContri}
\end{figure*}

We next turn to the one-loop functions $J^{\mu}_{\ldots}$ introduced in Eq. \eqref{res-FA}. They have the general structure
\begin{equation}
J^{\mu}_{\ldots}(\bar{p},p,x_n) = \ii\,\mu^{4-d}\,\IntdDPi{l}\,e^{\ii\,\kl\cdot\kx}\,\Big[ K^\nu_{\ldots}(\ppf,\ppi;l)\big(\delta^{\,\mu}_{\nu}-q_{\nu}\,q^{\mu}/(q^2-m_\pi^2)\big) + L^{\mu}_{\ldots}(\ppf,\ppi;l)\Big]\,,
\label{def-Js}
\end{equation}
for $\ldots\in\{\pi,L\pi,\pi R,L \pi R\}$ and $L,R\in\{N,\Delta\}$. While in our previous works \cite{Hermsen:2024eth,Hermsen:2025vds}, the functions $K^\nu_{\ldots}(\ppf,\ppi;l)$ were sufficient, now also the functions $L^{\mu}_{\ldots}(\ppf,\ppi;l)$ are needed. This originates in terms that contribute in the finite box, but not in the infinite volume \cite{Hermsen:2024eth}. We discuss this matter further in Appendix \ref{app:AxialAmplitudes}, where we also specify the functions $K^{\mu}_{\ldots}(\ppf,\ppi;l)$ and $L^{\mu}_{\ldots}(\ppf,\ppi;l)$. Note that both functions depend on the two-body LECs $g_S, \, g_V, \, g_T, \, f_\pm,\,f_M, \, b_{\chi}, \,g_F$, and the subleading axial $\Delta N$ transition LEC $f_E$. The LECs in our current study are taken from Ref. \cite{Hermsen:2024eth} and recalled in Table \ref{tab:listLECs}.  \\\par 
\begin{table*}
\centering
\begin{tabular}{lcp{3em}lcp{3em}lcp{3em}lc}
\toprule
LEC & \hspace{.75em} Fit result\,[MeV] & & LEC & \hspace{.75em} Fit result & & LEC & \hspace{.75em} Fit result\,[GeV$^{-1}$] & & LEC & \hspace{.75em} Fit result\,[GeV$^{-1}$] \\ 
\midrule
$f$  & $83.43$  && $g_A$ & $1.1449$ && $b_\chi$& $-0.6805$ && $f^+_A$ &  $-0.0617 $\\[.2em]
$M$  & $893.79$ && $f_S$ & $1.5857$ && $g_S$&$0.9163$ && $f_M $ &$-0.7335$\\[.2em]
$\Delta$ & $306.63$ && $h_A$&$0.7893$ && $g_T$&$1.5035$ && $f_E$ &$0.6949$
\\\bottomrule
\end{tabular}
\caption{Mean values of the LECs as determined in the fit from Ref. \cite{Hermsen:2024eth}. $M$ is the mass of the nucleon in the chiral limit and $\Delta$ is the mass difference of the nucleon and the $\Delta$-isobar in the chiral limit. We use these values in our numerical study in Section \ref{sec:Numerics}. }
\label{tab:listLECs}
\end{table*}

It is convenient to split the form factors $G_{A}(\bar p, p, x_n), \, G_P(\bar p, p, x_n), \, G_T(\bar p, p, x_n)$, and $G_X(\bar p, p, x_n)$ into corresponding tree-level and one-loop contributions. Formally, they follow by identifying $F^\mu_A(\bar p, p, x_n)$ with  $F^\mu_{\text{ tree} }(\bar p, p)\,\delta_{x_n, 0}$ and 
$F_{\text{loop}}^{\mu }(\bar p, p, x_n) $ in the projection Eqs. \eqref{res:projection} and \eqref{def:H}.
The tree-level contributions from our previous works read \cite{Hermsen:2024eth,Lutz:2020dfi}
\begin{equation}\begin{aligned}[b]
&G_{A}^{\text{tree}}(\bar{p},p)=
g_A\,Z_N+ 4\,g^+_\chi \,m_\pi^2  + g_R \,t \,, \hspace{4em}
G_{T}^{\text{tree}}(\bar{p},p) = 0 = G_{X}^{\text{tree}}(\bar{p},p)\,, \\
%%%
&\frac{t-m_\pi^2}{4\,M_N^2}\,G_P^{\text{tree}}(\ppf,\ppi) = 
-\,g_A\,\big(Z_N + \sqrt{Z_\pi}+  f_\pi/f - 2 \big)- \big(4\,g^+_\chi + g^-_\chi\big)\,m_\pi^2  - g_R \,\big(t-m_\pi^2\big) \,.
\label{res-tree} 
\end{aligned}\end{equation}
It is straightforward to recover these results with our projectors in Eq. \eqref{res:projection}, by applying the relation
\begin{equation}
\Big\{4\,M_N^2\,(q \cdot x_n)\,q_{\mu} + 2\,t\,\big( (p\cdot x_n)\,\bar{p}_{\mu} + (\bar{p}\cdot x_n)\,p_{\mu} \big) + (t-4\,M_N^2)\,t\,x_{n,\mu}
\Big\}\,H^{\mu}
= 0 \quad \text{for} \quad  F^{\mu}_A(\ppf,\ppi,x_n) = F_{\text{tree}}^{\mu}(\ppf,\ppi)\,,
\label{eqn:TraceTreeZero}
\end{equation}
at the on-shell conditions $\bar{p}^2 = p^2 =M_N^2$ after evaluating $H^{\mu}$. The $x_n$ dependence of the tree-level is factorized with $\delta_{x_n,0}$ as shown in  Eq. \eqref{res-FA}.  Therefore, the tree-level results do not have a term proportional to $x_n^{\mu}$, and the projectors can never generate a linear combination to remove the singular term for $x_n=0$ in Eq. \eqref{res:projection}.  Thus, Eq. \eqref{eqn:TraceTreeZero} is the only possibility to get rid of the $x_n=0$ singularity at tree level. This constitutes a useful cross-check for the projection in Eq. \eqref{res:projection}. \\\par

\section{Loop contributions to the form factors} \label{sec:Loop}
While it is convenient to introduce three auxiliary sums, 
\begin{equation}
G_A^{\text{loop}} (\bar p, p) = \sum_{\vec n \in \mathbb{Z}^3} 
G_A^{\text{loop}} (\bar p, p, x_n) \,,\quad 
G_P^{\text{loop}} (\bar p, p) = \sum_{\vec n \in \mathbb{Z}^3} 
G_P^{\text{loop}} (\bar p, p, x_n) \,,\quad 
G_T^{\text{loop}} (\bar p, p) = \sum_{\vec n \in \mathbb{Z}^3} 
G_T^{\text{loop}} (\bar p, p, x_n)\,,
\label{def:GAPTloop}
\end{equation}
a corresponding sum over  
$ G_X^{\text{loop}} (\bar p, p, x_n) $ 
is less useful. Here the accompanied Lorentz structure in the decomposition of Eq. \eqref{eqn:AxialMatrixElement} is proportional to $x_n^{\mu}$ which depends on the summation vector $\vec{n}$. The functions from Eq. \eqref{def:GAPTloop} can further be split into the different types of one-loop contributions,
\begin{equation}\begin{aligned}[b]
&G^{\text{loop}}_A(\bar{p},p)=\frac{g_A}{f^2}\,\Big\{J^A_{\pi}(\bar{p},p) + J_{ N \pi }^A(\bar{p},p) +  J_{\pi N}^A(\bar{p},p)\Big\}
+ \frac{f_S}{3\,f^2}\,  \Big\{J_{ \Delta \pi }^A(\bar{p},p) +  J_{\pi \Delta}^A (\bar{p},p)\Big\} 
\\ &   \hspace{3em} 
+ \frac{g_A^3}{4\,f^2}\,J^A_{N\pi N}(\bar{p},p) 
+ \frac{2\,g_A\,f_S}{3\,f^2}\,\Big\{J^A_{\Delta\pi N }(\bar{p},p)+ J^A_{N \pi \Delta }(\bar{p},p)\Big\}
+ \frac{5\,h_A\,f_S^2}{9\,f^2}\,J^A_{\Delta \pi \Delta}(\bar{p},p) + {\mathcal O} \big( Q^4\big) \,, \\[1em]
%%%
% GP
%%%
&\frac{t-m_\pi^2}{4\,M_N^2}\,G^{\text{loop}}_P(\ppf,\ppi) = \frac{g_A}{f^2}\,  \Big\{J^P_{\pi}(\ppf,\ppi) + J_{ N \pi }^P(\ppf,\ppi) +  J_{\pi N}^P(\ppf,\ppi) \Big\}
+ \frac{f_S}{3\,f^2}\,  \Big\{ J_{ \Delta \pi }^P(\ppf,\ppi) + J_{\pi \Delta}^P(\ppf,\ppi)\Big\}
\\ &   \hspace{3em}  
+ \frac{g_A^3}{4\,f^2}\,J^P_{N\pi N}(\ppf,\ppi) 
+ \frac{2\,g_A\,f_S}{3\,f^2}\,\Big\{J^P_{\Delta \pi N }(\ppf,\ppi)+ J^P_{N \pi \Delta }(\ppf,\ppi)\Big\}
+ \frac{5\,h_A\,f_S^2}{9\,f^2}\,J^P_{\Delta \pi \Delta}(\ppf,\ppi)
+ {\mathcal O} \big( Q^4\big) \,,
\\[1em]
%%%
%
%%%
& \frac{t}{2\,M_N^2}\,G^{\text{loop}}_T(\ppf,\ppi) = 
\frac{g_A}{f^2}\,  \Big\{J_{\pi}^T(\bar{p},p) + J_{ N \pi }^T(\ppf,\ppi)  + J_{\pi N}^T(\ppf,\ppi)\Big\}
+\, \frac{f_S}{3\,f^2}\,  \Big\{J_{ \Delta \pi }^T(\ppf,\ppi) + J_{\pi \Delta}^T(\ppf,\ppi)\Big\}
\\ &   \hspace{3em}  
+ \frac{g_A^3}{4\,f^2}\,J^T_{N\pi N}(\ppf,\ppi)
+ \frac{2\,g_A\,f_S}{3\,f^2}\,\Big\{J^T_{\Delta \pi N }(\ppf,\ppi)+ J^T_{N \pi \Delta }(\ppf,\ppi)\Big\} 
+ \frac{5\,h_A\,f_S^2}{9\,f^2}\,J^T_{\Delta \pi \Delta}(\ppf,\ppi)
+ {\mathcal O} \big( Q^5\big) \,,
\label{def:GloopToJs}
\end{aligned}\end{equation}
which may be viewed as a definition of the functions $J^{A}_{\ldots}(\bar{p},p),\,J^{P}_{\ldots}(\bar{p},p)$, and $J^{T}_{\ldots}(\bar{p},p)$. The loop functions $J^A_{\ldots}(\bar{p},p)$ of the axial-vector form factor were already determined by us in Ref. \cite{Hermsen:2025vds}.
For the loop function $G_X(\ppf,\ppi,x_n)$ a similar 
decomposition is useful with
\begin{equation}\begin{aligned}[b]
&G^{\text{loop}}_X(\ppf,\ppi, x_n) = 
\frac{g_A}{f^2}\,  \Big\{J^X_{\pi}(\ppf,\ppi,x_n) + J_{ N \pi }^X(\ppf,\ppi, x_n) + J_{\pi N}^X(\ppf,\ppi, x_n)\Big\}
+ \frac{f_S}{3\,f^2}\,  \Big\{J_{ \Delta \pi }^X(\ppf,\ppi, x_n) + J_{\pi \Delta}^X(\ppf,\ppi, x_n)\Big\}
\\ &   \hspace{2em} 
+ \frac{g_A^3}{4\,f^2}\,J^X_{N\pi N}(\ppf,\ppi, x_n)
+ \frac{2\,g_A\,f_S}{3\,f^2}\,\Big\{J^X_{\Delta \pi N }(\ppf,\ppi, x_n)+ J^X_{N \pi \Delta }(\ppf,\ppi, x_n)\Big\}
+ \, \frac{5\,h_A\,f_S^2}{9\,f^2}\,J^X_{\Delta \pi \Delta}(\ppf,\ppi, x_n)
+ {\mathcal O} \big( Q^4\big) \,.
\label{def:GXloopToJXs}
\end{aligned}\end{equation}

It is useful to recall 
our novel minimal set of in-box functions \cite{Hermsen:2025vds}, to be used in a decomposition of the in-box form factors. 
As an example, we give the renormalized $N\pi$ bubble basis elements,
\begin{equation}\begin{aligned}[b]
&\bar I^{(\bar a,\, h , \,a)}_{N \pi}(\vec{ \bar p}, \vec{ p}\,) = \delta_{\bar a,0}\,\delta_{ h,0}\,\delta_{ a,0}\, \bar I_{ N \pi}
+ 
\sum^{\vec n \neq 0}_{\vec{n}\in\mathbb{Z}^3} \,X_{N \pi}^{(\bar{a} + 2\,h +  a)}[\,\vec{\bar{p}}, \, \vec{x}_n\,]\,
 \big(\vec{\bar p}\cdot \vec{x}_n \big)^{\bar{a}}\, \big(\vec{x}_n^{\,2}\big)^h\, \big(\vec{p}\cdot\vec{x}_n \big)^a
 \,, \\
&\bar{I}_{N\pi} = \frac{-\,2}{16\,\pi^2} - \frac{1}{16\,\pi^2}\,\int_0^1\,\dd{z} \,\log\,\frac{F_{N\pi}(z)}{M_N^2} \sim Q
\,,  \\
&X_{N\pi}^{(n_x)}[\,\vec{\bar{p}},\, \vec{x}_n\,] =  \frac{(-1)^{\lfloor (1-n_x ) / 2 \rfloor}}{8\,\pi^2}\,\int_0^1 \dd{z}\,K_{n_x}\big[\, F_{N\pi}(z),\, \vert \vec{x}_n \vert \big]\,
 {\text{cs} }_{n_x}\big( z\,\vec{ \bar p}\cdot \vec{x}_n \big)
 \,,  \\
&F_{N\pi}(z) = z\,M_N^2 + (1-z)\,m_\pi^2- (1-z)\,z\,\bar p^2\,,\qquad
K_{n} [\,m^2, \,x ] = \bigg( \frac{m }{x} \bigg)^n\,K_n (\,m \, x\,)\,,
\label{eqn:NpiBasisElements}
\end{aligned}\end{equation}
where $p^2_0= M_N^2+\vec{ p\,}^2$, $\bar{p}^2_0= M_N^2+\vec{\bar  p}\,^2$,   and $K_n(m\,x)$ denotes the modified Bessel function, and we use the short-hand notation ${\text{cs}}_n(x)=\cos (x)$ for $n$ even, but ${\text{cs}}_n(x)=\sin (x)$ for $n$ odd; the brackets $\lfloor x \rfloor := \operatorname{max}\{k\in\mathbb{Z}\,\vert\,k\leq z\}$ denote the floor function. The complete set of the renormalized in-box basis functions is recalled in Appendix \ref{app:basisElements} where in our convention renormalized quantities always come with an extra bar.  \\\par 

We continue with the loop functions $J^{P}_{\ldots}(\bar{p},p)$ and $J^{T}_{\ldots}(\bar{p},p)$, properly  written in our in-box basis. To be able to achieve this, our previous reduction scheme from Ref. \cite{Hermsen:2025vds} needs to be extended. We delegate this technical development to Appendix \ref{app:redInBox} and focus here on the final results.  
Following the strategy of our previous works \cite{Hermsen:2025vds,Hermsen:2024eth,Sauerwein:2021jxb,Lutz:2020dfi}, a kinematical  coefficient function in front of an in-box basis element is expanded in chiral powers with 
\begin{align}
t\sim  m_\pi^2 \sim Q^2 \,,\qquad \delta= M_\Delta - M_N \,\Big( 1+ \frac{\Delta}{M} \Big) \sim Q^2 \,.
\label{def-counting-rule}
\end{align}
In contrast, the on-shell hadron masses $m_\pi,\,M_N,\, M_{\Delta}$, and the renormalized in-box basis elements $\bar{I}^{(\bar{a},h,a)}_{\ldots}$ are kept unexpanded. They are assigned a chiral power, which is determined by their leading behavior in a strict chiral expansion. \\\par  

We emphasize that after the sum over $\vec{n}$ is performed the only kinematic dependence of the form factors and their loop functions is given by the four-momenta $\ppf$ and $\ppi$. While the kinematic terms may depend on $t$, the basis elements are functions of the three-momenta $\kf$ and $\ki$, as shown for instance in Eq.~\eqref{eqn:NpiBasisElements}. The dependence on the box length $L$ is implied through the space-like components of the momenta, given as $\ki = (2\,\pi/L)\,\vec{m}$ for $\vec{m}\in\mathbb{Z}^3$. Thus we write for example $G_A(\ppf,\ppi)$ for the form factors, but $\bar I^{(\bar a,\, h , \,a)}_{N \pi}(\vec{ \bar p}, \vec{ p}\,)$ for the basis elements.   \\\par 

We are now prepared to present the renormalized loop amplitudes. We start with the renormalized tadpole contributions
\begin{equation}
\bar{J}_\pi^{A}(\bar p, p)  = -\,\bar{I}_\pi^{(0,0,0)}(\kf,\ki)\,,\hspace{3em}
\bar  J^P_\pi(\bar p, p) = \frac{1}{3}\,\bar I^{(0 , 0,  0)}_{\pi}(\vec{ \bar p},\, \vec{ p}\,) \,,\hspace{3em} \bar{J}^T_\pi(\bar p, p) = 0\,,\hspace{3em}
\bar{J}^{X}_\pi(\bar{p}, p, x_n) = 0\,,
\label{res:tadpoles}
\end{equation}
where we recover the infinite-volume limit from our previous work \cite{Hermsen:2024eth}. While we display in Eq. \eqref{res:tadpoles} the momentum dependence explicitly, we will not do so  in the following, as long as there is no confusion possible. \\\par 
We turn to the renormalized $N\pi$  and $\pi N$ bubble contributions:
\begin{equation}\begin{aligned}[b]
&\bar{J}^A_{N\pi} + \bar{J}^A_{\pi N} = 
-\,m_\pi^2\,\Big(\renLQ[N \pi]{0}{0}{0} + \renQR[\pi N]{0}{0}{0}\Big) 
+ \renLQ[N \pi]{1}{0}{0}
- \renLQ[N \pi]{0}{0}{1} 
- \renQR[\pi N]{1}{0}{0}
+ \renQR[\pi N]{0}{0}{1} 
\\ &\hspace{3em}
+\frac{4}{3}\,\big(g_S - 2\,g_T\big)\,M_N\,m_\pi^2\,\Big(\renLQ[N \pi]{0}{0}{0} + \renQR[\pi N]{0}{0}{0}\Big) 
-\frac{2}{3}\,\big(g_S+g_T\big)\,M_N\,\Big(\renLQ[N \pi]{0}{1}{0} + \renQR[\pi N]{0}{1}{0}\Big) 
\\ &\hspace{3em}
-2\,\big(g_S+g_T\big)\,M_N\,\Big(\renSLQ[N \pi]{2} + \renSQR[\pi N]{2}\Big)
+ \, 8\,g_F\,M_N\,\Big(
\renLQ[N \pi]{1}{0}{0} - \renLQ[N\pi]{0}{0}{1}
- \renQR[\pi N]{1}{0}{0}+\renQR[\pi N]{0}{0}{1} 
\Big) + \mathcal{O}(Q^4)\,, \\[.3em]
%%%
% induzierter pseudoskalar
%%%
&\bar{J}^P_{N\pi} + \bar{J}^P_{\pi N} = 
m_\pi^2\,\Big(\renLQ[N \pi]{0}{0}{0} + \renQR[\pi N]{0}{0}{0}\Big) 
-\frac{4}{3}\,\big(g_S - 2\,g_T\big)\,M_N\,m_\pi^2\,\Big(\renLQ[N \pi]{0}{0}{0} + \renQR[\pi N]{0}{0}{0}\Big) 
 \\ &\hspace{3em}
+\frac{2}{3}\,\big(g_S+g_T\big)\,M_N\,\bigg[1 - 3\,\frac{m_\pi^2}{t}\bigg]\,\Big(\renLQ[N \pi]{0}{1}{0} + \renQR[\pi N]{0}{1}{0}\Big) 
-\,2\,\big(g_S+g_T\big)\,M_N\,\bigg[1 + \frac{m_\pi^2}{t}\bigg]\,\Big(\renSLQ[N \pi]{2} + \renSQR[\pi N]{2}\Big)
\\ &\hspace{3em}
+ 16\,b_{\chi}\,M_N\,\frac{B_0\,m}{t}\,\Big(\renLQ[N\pi]{1}{0}{0} - \renLQ[N\pi]{0}{0}{1} - \renQR[\pi N]{1}{0}{0} + \renQR[\pi N]{0}{0}{1}  \Big)
+ \mathcal{O}(Q^4)\
\,, \\[.3em]
%%%
% indzuzierter Tensor
%%%
&\bar{J}^T_{N\pi} + \bar{J}^T_{\pi N} = 
2\,\Big(\renLQ[N\pi]{1}{0}{0} - \renLQ[N\pi]{0}{0}{1} + \renQR[\pi N]{1}{0}{0} - \renQR[\pi N]{0}{0}{1}\Big) + \mathcal{O}(Q^5)\,, \\
%%%
% X Form factor
%%%
&\bar{J}^X_{N\pi} + \bar{J}^X_{\pi N} = \big(1+8\,g_F\,M_N\big)\,\Big(X_{N\pi}^{(1)} - X_{\pi N}^{(1)}\Big) 
-\,4\,\big(g_S+g_T\big)\,\big(\vec{q}\cdot \vec{x}_n\big)\,\frac{M_N}{t}\Big(X_{N\pi}^{(2)} + X_{\pi N}^{(2)}\Big) + \mathcal{O}(Q^4)\,.
\label{res:Npibubbles}
\end{aligned} \end{equation}
For $\cdots\in\{N\pi,\pi N\}$ we recall the definition of the basis element $\bar{S}_{\ldots}^{(2)}$ and its limit expression  at $t=0$,
\begin{equation}
\bar{S}_{\ldots}^{(2)} = \frac{\renQR[\ldots]{2}{0}{0} - 2\,\renQR[\ldots]{1}{0}{1} + \renQR[\ldots]{0}{0}{2}}{t}\,,\qquad
\bar{S}_{\ldots}^{(2)}\Big\vert_{t=0} =  -\,\frac{1}{3}\,\bar{I}_{\ldots}^{(0,1,0)} - \frac{1}{3\,M_N^2}\bar{I}_{\ldots}^{(1,0,1)} \,.
\label{eqn:defS2}
\end{equation}

Some comments on Eq. \eqref{res:Npibubbles} are in order. While we observe that $\bar{J}^{P}_{N\pi} + \bar{J}^{P}_{\pi N}$ has no $t=0$ singularity in the infinite-volume limit, it exhibits one in the finite-box case. To our best knowledge  lattice works e.g. Refs. \cite{Capitani:2017qpc,RQCD:2019jai,Bali:2023sdi,Alexandrou:2023qbg} do not determine $G_P(\ppf,\ppi)$ at equal three-momenta $\kf=\ki$, so we do not consider this an issue at this stage. 
(We point out a misprint in Ref. \cite{Hermsen:2024eth}.  The first term in the induced pseudoscalar loop contributions should come with an extra factor two \cite{Hermsen:2024eth}.) 
%Since this corrections is numerical small, we are confident, that the fit in our previous work \cite{Hermsen:2024eth} remains qualitatively unchanged and therefore the conclusions and observations therein remain true. 
Furthermore, we note in $\bar{J}^P_{N\pi} + \bar{J}^P_{\pi N}$ that the term proportional to $b_{\chi}$ only contributes in the finite volume. Thus we have now identified two LECs, $b_{\chi}$ and $g_F$, which only contribute in the finite volume. To our knowledge, this was not reported before in the literature, as only the axial charge was studied in the finite volume,  \cite{Beane:2004rf,Khan:2006de} and not the momentum dependence of the form factors. In closing, we want to emphasize that the loop contributions to the induced pseudotensor form factor are zero in the infinite volume. This is because the infinite-volume parts in the  integrals $\renQR[\ldots]{\aaf}{h}{\aai}$ with $\aaf+h+\aai>0$ are always zero.      \\\par 

We turn to the $N\pi N$-triangles,
\begin{align}
&\bar{J}^A_{N\pi N} = \frac{1}{3}\,\bigg\{\,\renQ[\pi]{0}{0}{0} + m_\pi^2\,\Big(\renLQ[N\pi ]{0}{0}{0} + \renQR[\pi N]{0}{0}{0}\Big)
+2\,\Big(\renLQ[N\pi]{1}{0}{0} - \renLQ[N\pi]{0}{0}{1} 
-\,\renQR[\pi N]{1}{0}{0} + \renQR[\pi N]{0}{0}{1} \Big)\bigg\}
 \notag \\ &\hspace{1em}
-\, \frac{4}{3}\,M_N^2\,\bigg\{\,m_\pi^2\,\renLQR[N-\pi-N]{0}{0}{0}{0}{0} 
+ \renLQR[N-\pi-N]{0}{0}{0}{1}{0}
+3\,\renSLQR[N-\pi-N]{0}{0}{2}
+\, 2\,\Big(\renLQR[N-\pi-N]{1}{0}{1}{0}{0} - \renLQR[N-\pi-N]{1}{0}{0}{0}{1}
-\, \renLQR[N-\pi-N]{0}{1}{1}{0}{0} 
+ \renLQR[N-\pi-N]{0}{1}{0}{0}{1} \Big)
\notag \\ &\hspace{5em}
+ t\,\Big(\renLQR[N-\pi-N]{2}{0}{0}{0}{0} + \renLQR[N-\pi-N]{0}{2}{0}{0}{0}\Big)\bigg\} + \mathcal{O}(Q^4)\,,  \notag \\[.3em]
%%%
% JP Beitrag
%%%
&\bar{J}^P_{N\pi N} = -\, \frac{1}{3}\,\bigg\{\renQ[\pi]{0}{0}{0} + m_\pi^2\,\Big(\renLQ[N\pi ]{0}{0}{0} + \renQR[\pi N]{0}{0}{0}\Big)
-\,\bigg[1 + 3\,\frac{m_\pi^2}{t}\bigg]\,\Big(\renLQ[N\pi]{1}{0}{0} - \renLQ[N\pi]{0}{0}{1} 
- \renQR[\pi N]{1}{0}{0} + \renQR[\pi N]{0}{0}{1} \Big)\bigg\}
 \notag \\ &\hspace{1em}
+\,\frac{4}{3}\,M_N^2\,\bigg\{m_\pi^2\,\renLQR[N-\pi-N]{0}{0}{0}{0}{0} 
+\bigg[1 - 3\,\frac{m_\pi^2}{t}\bigg]\,\renLQR[N-\pi-N]{0}{0}{0}{1}{0}
- 3\,\bigg[1 + \frac{m_\pi^2}{t}\bigg]\,\renSLQR[N-\pi-N]{0}{0}{2}
\notag \\ &\hspace{5em}
-\bigg[1 + 3\,\frac{m_\pi^2}{t}\bigg]\,\Big(\renLQR[N-\pi-N]{1}{0}{1}{0}{0} - \renLQR[N-\pi-N]{1}{0}{0}{0}{1}
- \renLQR[N-\pi-N]{0}{1}{1}{0}{0} + \renLQR[N-\pi-N]{0}{1}{0}{0}{1} \Big)
+\Big[t-3\,m_\pi^2\Big]\,\Big(\renLQR[N-\pi-N]{2}{0}{0}{0}{0} 
+ \renLQR[N-\pi-N]{0}{2}{0}{0}{0}\Big)
\bigg\}+ \mathcal{O}\big(Q^4\big)\,,
\notag \\[.3em]
%%%
% JT Beitrag
%%%
&\bar{J}^T_{N\pi N} = 4\,m_\pi^2\,\Big(\renLQR[N-\pi-N]{0}{0}{1}{0}{0} - \renLQR[N-\pi-N]{0}{0}{0}{0}{1}\Big)
- 4\,\Big(\renLQR[N-\pi-N]{0}{0}{2}{0}{0} - \renLQR[N-\pi-N]{0}{0}{0}{0}{2}\Big)
+2\,t\,m_\pi^2\,\Big(\renLQR[N-\pi-N]{1}{0}{0}{0}{0} - \renLQR[N-\pi-N]{0}{1}{0}{0}{0}\Big)
\notag \\ &\hspace{5em}
-\, 2\,t\, \Big(\renLQR[N-\pi-N]{1}{0}{1}{0}{0} + \renLQR[N-\pi-N]{1}{0}{0}{0}{1} - \renLQR[N-\pi-N]{0}{1}{1}{0}{0} - \renLQR[N-\pi-N]{0}{1}{0}{0}{1}\Big) + \mathcal{O}\big(Q^5\big)\,,\notag \\[.3em]
%%%
%
%%%
&\bar{J}^X_{N\pi N} = 
X_{N\pi}^{(1)} - X_{\pi N}^{(1)}
-4\,M_N^2\,\Big(X_{N^1\pi N^0}^{(1)} - X_{N^0\pi N^1}^{(1)}\Big)
-8\,\big(\vec{q}\cdot\vec{x}_n\big)\,\frac{M_N^2}{t}\,X_{N^0\pi N^0}^{(2)}
+\mathcal{O}(Q^4)\,,
\label{res:NpiNTri}
\end{align}
 where we recalled the axial-vector contributions for convenience.  The remaining bubble and triangle contributions involving one or two decuplet propagators  are given in Appendix \ref{res:JPJT}.  
\clearpage

\section{Relevance of the complete set of in-box form factors} \label{sec:complete}
Eq.~\eqref{eqn:axialWard:chiralLimit} gives the in-box axial Ward identity in the chiral limit. Away from the chiral limit, for non-vanishing quark masses, this relation is generalized to
\begin{equation}
\sum_{\vec{n}\in\mathbb{Z}^3}\bigg[(m_\pi^2-t)\,G_A(\ppf,\ppi,\xn) - t\,\frac{t-m_\pi^2}{4\,M_N^2}\,G_P(\ppf,\ppi,\xn)
 + (m_\pi^2-t)\,\big(q\cdot \xn\big)\,G_{X}(\ppf,\ppi,\xn)\bigg] = m_\pi^2\,\sum_{\vec{n} \in\mathbb{Z}^3}\,G_{PS}(\ppf,\ppi,\xn)\,.
\label{chap:PPFF:eqn-AxialWardbox} 
\end{equation}
On the right-hand side, we introduced the pseudoscalar form factor, which we will discuss in to more detail shortly. The masses $m_\pi$ and $M_N$ appearing in Eq.~\eqref{chap:PPFF:eqn-AxialWardbox} are the on-shell in-box masses of the respective hadrons.  While in the infinite-volume limit of the axial Ward identity, which is given with $G_X(\ppf,\ppi,\xn)~\to~0$, only three form factors contribute, in the finite-box case the $X$ form factor also contributes, while the induced pseudotensor one does not. This implies that $G_T$ is independent of all four form factors present in Eq.~\eqref{chap:PPFF:eqn-AxialWardbox}. This supports our findings from Appendix~\ref{app:decompDirac}, where we argue that $G_T$ cannot be absorbed into $G_P$. Therefore, the induced pseudotensor form factor is relevant for the determination of the axial-vector matrix element in Eq.~\eqref{eqn:AxialMatrixElement}, but it does not contribute to the axial Ward identity. This stands in contrast to the in-box form factor $G_X$, which is required in both cases, the decomposition of the matrix element and the axial Ward identity. An explicit determination of the pseudoscalar form factor then allows a verification of the renormalized results of the three form factors $G_A,\,G_P,\,G_{PS}$ and the contraction $(q\cdot\xn)\,G_{X}$.  \\\par 

In lattice works in which not only the axial-vector matrix element, but also the pseudoscalar matrix element is determined, the axial Ward identity is also used to verify results. To connect to these works, it is convenient to define \cite{Bali:2018qus,Alexandrou:2023qbg} the ratio 
\begin{equation}
    r_{\text{PCAC}} = \frac{4\,M_N^2\,m_\pi^2\,G_{PS}(\ppf,\ppi) + t\,\big(t-m_\pi^2\big)\,G_P(\ppf,\ppi)}{(m_\pi^2-t)\,G_{A}(\ppf,\ppi)} \ , 
\end{equation}
 and rewrite Eq.~\eqref{chap:PPFF:eqn-AxialWardbox} as
\begin{equation}\begin{aligned}[b]
&r_{\text{PCAC}} - 1 = 
\frac{1}{G_{A}(\ppf,\ppi)}\,\sum_{\vec{n} \in\mathbb{Z}^3}\,\big(q\cdot \xn\big)\,G_{X}(\ppf,\ppi,\xn)\,.
\label{chap:PSFF:eqn-AxialWardRatio}
\end{aligned}\end{equation}
If the axial matrix element in the finite volume is decomposed into $G_A(q^2)$ and $G_P(q^2)$ only, as done e.g. in the lattice works \cite{RQCD:2019jai,Alexandrou:2023qbg,Bali:2018qus,Gupta:2017dwj}, the right-hand side of Eq.~\eqref{chap:PSFF:eqn-AxialWardRatio} vanishes and $r_{\text{PCAC}}=1$ needs to be satisfied. This is discussed for example in more detail in the review article~\cite{Gupta:2024qip}, where it is argued that up to discretizations errors $r_{\text{PCAC}}=1$ should hold for all lattice ensembles. While this constraint helped to identify excited-state contamination in the past, see e.g. Refs. \cite{Bali:2018qus,Jang:2019vkm}, lattice works, e.g. Refs. \cite{Alexandrou:2023qbg,RQCD:2019jai}, still report deviations in the small percentage range from $r_{\text{PCAC}}=1$. 
However, if the full set of in-box form factors is considered, it becomes clear from Eq.~\eqref{chap:PSFF:eqn-AxialWardRatio} that the ratio $r_{\text{PCAC}}$ can only be 1 in the infinite volume and not in a finite box. Thus, the additional in-box form factor $G_X$ could play a part to resolve the deviation of $r_{\text{PCAC}}$ from 1 in lattice simulations. Consequently, for precise results, the $X$ form factor should be included in the decomposition in the matrix element of the axial-vector current between two nucleons.\\\par 

After emphasizing the relevance of our set of in-box form factors, we will now continue with the verification of the axial Ward in terms of our renormalized form factors results.  For that, we first need to introduced the pseudoscalar form factor $G_{PS}(\ppf,\ppi)$, which is defined through the matrix element of the QCD pseudoscalar density $P_i(x) = i\,\bar{q}(x)\,\gamma_5\,\tau_{i}\,q(x)$ \cite{Gasser:1983yg} between two nucleon states:
\begin{equation}\begin{aligned}[b]
&\langle N(\ppf)\,\vert\,m\,P_i(0)\,\vert\,N(\ppi)\rangle = \bar{u}(\ppf)\,F_{PS}(\ppf,\ppi)\,\tau_{i}\,u(\ppi)\,,\\
&F_{PS}(\ppf,\ppi) = \frac{i \,m_\pi^2\,M_N}{m_\pi^2-t}\,G_{PS}(\ppf,\ppi)\,\gamma_{5}\,, \quad 
G_{PS}(\ppf,\ppi) = \sum_{\vec{n} \in\mathbb{Z}^3}\,G_{PS}(\ppf,\ppi,\xn)\,.
\label{chap:PseudoFF-startDecomp}
\end{aligned}\end{equation}
We emphasize that the decomposition in Eq.~\eqref{chap:PseudoFF-startDecomp} holds in the infinite volume but also in a finite box. No additional in-box form factors are needed, as it was the case for the axial-vector matrix element in Eq.~\eqref{eqn:AxialMatrixElement}. Due to $G$-parity we have $G_{PS}(\ppf,\ppi) = G_{PS}(\ppi,\ppf)$. The decomposition of the matrix element in Eq.~\eqref{chap:PseudoFF-startDecomp} and therefore the definition of $G_{PS}(\ppf,\ppi)$ is ambiguous in the literature \cite{Chen:2012nx,Alexandrou:2023qbg,Schindler:2006it,Fuchs:2003vw,Goity:1999by,Bernard:1992qa}. For example, the convention in Refs.~\cite{Schindler:2006it,Fuchs:2003vw} is recovered by replacing $M_N$ with $f_\pi$ in Eq.~\eqref{chap:PseudoFF-startDecomp}. We do not choose this parametrization, since we do not want to factorize the physical value of the  pion-decay coupling $f_\pi$. Instead, we factorized the on-shell in-box nucleon mass, since we keep the hadron masses in our renormalization scheme and do not expand them. Thus, $M_N$ naturally appears in the amplitude $F_{PS}(\ppf,\ppi)$ and can therefore be factorized.  \\\par  

We explicitly derived the amplitudes contributing to the pseudoscalar form factor from the chiral Lagrangian density, as documented in Appendix~\ref{app:resultsPseudoscalarFF}. Here we only define the tree and one-loop functions of the pseudoscalar form factor,
\begin{align}
&G_{PS}(\ppf,\ppi) = G_{PS}^{\text{tree}}(\ppf,\ppi) + G_{PS}^{\text{loop}}(\ppf,\ppi)\,, \notag \\
&G^{\text{loop}}_{PS}(\bar{p},p)=\frac{g_A}{f^2}\,\Big\{J^{PS}_{\pi}(\bar{p},p) + J_{ N \pi }^{PS}(\bar{p},p) +  J_{\pi N}^{PS}(\bar{p},p)\Big\}
+ \frac{f_S}{3\,f^2}\,  \Big\{J_{ \Delta \pi }^{PS}(\bar{p},p) +  J_{\pi \Delta}^{PS} (\bar{p},p)\Big\} 
\notag\\ &   \hspace{7em} 
+ \frac{g_A^3}{4\,f^2}\,J^{PS}_{N\pi N}(\bar{p},p) 
+ \frac{2\,g_A\,f_S}{3\,f^2}\,\Big\{J^{PS}_{\Delta\pi N }(\bar{p},p)+ J^{PS}_{N \pi \Delta }(\bar{p},p)\Big\}
+ \frac{5\,h_A\,f_S^2}{9\,f^2}\,J^{PS}_{\Delta \pi \Delta}(\bar{p},p) + {\mathcal O} \big( Q^4\big)\,,
\label{def:GPSloopToJPS}
\end{align}
where their renormalized forms are documented in Eqs.~\eqref{app:PseudoScalar:bubbles} and
\eqref{app:PseudoScalar:triangles}. For a more detailed discussion of the pseudoscalar form factor, we refer to Ref.~\cite{felix:PhD}.  \\\par

As we have explicitly determined the pseudoscalar form factor of the nucleon, we can validate our results rather then using Eq.~\eqref{chap:PPFF:eqn-AxialWardbox} to express one of the four form factors in terms of the other. First we recall, that at zero momentum transfer, the decomposition of the axial-vector matrix element is given $G_A(\ppf,\ppi)$ only, which follows from 
\begin{equation}
G_A(\ppf,\ppi,\xn)\Big\vert_{t=0} = G_{PS}(\ppf,\ppi,\xn)\Big\vert_{t=0}\,.
\label{chap:PSFF:WardID-zeroMomTransfer}
\end{equation}
As we discussed in the previous section terms proportional to $1/t^2$ appear in the form factor $G_P(\ppf,\ppi)$. Comparing Eq.~\eqref{chap:PSFF:WardID-zeroMomTransfer}  to Eq.~\eqref{chap:PPFF:eqn-AxialWardbox} thus enforces the relation
\begin{equation}
t\,\frac{t-m_\pi^2}{4\,M_N^2}\,G_P(\ppf,\ppi,\xn)\bigg\vert_{t=0} =  m_\pi^2\,\big(q\cdot \xn\big)\,G_{X}(\ppf,\ppi,\xn)\bigg\vert_{t=0}\,.
\label{chap:AWid:eqn-ratioFiniteBox}
\end{equation}
For the tree and $\pi$-tadpole contributions this relation is easily checked as the corresponding terms in the induced pseudoscalar form factor $G_P(\ppf,\ppi)$ does not have any singularity and in the $G_{X}(\ppf,\ppi,\xn)$ form factor the tree-level and $\pi$-tadpole contributions are zero. The case of the remaining loop contributions will be discussed shortly. \\\par  

We now turn to the case of non-vanishing momentum transfer. We note that Eq.~\eqref{chap:PPFF:eqn-AxialWardbox} needs to hold for any LEC combination in the form factors separately. Therefore, we start with the tree-level and $\pi$-tadpole contributions, since they share the same LEC combination. First, we recall from Eqs.~\eqref{res-tree} and \eqref{res:tadpoles} that $G_X^{\text{tree}}(\ppf,\ppi) = 0 = J_\pi^{X}(\ppf,\ppi)$. Then we replace the form factors in Eq.~\eqref{chap:PPFF:eqn-AxialWardbox} with their respective tree-level results from Eq.~\eqref{res-tree}, and obtain
\begin{align}
(m_\pi^2-t)\,G^{\text{tree}}_{A}(\ppf,\ppi) - t\,\bigg[\frac{t-m_\pi^2}{4\,M_N^2}\,G_P^{\text{tree}}(\ppf,\ppi)\bigg]  - m_\pi^2\,G^{\text{tree}}_{PS}(\ppf,\ppi) &= 
\frac{g_A}{f}\,\big(
m_\pi^2-t\big)\big(f\,\big(2-\sqrt{Z_\pi}\big) - f_{\pi}\big)\,.
\label{chap:AxialWardFV:WardIDTree}
\end{align}
We note that the tree-level contributions in the three form factors proportional to  the counterterms $g_{\chi}^{\pm},g_R$ fulfill the axial Ward identity, since they vanish in the difference. The same holds for the terms proportional to the nucleon wave renormalization function $Z_N$, for which we need its explicit expression to confirm the axial Ward identity. However, the mesonic quantities $\sqrt{Z_\pi}$ and $f_\pi$ do not vanish. This does not imply that the axial Ward identity is broken, since we find for the $\pi$-tadpole contributions
\begin{align}
\frac{g_A}{f^2}\,\Big\{(m_\pi^2-t)\,J^{A}_{\pi}(\ppf,\ppi) - t\,J^{P}_{\pi}(\ppf,\ppi) - m_\pi^2\,J^{PS}_{\pi}(\ppf,\ppi)\Big\}
=
-\,\frac{2\,g_A\,}{3\,f^2}\,\big(m_\pi^2-t\big)\,\renQ[\pi]{0}{0}{0}\,.
\label{chap:AxialWardFV:WardIDTreeTadpole}
\end{align}
Because only the sum of the tree and $\pi$-tadpole contributions needs to vanish, Eqs.~\eqref{chap:AxialWardFV:WardIDTree} and \eqref{chap:AxialWardFV:WardIDTreeTadpole} imply
\begin{equation}
    f\,\big(2-\sqrt{Z_\pi}\big) - f_{\pi} = \frac{2}{3}\,\basisQ[\pi]{0}{0}{0}\,.
    \label{eqn:mesonicToTadpole}
\end{equation}
By using the explicit in-box results for $\sqrt{Z_\pi}$ and $f_\pi$, which we recall in Eq.~\eqref{app:PseudoScalar:mesonicObservables}, we can confirm Eq.~\eqref{eqn:mesonicToTadpole} explicitly. This constitutes a powerful verification of our results not only for the baryonic form factors but also for the mesonic quantities. We emphasize that we did not observe this in our previous work \cite{Hermsen:2024eth}, where we considered the chiral limit in the infinite volume only, because $\renQ[\pi]{0}{0}{0}$ vanishes in the case of zero quark masses.     \\\par

To verify the axial Ward identity for the remaining loop contributions, it is convenient to introduce the loop functions
\begin{equation}
J^{X_q}_{\ldots}(\bar{p}, p) = \sum_{n\in\mathbb{Z}^3} (q\cdot x_n)\,J^X_{\ldots}(\bar{p},p,x_n)\,.
\label{def:JXq}
\end{equation}
With the already derived loop functions $J^{X}_{\ldots}(\ppf,\ppi,\xn)$ from Eqs.  \eqref{res:Npibubbles}, \eqref{res:NpiNTri}, \eqref{res:DpiBub},
\eqref{res:DpiNTri}, and \eqref{res:DpiDTri}, it is then straightforward to find
\begin{align}
&\bar{J}^{X_q}_{\pi} = 0\,,\notag \\[.3em]
%%%
&\bar{J}_{N\pi}^{X_q} + \bar{J}_{\pi N}^{X_q} = 
-\,\big(1+8\,g_F\,M_N\big)\,\Big(\renLQ[N\pi]{1}{0}{0} - \renLQ[N\pi]{0}{0}{1} 
- \renQR[\pi N]{1}{0}{0} + \renQR[\pi N]{0}{0}{1} \Big) + 4\,\big(g_S + g_T\big)\,M_N\,\Big(\renSLQ[N\pi]{2} + \renSQR[\pi N]{2}\Big) + \mathcal{O}\big(Q^4\big)\,,\notag \\[.3em]
%%%
&\bar{J}^{X_q}_{\Delta\pi} + \bar{J}^{X_q}_{\pi \Delta} = 
\frac{2}{3}\,M_N\,\Big(
5\,f_A^{+}\,r\,\alpX{1}{1}{1} - f_A^{-}\,r\,\alpX{2}{1}{1} + 4\,f_M\,\alpX{3}{1}{1} \Big)
\Big(\renLQ[\Delta\pi]{1}{0}{0} - \renLQ[\Delta\pi]{0}{0}{1} 
- \renQR[\pi \Delta]{1}{0}{0} + \renQR[\pi \Delta]{0}{0}{1} \Big)
\notag \\ &\hspace{6em}
-\,\frac{2}{3}\,M_N\,\Big(5\,f_A^{+}\,\alpX{1}{0}{4} - f_A^{-}\,\alpX{2}{0}{4}\Big)\Big(\renSLQ[\Delta\pi]{2} + \renSQR[\pi \Delta]{2}\Big) + \mathcal{O}(Q^4)\,,\notag \\[.3em]
%%%
% N pi N triangle
%%%
&\bar{J}^{X_q}_{N\pi N} = -\,\renLQ[N\pi]{1}{0}{0} + \renLQ[N\pi]{0}{0}{1} + \renQR[\pi N]{1}{0}{0} - \renQR[\pi N]{0}{0}{1}  
\notag \\ &\hspace{6em}
+ 4\,M_N^2\,\Big( \renLQR[N-\pi-N]{1}{0}{1}{0}{0} - \renLQR[N-\pi-N]{1}{0}{0}{0}{1} 
- \,\renLQR[N-\pi-N]{0}{1}{1}{0}{0} + \renLQR[N-\pi-N]{0}{1}{0}{0}{1}\Big) 
+8\,M_N^2\,\renSLQR[N-\pi-N]{0}{0}{2} + \mathcal{O}(Q^4)\,, \notag \\[.3em]
%%%
% Npi Delta Triangle 
%%%
& \bar{J}^{X_q}_{\Delta\pi N} +  \bar{J}^{X_q}_{N\pi\Delta}
= -\,\frac{2}{3}\,\Big(f_S\,\alpX{4}{0}{1} + 3\,f_E\,M_N\,r\,\alpX{8}{0}{1}\Big)\,\Big(\renLQ[\Delta\pi]{1}{0}{0} - \renLQ[\Delta\pi]{0}{0}{1} 
- \renQR[\pi \Delta]{1}{0}{0} + \renQR[\pi \Delta]{0}{0}{1} \Big)
\notag \\ &\hspace{6em}
- \frac{1}{3}\,f_S\,\bigg\{\alpX{4}{0}{4}\,\Big(\renSLQ[\Delta\pi]{2} + \renSQR[\pi\Delta]{2}\Big)
+ 4\,M_N^2\,\alpX{5}{0}{4}\,\Big(\renSLQR[\Delta-\pi-N]{0}{0}{2} + \renSLQR[N-\pi-\Delta]{0}{0}{2}\Big)\bigg\}
\notag \\ &\hspace{6em}
- \frac{4}{3}\,f_S\,M_N^2\,\alpX{6}{0}{1}\,\Big(
\renLQR[\Delta-\pi-N]{1}{0}{1}{0}{0} - \renLQR[\Delta-\pi-N]{1}{0}{0}{0}{1}
-\renLQR[N-\pi-\Delta]{0}{1}{1}{0}{0} + \renLQR[N-\pi-\Delta]{0}{1}{0}{0}{1}
\Big) + \mathcal{O}\big(Q^4\big)\,,\notag\\[.25em]
%%%
%
%%%
&\bar{J}^{X_q}_{\Delta\pi\Delta} = -\,\frac{1}{9}\,\alpX{9}{1}{1}\,\Big(
\renLQ[\Delta\pi]{1}{0}{0} - \renLQ[\Delta\pi]{0}{0}{1} - \renLQ[\pi\Delta]{1}{0}{0} + \renQR[\pi\Delta]{0}{0}{1} \Big) 
+\,\frac{2}{9}\,\bigg\{ \alpX{9}{0}{4}\,\Big(\renSLQ[\Delta\pi]{2} + \renSQR[\pi\Delta]{2}\Big) + 4\,M_N^2\,\alpX{10}{0}{4}\,\renSLQR[\Delta-\pi-\Delta]{0}{0}{2}\bigg\}
\notag \\ &\hspace{6em}
+ \frac{4}{9}\,M_N^2\,\alpX{11}{1}{1}\,\Big(\renLQR[\Delta-\pi-\Delta]{1}{0}{1}{0}{0} - \renLQR[\Delta-\pi-\Delta]{1}{0}{0}{0}{1} - \renLQR[\Delta-\pi-\Delta]{0}{1}{1}{0}{0} + \renLQR[\Delta-\pi-\Delta]{0}{1}{0}{0}{1} \Big)
+\mathcal{O}\big(Q^4\big) \, .
\label{res:JXq}
\end{align}
Due the summation in Eq. \eqref{def:JXq}, we can write the results in Eq. \eqref{res:JXq} in terms of our in-box basis elements. Then, the in-box axial Ward identity from Eq. \eqref{chap:PPFF:eqn-AxialWardbox} implies the following eight equations:
\begin{align}
\label{chap:AxialWardFV:WardIDLoops}
&(m_\pi^2-t)\,J^{A}_{N\pi}(\ppf,\ppi) - t\,J^{P}_{N\pi}(\ppf,\ppi) + (m_\pi^2-t)\,J^{X_q}_{N\pi}(\ppf,\ppi) - m_\pi^2\,J^{PS}_{N\pi}(\ppf,\ppi) = 0\,,\hspace{-.3em} \notag\\
&(m_\pi^2-t)\,J^{A}_{\pi N}(\ppf,\ppi) - t\,J^{P}_{\pi N}(\ppf,\ppi) + (m_\pi^2-t)\,J^{X_q}_{\pi N}(\ppf,\ppi) - m_\pi^2\,J^{PS}_{\pi N}(\ppf,\ppi) = 0\,,\notag \\
%%%
&(m_\pi^2-t)\,J^{A}_{\Delta\pi}(\ppf,\ppi) - t\,J^{P}_{\Delta\pi}(\ppf,\ppi) + (m_\pi^2-t)\,J^{X_q}_{\Delta\pi}(\ppf,\ppi) - m_\pi^2\,J^{PS}_{\Delta\pi}(\ppf,\ppi) = 0\,,\notag \\
%%%
&(m_\pi^2-t)\,J^{A}_{\pi\Delta}(\ppf,\ppi) - t\,J^{P}_{\pi\Delta}(\ppf,\ppi) + (m_\pi^2-t)\,J^{X_q}_{\pi\Delta}(\ppf,\ppi) - m_\pi^2\,J^{PS}_{\pi\Delta}(\ppf,\ppi) = 0\,,\notag \\
%%%
&(m_\pi^2-t)\,J^{A}_{N\pi N}(\ppf,\ppi) - t\,J^{P}_{N\pi N}(\ppf,\ppi) + (m_\pi^2-t)\,J^{X_q}_{N\pi N}(\ppf,\ppi) - m_\pi^2\,J^{PS}_{N\pi N}(\ppf,\ppi) = 0\,,\notag \\
%%%
&(m_\pi^2-t)\,J^{A}_{\Delta\pi N}(\ppf,\ppi) - t\,J^{P}_{\Delta\pi N}(\ppf,\ppi) + (m_\pi^2-t)\,J^{X_q}_{\Delta\pi N}(\ppf,\ppi) - m_\pi^2\,J^{PS}_{\Delta\pi N}(\ppf,\ppi) = 0\,,\notag \\
%%%
&(m_\pi^2-t)\,J^{A}_{N \pi \Delta}(\ppf,\ppi) - t\,J^{P}_{N \pi \Delta}(\ppf,\ppi) + (m_\pi^2-t)\,J^{X_q}_{N \pi \Delta}(\ppf,\ppi) - m_\pi^2\,J^{PS}_{N\pi\Delta}(\ppf,\ppi) = 0\,,\notag \\
%%%
&(m_\pi^2-t)\,J^{A}_{\Delta\pi\Delta}(\ppf,\ppi) - t\,J^{P}_{\Delta\pi\Delta}(\ppf,\ppi) + (m_\pi^2-t)\,J^{X_q}_{\Delta\pi\Delta}(\ppf,\ppi) - m_\pi^2\,J^{PS}_{\Delta\pi\Delta}(\ppf,\ppi) = 0\,.
\end{align}
For $t\neq 0$ it is straightforward to verify these equations by using the renormalized results from our previous work \cite{Hermsen:2025vds} and this work. However, as previously mentioned, for $t=0$ the matter is a bit more complicated, because the $J^{P}_{\ldots}(\ppf,\ppi)$ contributions can have $1/t^2$ terms. Thus, we explicitly checked that the limit
\begin{equation}
t\,J^P_{\ldots}(\ppf,\ppi)\Big\vert_{t=0} = m_{\pi}^2\, J^{X_q}_{\ldots}(\ppf,\ppi)\Big\vert_{t=0}\,,
\label{chap:AWid:eqn-ratioFiniteBox-loopVersion}
\end{equation}
implied from Eq.~\eqref{chap:AWid:eqn-ratioFiniteBox}, holds for our results. Most importantly for these relations are the integrals $\bar{S}_{\ldots}^{(2)}$ and their limit recalled in Eq.~\eqref{eqn:defS2}. Furthermore, Eq.~\eqref{chap:AWid:eqn-ratioFiniteBox-loopVersion} implies that in the chiral limit the induced pseudoscalar loop contribution do not contain any singularities int $t=0$, except the integral $\bar{S}_{\ldots}^{(2)}$. It is straightforward to check that our renormalized results fulfill this.   \\\par 

It is clear from the relation in Eq.~\eqref{chap:AxialWardFV:WardIDLoops} that Eq.~\eqref{chap:PSFF:eqn-AxialWardRatio} holds already on the level of the different loop contributions,
\begin{equation}
1 = \frac{m_\pi^2\,J^{PS}_{\ldots}(\ppf,\ppi) + t\,J^{P}_{\ldots}(\ppf,\ppi)}{(m_\pi^2-t)\,J^{A}_{\ldots}(\ppf,\ppi)} - \frac{J^{X_q}_{\ldots}(\ppf,\ppi)}{J^{A}_{\ldots}(\ppf,\ppi)}\,,
\label{chap:AxialWardFV:WardRatioLoops}
\end{equation}
with $\ldots\in\{L\pi, \pi R, L\pi R\}$ and $L,R\in\{N,\Delta\}$. \\\par 

So far we discussed and emphasized the relevance of the two additional in-box form factors. However, this raises the question whether the form factors can be determined, and if so, how? First, it is clear that the matrix element of the pseudoscalar density does not pose any problems, because we do not have any additional in-box form factors. For the axial-vector matrix element, the situation is more challenging. To explain this in more detail, we recall that in lattice simulations the axial-vector form factors are determined through the three-point function where a nucleon is created at the source, the axial-current operator $A^{\mu}$ is inserted at time $\tau$, and the nucleon is destroyed at the sink, see e.g. Ref.~\cite{RQCD:2019jai}:
\begin{equation}
    C^{(\mu,\nu)}_{t,\tau}(\kf,\ki) = a^6\,\Gamma^{\nu}_{\alpha\bar{\alpha}}\,\sum_{\vec{x},\vec{y}}\,e^{-i\,\kf\cdot \vec{x} + i\,(\kf-\ki)\,\vec{y}}\big\langle N^{\bar{\alpha}}(\vec{x},t)\,A^{\mu}\,(\vec{y},\tau)\,\bar{N}^{\alpha}(0,0)\big\rangle\,,
    \label{three-point-axial}
\end{equation}
where $a$ is the lattice scale. In Eq.~\eqref{three-point-axial} we used $\Gamma^{\nu} = (1+\gamma_0)\,\gamma^{\nu}\,\gamma_5/2$  and $N$ is the three-quark nucleon interpolating current of a nucleon. Following Ref.~\cite{RQCD:2019jai}, after inserting identity operators and evaluated at large Euclidean times $t,\tau $ and $t-\tau$  the correlation functions can be approximated by the ground-state contributions, which is proportional to the trace \cite{Liang:2016fgy,RQCD:2019jai}
\begin{equation}
    C^{(\mu,\nu)}_{\infty}(\kf,\ki) = \operatorname{tr}\Big[\Gamma^{\nu}\,\frac{\slashed{\ppf}+M_N}{2\,M_N}\,F^{\mu}(\ppf,\ppi)\,\frac{\slashed{p}+M_N}{2\,M_N}\Big]\,.
\end{equation}
In previous lattice works , e.g. Refs.~\cite{RQCD:2019jai,Liu:1992ab,Alexandrou:2010hf,Gupta:2017dwj,Capitani:2017qpc,Jang:2019vkm}, where the  axial-vector matrix element was only decomposed into the two form factors $G_A(q^2)$ and $G_P(q^2)$, both form factors could be projected out by specifying $\Gamma$ properly. However, if we consider the six in-box form factors $G_A(\ppf,\ppi),\,G_P(\ppf,\ppi),\,G_T(\ppf,\ppi)$ and 
\begin{equation}
 G_X^{(k)}(\ppf,\ppi) = \sum_{\vec{n}\in\mathbb{Z}^3}\,x_n^{(k)}\,G_X(\ppf,\ppi,\xn) \,,\quad k\in\{1,2,3\}\,,
 \label{def:GXloop}
\end{equation}
where $x_n^{(k)}$ is the $k$-th component of the four vector $\xn$, 
a simple projection technique is not possible anymore. Instead, we need to consider the correlation function evaluated at different external three-momenta but with the constraint that the form factors remain the same. Thus, for fixed external three-momenta we define
\begin{align}
C_{+}^{(\mu,\nu)} = C_{\infty}^{(\mu,\nu)}(\kf,\ki)  + C_{\infty}^{(\mu,\nu)}(-\kf,-\ki)\,,\qquad
C_{-}^{(\mu,\nu)} = C_{\infty}^{(\mu,\nu)}(\kf,\ki)  - C_{\infty}^{(\mu,\nu)}(-\kf,-\ki)\,.
\end{align}
To extract the form factors $C_{\pm}^{(\mu,\nu)}$ we need to prove that the form factors are equal in both three-momentum combinations. First, we note that the momentum transfer $t = q^2$ remains unchanged. We recall that an explicit dependence on the three momenta $\kf,\ki$ is given by our in-box basis elements in Appendix~\ref{app:basisElements}. However, we observe that the finite-box corrections are invariant under the exchange $(\kf,\ki)\to(-\kf,-\ki)$. It is then convenient to consider the following six linear combinations
\begin{align}
&\sum_{i = 1}^3\,\kq_i\,C_{-}^{(0,i)} = \frac{G_A(\ppf,\ppi)}{2\,M_N^2}\,\big(M_N\,(\kq\cdot\kQ)  + \bar{E}\,(\kq\cdot\ki) + E\,(\kq\cdot\kf)\big)
\notag \\ &\hspace{8em}
- \frac{\big(\bar{E}-E\big)\,G_P(\ppf,\ppi) + \big(\bar{E}-E\big)\,G_T(\ppf,\ppi)}{4\,M_N^3}\,\big(M_N\,\kq[2]  - \bar{E}\,(\kq\cdot\ki) + E\,(\kq\cdot\kf)\big)\,, \notag\\
%%%
&\sum_{i,j=1}^3\,\delta_{i,j}\,C_{+}^{(i,j)} = \frac{G_A(\ppf,\ppi)}{M_N^2}\,\Big(3\,\big(M_N+\bar{E}\big)\big(M_N+E\big) - \kf\cdot\ki\Big)
\notag \\ &\hspace{8em}
-\frac{G_P(\ppf,\ppi)}{2\,M_N^3}\,\Big(M_N\,\kq[2] - \bar{E}\,\big(\kq\cdot\ki\big) + E\,\big(\kq\cdot\kf\big)\Big)
-\frac{G_T(\ppf,\ppi)}{2\,M_N^3}\,\Big(M_N\,\big(\kq\cdot\kQ\big) - \bar{E}\,\big(\kQ\cdot\ki\big) + E\,\big(\kQ\cdot\kf\big)\Big)\,, \notag\\
%%%
&\sum_{i,j=1}^3\,\kq_{i}\,\kq_{j}\,C_{+}^{(i,j)}
= \frac{G_A(\ppf,\ppi)}{M_N^2}\,\Big(\big(M_N + \bar{E}\big)\big(M_N + E\big) + \kf[2]\,\big(\kq\cdot\ki\big) - \ki[2]\,\big(\kq\cdot\kf\big)\Big)
\notag \\ &\hspace{8em}
-\frac{G_P(\ppf,\ppi)\,\kq[2] + G_T(\ppf,\ppi)\,\big(\kq\cdot\kQ\big)}{2\,M_N^3}\,\Big(M_N\,\kq[2] - \bar{E}\,\big(\kq\cdot\ki\big) + E\,\big(\kq\cdot\kf\big)\Big)\,, \notag \\
%%%
&\sum_{i = 1}^3\,\kq_{i}\,C_{-}^{(k,i)} = -\,\frac{2}{M_N}\,G_X^{(k)}(\ppf,\ppi)\,\Big(M_N\,\kq[2] - \bar{E}\,\big(\kq\cdot\ki\big) + E\,\big(\kq\cdot\kf\big)\Big)\,, \quad\text{ for } k\in\{1,2,3\}
\label{eqn:systemCorrAndFF}
\end{align}
where we introduced the short-hand notations $\bar{E} = \bar{p}^0$ and $E = p^0$. The inversion of Eq.~\eqref{eqn:systemCorrAndFF} is given in Appendix~\ref{app:FFinCorrFunc}.

\clearpage

\clearpage
\begin{table*}
    \begin{tabular}{lp{3em}ccp{5em}cc}
    \toprule
        && \,in-box & $\infty$  && \,in-box & $\infty$\\\midrule
        $L\,m_\pi$ && 4.04 & - && 4.67 &-   \\
        $m_\pi\,$[\si{\mega\electronvolt}] && 357 & 360 &&  342 & 344   \\
        $M_N\,$[\si{\giga\electronvolt}] && 1.167 & 1.186 && 1.148 & 1.159    \\
        $M_\Delta\,$[\si{\giga\electronvolt}] &&1.511 & 1.496 && 1.486 & 1.477   \\
        $M_{\Delta}-M_N\,$ [MeV] && 344 & 310 && 338 & 318    \\
        Fitted in \cite{Lutz:2020dfi} &&\multicolumn{2}{c}{yes} && \multicolumn{2}{c}{yes}
         \\  \bottomrule\\
    \end{tabular}
    \caption{Hadron masses of the two ensembles used in our numerical analysis in Section \ref{sec:Numerics}. The in-box masses are taken from Ref. \cite{Lutz:2020dfi}. The pion mass in the infinite volume was calculated and the corresponding nucleon and $\Delta$-isobar masses are also taken from Ref. \cite{Lutz:2020dfi}.
    The last line indicates if the corresponding axial-vector form factor data were used in the global fit of Refs. \cite{Lutz:2020dfi}. The first ensemble is from the ETMC collaboration \cite{EuropeanTwistedMass:2008pab,Alexandrou:2010hf}, while the second is a CLS ensemble \cite{Capitani:2017qpc,Capitani:2015sba}. }
    \label{tab:dataForNumeric}
\end{table*}

\section{Illustration of finite-box effects on two lattice ensembles}\label{sec:Numerics}
In this Section, we discuss the one-loop contributions to  the form factors evaluated on the two ensembles summarized in Table \ref{tab:dataForNumeric}. For both ensembles the $\Delta$-isobar is stable in the finite box and in the infinite-box limit, with 
\begin{equation}
    M_{\Delta}^{\text{box}} - M_N^{\text{box}}<m_\pi^{\text{box}} \quad \text{ and } \quad 
    M_{\Delta}^{\text{inf}} - M_N^{\text{inf}}<m_\pi^{\text{inf}}   \,.
\end{equation}

We compute the loop functions in  terms of the in-box hadron masses as given in Table \ref{tab:dataForNumeric}.
Their values in the finite boxes of the two considered ensembles are given in Table \ref{tab:fullFiniteBoxCorr} for the particular momentum transfer $t \simeq  - 4\,\pi^2 / L^2$. 
\\\par \enlargethispage{4\baselineskip}
Significant finite-box effects in the loop functions on both ensembles are found. 
A clear pattern is seen with the largest and smallest relative changes in  the pseudoscalar and tensor loop functions, respectively.
The largest effect is in the pseudoscalar $N\pi N$-triangle loop with a relative change of almost $60\%$ in the smaller box and still $24\%$ in the larger box. These are significantly larger than those in the axial  $N\pi N$-triangles with effects smaller than $1\%$. 
The second-largest finite-box effects are generated in the loop contributions with the $\Delta$-isobar. Here the relative changes in the $\Delta\pi\Delta$-triangles are approximately 1.5 times larger than those in the $\Delta\pi$-bubbles. Even for the larger box, they are larger than $10\%$ for the $\Delta\pi\Delta$-triangles and larger than $5\%$ for the $\Delta\pi$-bubbles. This implies the relevance of finite-box effects even in the larger box. 
On the other hand, the finite-box effects in the axial-vector $N\pi$-bubbles are about $\sim 7\%$ in the smaller box, and about $\sim 3\%$ in the larger box. The relative changes in the pseudoscalar $N\pi$-bubbles are approximately only half that large. We note that the finite-box effects of the $\Delta\pi N$-triangles are comparable to those in the $N\pi$-bubbles, where the pseudoscalar contributions are slightly larger than the axial contributions.  
\\\par \enlargethispage{4\baselineskip}

We confirm our previous findings in Refs.
\cite{Lutz:2020dfi,Hermsen:2024eth,Hermsen:2025vds} that the use of an in-box mass of the $\Delta$-isobar plays a crucial role in the correct determination of the form factors confined in a box.  We further consolidate this claim by more detailed studies of the loop functions, where the relative importance of different volume effects is illustrated.

%\enlargethispage{2\baselineskip}

\begin{table*}[b]
    \begin{tabular}{c| *{5}{c@{\hspace{1em}}} }
    \toprule
    $L\,m_\pi = 4.04$ & $N\pi$ & $ \Delta\pi$ & $N\pi N$ & $\Delta\pi N$ & $\Delta\pi\Delta$ \\
        \midrule
        $\Delta_V \, \bar{J}_{\ldots}^{A} \,[\%]$ & $ - 6.6 $ & $ - 13.0 $ & $  -0.2 $ & $ -4.2 $ & $ - 19.5 $\\[.3em]
        $\Delta_V\,\bar{J}_{\ldots}^{P} \,[\%]$ & $3.2$ & $-11.3$ & $-60.0$ & $-5.6$ & $-20.2$ \\[.3em]
        $\bar{J}_{\ldots}^{T}/\bar{J}_{\ldots}^{P} \,[\%]$ & $0.5 $ & $-0.03$ & $-15.2$ & $-0.3 $ & $0.04$ \\\midrule
        $L\,m_\pi = 4.67$ &&&&& \\\midrule
        $\Delta_V\,\bar{J}_{\ldots}^{A}\,[\%]$ & $ -3.4$ & $-7.9$ & $ 0.02 $ & $-2.1$ &  $-11.8$\\[.3em]
        $\Delta_V\,\bar{J}_{\ldots}^{P} \,[\%] $ & $1.2$ & $-6.7 $ & $-24.8$ & $-2.8$ & $-12.4$ \\[.3em]
        $\bar{J}_{\ldots}^{T}/\bar{J}_{\ldots}^{P}\,[\%] $ & $0.2$ & $-0.01$ & $-2.2$ & $-0.1$ & $0.01$\\
        \bottomrule
    \end{tabular}
    \caption{In the first two lines of each ensemble, we give the finite-box effects for the axial and pseudoscalar loop-functions at $t \simeq - 4\,\pi^2 / L^2$. They are determined as the ratio of the finite-box corrections relative to the infinite-volume limit.  In the third line of each ensemble, we give the ratio of the tensor to the pseudoscalar loop-contributions. }
    \label{tab:fullFiniteBoxCorr}
\end{table*}

\begin{figure*}
    \centering
    \includegraphics[width=\textwidth]{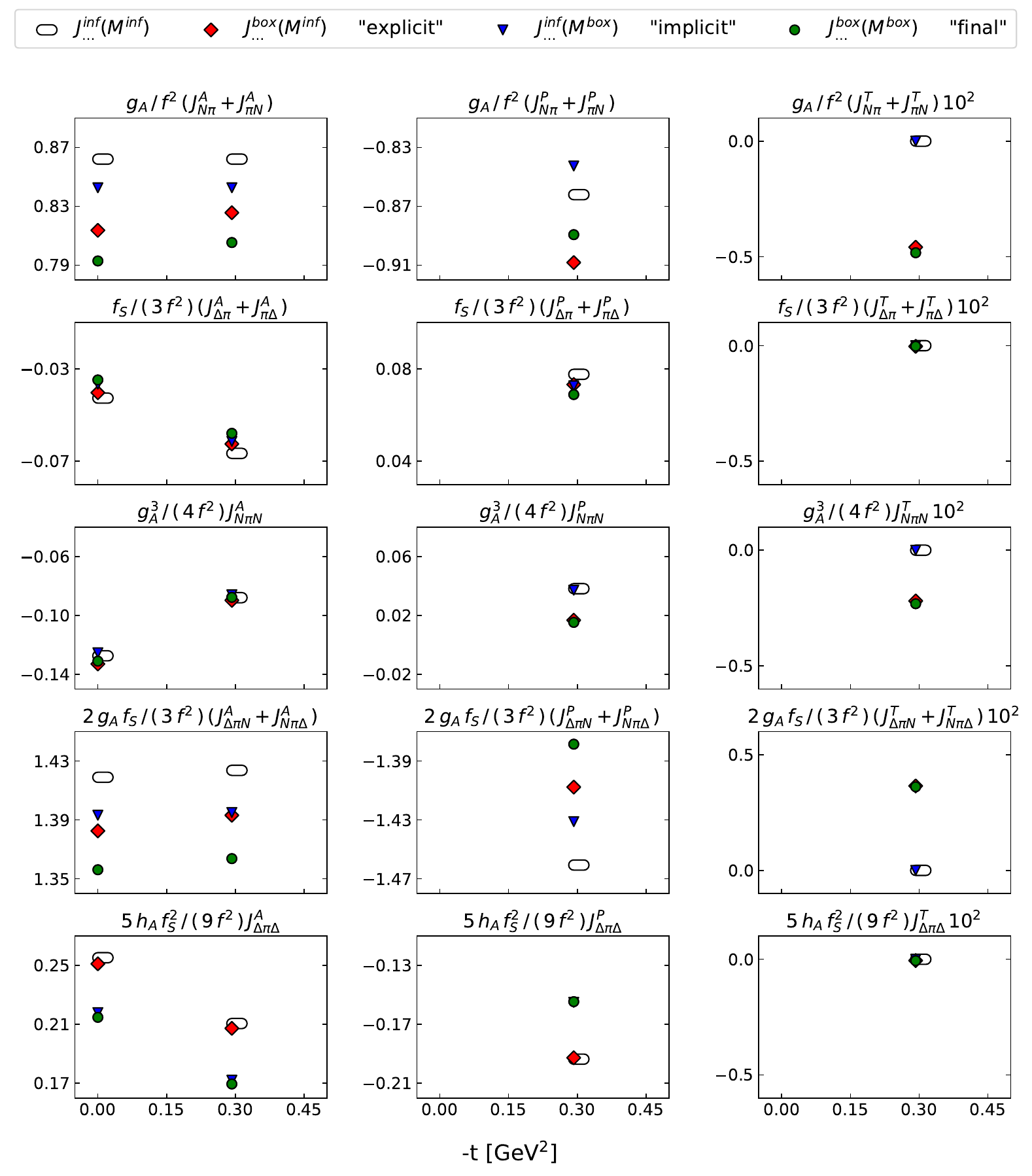}
	\caption{Loop contribution to the axial-vector, induced pseudoscalar, and induced pseudotensor form factor for the first ensemble with $L\,m_\pi = 4.04$ . The white stretched circles correspond to the infinite-volume case, with no finite-box corrections in the masses or in the form factor. The blue triangles contain only the finite-box corrections of the hadron masses. These \enquote{implicit} finite-box corrections were already considered in Ref. \cite{Lutz:2020dfi}. The red diamonds contain only the \enquote{explicit} finite-box corrections to the form factor. The green dots give the full finite-box results, taking both types of finite-box corrections into account \cite{Hermsen:2025vds}.}
	\label{fig:loopContributionsmpi404}
\end{figure*}

\clearpage 

In Figs. \ref{fig:loopContributionsmpi404} and \ref{fig:loopContributionsmpi467}, we display the size of various loop contributions to the three form factors 
\begin{equation}
    G_A^{\text{loop}}(\ppf,\ppi)\,,\qquad
    \frac{t-m_\pi^2}{4\,M_N^2}\,G_P^{\text{loop}}(\ppf,\ppi)\,,\qquad \text{ and }\qquad
    \frac{t}{2\,M_N^2}\,G_T^{\text{loop}}(\ppf,\ppi)\ ,
    \label{eqn:defFFNumeric}
\end{equation}
for the $L\,m_\pi = 4.04$ and $L\,m_\pi = 4.67$ ensembles from Table \ref{tab:dataForNumeric}.
For each loop contribution and given momentum transfer $t$, we show four values, corresponding to distinct treatments of finite-box effects. 
The infinite-volume results are given by white stretched circles always. The red square symbols, instead, show the size of the in-box loop functions with only \enquote{explicit} volume effects considered, i.e. they are evaluated with infinite-volume limit hadron masses. This is an approach mostly used in the literature so far \cite{Djukanovic:2024krw,Bali:2023sdi,Bali:2022qja,Gupta:2018qil,Park:2025rxi}.  The blue triangles in the plots represent the \enquote{implicit} finite-box results, which stem from using the in-box hadron masses in the infinite-volume expressions for the form factors. 
Our most complete results are given by the green circles, which include both explicit and implicit finite-volume effects. \\\par 

In our previous works  \cite{Lutz:2020dfi,Hermsen:2024eth} 
only the implicit volume effects (blue triangle points) were considered. 
Such expressions were conjectured to approximate finite-volume effects in the form factors for $L\,m_\pi>4$. We assumed that only for $L\,m_\pi<4$ the explicit corrections become relevant. Thus, both ensembles in Table \ref{tab:dataForNumeric} were part of the global fits. \\\par 

First, we observe in Fig. \ref{fig:loopContributionsmpi404} that the explicit finite-box corrections show a $t$-dependence, even if the infinite-volume limit is constant. This behavior is the consequence of the explicit three-momentum dependence of our in-box basis integrals given in Eq. \eqref{eqn:NpiBasisElements} and in Appendix \ref{app:basisElements}. While in the $N\pi$-bubbles and $N\pi N$-triangles, the explicit finite-box effects are significantly larger than the implicit, in the $\Delta\pi$-bubbles and $\Delta\pi\Delta$-triangles, the significance of the explicit and implicit finite-box effects is mirrored. The contrasting significance of the implicit effects in loop contributions with and without the $\Delta$-isobar is due to the stronger volume dependence of the $\Delta$-isobar mass than the nucleon mass. In the $\Delta\pi N$-triangles neither the implicit nor the explicit finite-box effects are dominating. Instead, both corrections are of the same order.  \\\par 

In Fig.~\ref{fig:loopContributionsmpi467} we observe that the  explicit finite-box effects have decreased significantly and are  roughly the same as the implicit ones. Thus, the implicit effects are now even more relevant for all loop contributions and not only in the loop contributions with a $\Delta$-isobar. However, even in this larger box, the still not negligible full finite-box effects are an intricate combination of both the explicit and implicit effects.  Additionally, we can, although it is hard to detect, confirm a non-linear $t$-dependence of the in-box results.  \\\par 

 In our previous work \cite{Hermsen:2025vds}, we investigated the axial $N\pi$-bubbles and $\Delta\pi N$-triangles on three ensembles with approximately the same pion mass of $m_\pi^{\text{box}} \approx 250 $ MeV and $L\,m_\pi$ values from $3.74$ over $4.22$ to $5.15$. For the $N\pi$-bubbles, we reported finite-box effects of $\sim 9\%$ for the $L\,m_\pi = 3.74$ ensemble and $\sim 1\%$ for the $L\,m_\pi = 5.15$ ensemble. In Table \ref{tab:fullFiniteBoxCorr} we found finite-box effects for the $N\pi$-bubble of $\sim 7\%$ for the $L\,m_\pi = 4.04$ ensemble and $\sim 3\%$, for the $L\,m_\pi = 4.67$ ensemble. This supports a $\exp(-L\,m_\pi)$ type of behavior for the finite-box effects in $N\pi$-bubbles, which are dominated by the explicit effects. 
 In Ref. \cite{Hermsen:2025vds} we found for the $\Delta\pi N$-triangles finite-box effects of approximately $5\%$ for all ensembles and momentum transfers, which is in the same range as the results from Table \ref{tab:fullFiniteBoxCorr}. These almost constant finite-box effects are caused by the subtle interplay of the explicit and implicit effects in the $\Delta\pi N$-triangles. This is a warning that a flat dependence is not necessarily a sign of negligible finite-box effects.    \\\par 
 
Our illustrative examples demonstrate that the full finite-box corrections are an intricate combination of implicit and explicit finite-box effects. For the $N\pi$-bubbles and for the $N\pi N$-triangles the explicit finite-box effects are dominating and we still find non-vanishing effects in the larger box. When the $\Delta$-isobar propagator gets involved the matter becomes more complex. In the $\Delta\pi$-bubbles and $\Delta\pi\Delta$-triangles, the implicit finite-box effects are dominating the full effects. This is mostly due to the use of the on-shell in-box masses in the loop contributions. For the $\Delta\pi N$-triangles both the implicit and explicit effects are relevant. This indicates that for a precise description of all loop-contributions and thus the whole form factors both types of finite-box effect, explicit and implicit, need to be considered. 
\clearpage

\begin{figure*}
    \centering
    \includegraphics[width=\textwidth]{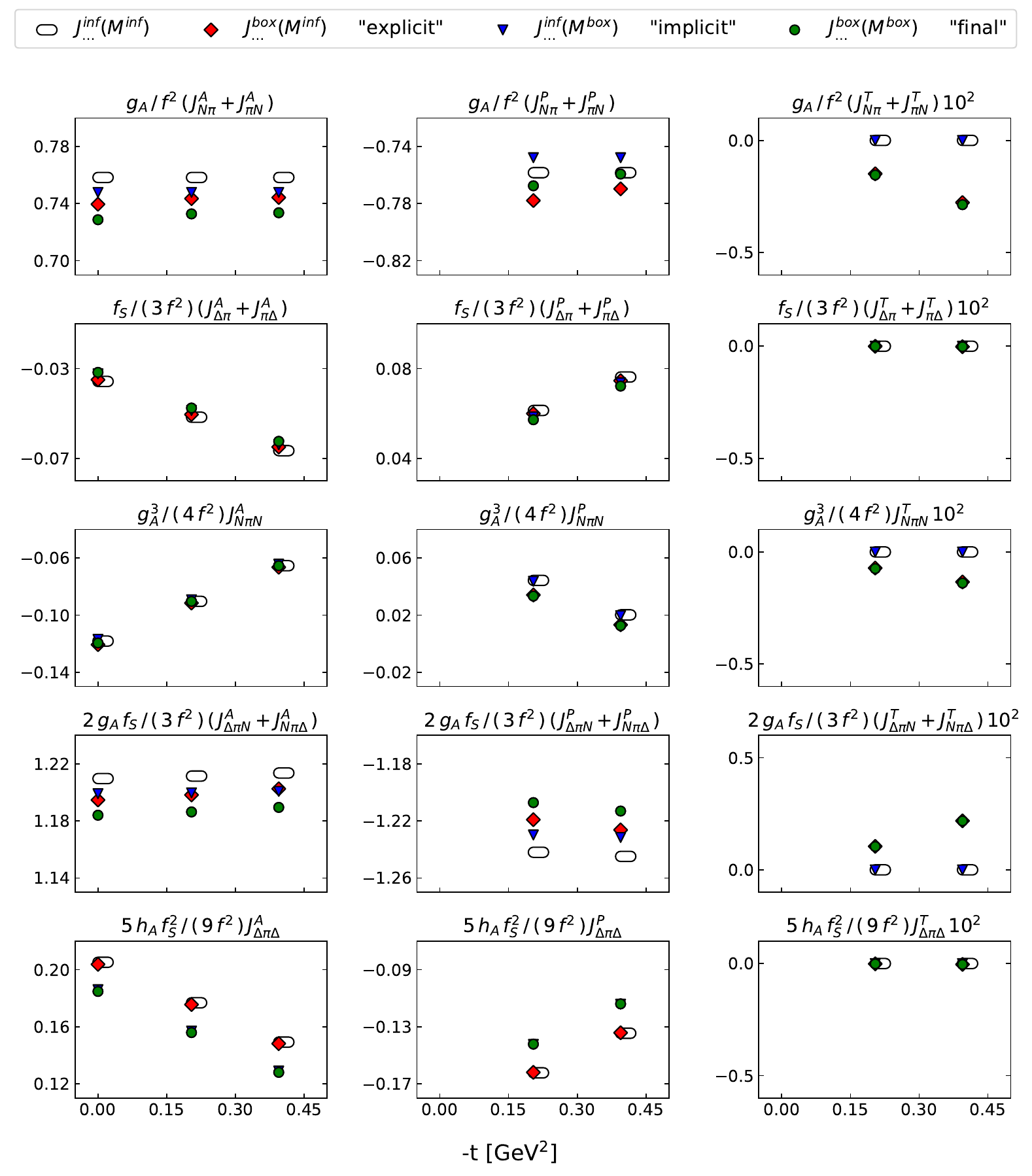}
	\caption{Loop contribution to the axial, pseudoscalar, and pseudotensor form factor for the $L\,m_\pi = 4.67$ ensemble. See caption of Fig. \ref{fig:loopContributionsmpi404}.}
	\label{fig:loopContributionsmpi467}
\end{figure*}

\clearpage
\section{Summary} \label{sec:Summary}
In this work we investigated the matrix element of the axial-vector current between two nucleon states in a finite box, assuming strict isospin symmetry. We showed that the widely-used decomposition into the axial-vector form factor $G_A$ and the induced pseudoscalar form factor $G_P$ no longer holds in a finite box. In addition, the induced pseudotensor form factor $G_T$ and the so-far unknown form factor $G_X$ are required.  We derived expressions for the three form factors $G_P, G_T$, and $G_X$ in terms of the on-shell hadron masses and our minimal in-box basis elements from Ref. \cite{Hermsen:2025vds}. We discussed the so far unknown in-box version of the axial Ward identity in terms of the full set of in-box form factors, where we observed that $G_X$, but not $G_T$, contributes. Furthermore, we explicitly derived the pseudoscalar form factor $G_{PS}$ at the one-loop level and expressed it in our minimal set of in-box basis elements. Together with the results for $G_A$ from our previous work \cite{Hermsen:2025vds}, we could then verify that  our renormalized form factors results fulfill the axial Ward identity.   \\\par 

We studied the finite-box effects in selected one-loop contributions on two flavor-SU(2) ensembles, for which the hadron masses are available and the $\Delta$-isobar is stable. Even for the ensemble with $L\,m_\pi=4.67$, we observed sizable finite-volume effects. The intricate interplay of implicit and explicit finite-box effects was studied in more detail for the different loop contributions. As soon as a $\Delta$-isobar propagator is involved the implicit effects become relevant. This is caused by the use of the in-box  $\Delta$-isobar mass inside the loop contributions, because the $\Delta$-isobar mass does not always have a simple  exponentially-suppressed finite-volume dependence. In general, our illustrative study showed that the full finite-box results are crucial for a precise determination of the form factors. This is in line with our previous work \cite{Hermsen:2025vds}.   \\\par

For zero momentum transfer,i.e. $t=0$, only $G_A$ contributes to the matrix element of the axial-vector current $A^{\mu}_i$ between two nucleons. However, in a finite box at $t\neq0$ the matrix element needs to be decomposed into six form factors instead of two in the infinite-volume limit. We showed that it is possible to derive this full set of axial-vector form factors from lattice simulations if  specific momentum combinations of the standard three-point correlation functions are considered. Since the corrections due to the additional in-box form factors only seem to be in the small-percentage range, high-statistics simulations are needed. Ideally, this would be done on flavor-SU(3) ensembles with fixed strange quark masses, such that the more precise flavor-SU(2) chiral extrapolation framework is applicable. Note, however, that for our scheme, the masses of the nucleon and the $\Delta$-isobar are needed both in  not too large boxes. To extrapolate lattice data to the physical pion mass and from the finite box to the infinite volume in a controlled and precise manner, our results can play a guiding role. \\\par 

While results for the axial-vector form factors for the  whole flavor-SU(3) baryon octet in the infinite volume were derived in Ref. \cite{Sauerwein:2021jxb}, a generalization to the finite box seems only fruitful if enough lattice data exist. However, the impact of the additional mass difference of the flavor-SU(3) baryon octet baryon masses onto the finite-box effects warrant a dedicated study either way. Furthermore, matrix elements with external $\Delta$-isobars should also be scrutinized in more detail in the finite box. As the finite-box effects are already significant in the nucleon case, one may speculate that they are even more important in a description of matrix elements involving external $\Delta$-isobar states. 

\clearpage
\appendix
\section{Decomposition of the Dirac structures}\label{app:decompDirac}
In this Section we will show that the decomposition of the Dirac structures of the axial-vector form factor given in Eq.
\eqref{eqn:AxialMatrixElement} is the minimal one.  After the loop momentum is taken care of, each possible Feynman amplitude contributing to the axial-vector matrix element between two nucleon states can be written as 
\begin{equation}
\langle\,N(\ppf)\,\vert\,A^\mu_i(0)\,\vert\,N(\ppi)\rangle = \bar{u}(\ppf)\,\sum_{\vec{n}\in\mathbb{Z}^3}\sum_{k}\,F_k[\ldots]\,b_k^\mu\,\frac{\tau_{i}}{2}\,u(\ppi)\,,
\end{equation}
where $F_k[\ldots]$ is a function that depends on the hadron masses and the products  $\ppf[2],\ppi[2],\big(\ppf\cdot\ppi\big),\big(\kf\cdot\kx\big),\big(\ki\cdot\kx\big)$, and $\kx^2$. 
%%%
% start new
%%%
{\color{Blue}
We recall that the sum over $\mathbb{Z}^3$ and the vector $\kx$ originate from the one-loop integrals evaluated in a finite box, which discretize the loop three-momentum:
\begin{equation}
\int\frac{\dd{l_0}}{2\,\pi}\,\int\frac{\dd[3]\vec{l}}{(2\,\pi)^3}\,f(l_0,\vec{l}) \to  \sum_{\vec{n}\in\mathbb{Z}^3}\,\int\,\frac{\dd[4]{l}}{(2\,\pi)^4}\,e^{i\,\vec{l}\cdot \vec{x}_n}\,f(l_0,\vec{l})\,, \quad \vec{x}_n = L\,\vec{n}\,.
\end{equation}
As this is commonly done in finite-volume calculations for hadronic observables, we refer to our previous work \cite{Hermsen:2025vds} for more details. 
}
%%%
% end new
%%%
Additionally, $\{b_k^\mu\}$ is the minimal set of independent Dirac structures which need to be determined and which defines the decomposition in Eq. \eqref{eqn:AxialMatrixElement}. We recall the Dirac equations,
\begin{equation}
\slashed{p}\,u(\ppi) = M_N\,u(\ppi)\,,\qquad
\bar{u}(\ppf)\,\slashed{\ppf} = \bar{u}(\ppf)\,M_N\,,\qquad
\sigma^{\mu\nu} = \frac{\ii}{2}\,\big(\gamma^\mu\,\gamma^\nu - \gamma^\nu\,\gamma^\mu\big) \,,\qquad
g^{\mu\nu} = \frac{1}{2}\,\big(\gamma^\mu\,\gamma^\nu + \gamma^\nu\,\gamma^\mu\big)\, ,
\label{app:eqn:sigmu-gmunu}
\end{equation}
and the basis set from which all possible Dirac structures can be constructed,
\begin{equation}
\gamma^\mu\,\gamma_5\,,\qquad 
\ppf[\mu]\,\gamma_5\,,\qquad
\ppi[\mu]\,\gamma_5\,,\qquad
q^{\mu}\,\gamma_5\,,\qquad
x_n^{\mu}\,\gamma_5\,.
\end{equation}

We will start with the following maximal set of combinations:
\begin{align}
&\gamma^\mu\,\gamma_5\,,  &&q^{\mu}\,\gamma_5\,,  &&Q^{\mu}\,\gamma_5\,, &&\hspace{-5em}x_n^\mu\,\gamma_5\,,&&\notag \\
%%%
&\gamma^\mu\,\slashed{q}\,\gamma_5\,, &&\gamma^\mu\,\slashed{Q}\,\gamma_5\,, &&\gamma^\mu\,\slashed{x}_n\,\gamma_5\,,&&\notag \\
%%%
&\slashed{q}\,\gamma^\mu\,\gamma_5\,, &&\slashed{Q}\,\gamma^\mu\,\gamma_5\,, &&\slashed{x}_n\,\gamma^\mu\,\gamma_5\,,&&\notag \\
%%%
&q^\mu\,\slashed{q}\,\gamma_5\,, &&q^\mu\,\slashed{Q}\,\gamma_5\,, &&q^\mu\,\slashed{x}_n\,\gamma_5\,,&&\notag \\
%%%
&Q^\mu\,\slashed{q}\,\gamma_5\,, &&Q^\mu\,\slashed{Q}\,\gamma_5\,, &&Q^\mu\,\slashed{x}_n\,\gamma_5\,,&&\notag \\
%%%
&x_n^\mu\,\slashed{q}\,\gamma_5\,, &&x_n^\mu\,\slashed{Q}\,\gamma_5\,, &&x_n^\mu\,\slashed{x}_n\,\gamma_5\,,&&\notag \\
%%%
&\slashed{a}\,\gamma^\mu\,\slashed{b}\,\gamma_5\,,
&&\slashed{a}\,b^\mu\,\slashed{c}\,\gamma_5
\,, &&a^\mu,\,b^\mu,\,c^\mu\in\{q^{\mu},\,Q^{\mu},x_n^\mu\}\,. &&
\label{app:eqn:listDiracKombis}
\end{align}
Here we used the four-vectors $x_n=(0,\kx)$ and  chose the momenta $q^\mu = \bar{p}^\mu - p^\mu,\,Q^\mu=\bar{p}^\mu + p^\mu$, instead of $\bar{p}^\mu$ and $p^\mu$\,.
For any vector $a^\mu$ we find the relations
\begin{align}
\label{app:eqn:decompVektor}
a^\mu = \gamma^\mu\,\slashed{a} + \ii\,\sigma^{\mu\nu}\,a_\nu\,,\qquad
%%%
a^\mu = \slashed{a}\,\gamma^{\mu} -\ii\,\sigma^{\mu\nu}\,a_{\nu}\,.
\end{align}
Therefore, we can absorb the second and third line, while adding the structures
\begin{equation}
\ii\,\sigma^{\mu\nu}\,q_{\nu}\,\gamma_5\,,\qquad 
\ii\,\sigma^{\mu\nu}\,Q_{\nu}\,\gamma_5\,,\qquad 
\ii\,\sigma^{\mu\nu}\,(x_n)_{\nu}\,\gamma_5\,.
\end{equation}
Next, we will show that $\ii\,\sigma^{\mu\nu}\,(x_n)_{\nu}\,\gamma_5$ is redundant. We expand $\sigma^{\mu\nu}$, use Eq. \eqref{app:eqn:decompVektor}, and put the term between the two spinors. We find 
\begin{equation}\begin{aligned}[b]
&\bar{u}(\ppf)\,\bigg(\frac{\ppf[\mu]\,\slashed{x}_n\,\gamma_5}{M_N} + \frac{\,p^\mu\,\slashed{x}_n\,\gamma_5}{M_N}\bigg)\,u(\ppi) = 
-2\,\ii\,\bar{u}(\ppf)\,\sigma^{\mu\nu}\,(x_n)_\nu\,\gamma_5\,u(\ppi) \\ 
&\qquad
+\frac{1}{2}\,\bar{u}(\ppf)\,\bigg[\frac{\gamma^\mu\,(\ppf\cdot x_n) - \slashed{\ppf}\,\gamma^\mu\,\slashed{x}_n}{M_N}
 - \frac{\slashed{x}_n\,\gamma^\mu\,\slashed{p} - (\ppi\cdot x_n)\,\gamma^{\mu}}{M_N}\bigg)\,\gamma_5\,u(\ppi) \ .
\end{aligned}\end{equation}
With the same argument we can absorb the structure $\sigma^{\mu\nu}\,Q_\nu\,\gamma_5$. 
Next, we turn to the structures in the last line of Eq. \eqref{app:eqn:listDiracKombis}.
First, for $a^\mu,\,b^\mu,\,c^\mu\in\{q^{\mu},\,Q^{\mu},x_n^\mu\}$  we find
\begin{equation}
\slashed{a}\,b^\mu\,\slashed{c} = \slashed{a}\,\slashed{c}\,b^\mu = (a\cdot c)\,b^\mu\,.
\end{equation}
Since the $a\cdot c$ dependence is encoded in the functions $F_k[\ldots]$ these type of structure can be absorbed in other terms. 
Using  the first relation from Eq. \eqref{app:eqn:decompVektor} we find 
\begin{equation}\begin{aligned}[b]
a^{\mu}\,\slashed{b} 
&= \gamma^\mu\,a\cdot b - \frac{1}{2}\,\big(\gamma^{\mu}\,a\cdot b - \slashed{a}\,\gamma^\mu\,\slashed{b}\big) \ ,
\end{aligned}\end{equation}
and we can therefore absorb the terms $\slashed{a}\,\gamma^\mu\,\slashed{b}\,\gamma_5$ with $a^\mu,\,b^\mu \in\{q^{\mu},\,Q^{\mu},x_n^\mu\}$. At this point, we shortened the list of possible structures to 
\begin{equation}\begin{aligned}[b]
&\gamma^\mu\,,\qquad q^{\mu}\,,\qquad Q^{\mu}\,,\qquad x_n^\mu\,,\qquad \ii\,\sigma^{\mu\nu}\,q_{\nu}\,,\\
%%%
&q^\mu\,\slashed{q}\,,\qquad q^\mu\,\slashed{Q}\,,\qquad q^\mu\,\slashed{x}_n\,,\\
%%%
&Q^\mu\,\slashed{q}\,,\qquad Q^\mu\,\slashed{Q}\,,\qquad Q^\mu\,\slashed{x}_n\,,\\
%%%
&x_n^\mu\,\slashed{q}\,,\qquad x_n^\mu\,\slashed{Q}\,,\qquad x_n^\mu\,\slashed{x}_n\,.
\end{aligned}\end{equation}
Using the Dirac equations only, we find the two relations
\begin{equation}\begin{aligned}[b]
\bar{u}(\ppf)\,\slashed{q} \,\gamma_5\,u(\ppi)
= 
2\,M_N\,\bar{u}(\ppf)\,\gamma_5\,u(\ppi)\,,\qquad
%%%
\bar{u}(\ppf)\,\slashed{Q} \,\gamma_5\,u(\ppi)= 0 \ ,
\label{app:DiracStruture:intermidate}
\end{aligned}\end{equation}
and we can therefore remove the two first structures in the second to last line, since they are proportional to the elements in the first line. By replacing $a^\mu\to p^\mu$ in the first Eq. of \eqref{app:eqn:decompVektor} and $a^\mu\to \bar{p}^\mu$ in the second Eq.
\eqref{app:eqn:decompVektor} and putting the difference between the Dirac spinors, we find 
\begin{equation}\begin{aligned}[b]
\label{chap:decompAxialFFOctet-gordonId02}
&\bar{u}(\ppf)\,q^\mu\,\gamma_5\,u(\ppi) = 
2\,M_N\,\bar{u}(\ppf)\,\gamma^\mu\,\gamma_5 \,u(\ppi)
- \bar{u}(\ppf)\,\ii\,\sigma^{\mu\nu}\,Q_{\nu}\,\gamma_5\,u(\ppi)\,,\\
&\qquad \Rightarrow 
\bar{u}(\ppf)\,(q\cdot x_n)\,\gamma_5\,u(\ppi) = 
2\,M_N\,\bar{u}(\ppf)\,\slashed{x}_n\,\gamma_5 \,u(\ppi)\,,
\end{aligned}\end{equation}
where we used the Dirac equations and contracted with with $(x_n)_\mu$.
We can multiply the second line with $q^\mu,\,Q^\mu$, or $(x_n)_{\mu}$, and thereby eliminate the last structures in the second to last line in Eq. \eqref{app:DiracStruture:intermidate} and reduce the set to the first line. 
Analogously to Eq. \eqref{chap:decompAxialFFOctet-gordonId02} we derive 
\begin{equation}
2\,\bar{u}(\ppf)\,\ii\,\sigma^{\mu\nu}\,q_\nu\,\gamma_5\,u(\ppi) = -\,\bar{u}(\ppf)\,Q^{\mu}\,\gamma_5\,u(\ppi) \ .
\end{equation}
Thus, we can either choose $Q^\mu$ or $\ii\,\sigma^{\mu\nu}\,q_\nu$. We find for the decomposition in the finite-box case that
\begin{equation}\begin{aligned}[b]
&\langle N(\bar{p})\vert\,A^{\mu}_i(0)\,\vert N(p)\rangle = \bar{u}(\bar{p})\,\sum_{\vec{n}\in\mathbb{Z}^3}\,F^\mu_A(\bar{p},p,\xn)\,\frac{\tau_i}{2}\,u(p)\,,\\
&F^\mu_A(\bar{p},p,\xn) = \Big(\gamma^\mu\,G_A(\bar{p},p,\xn) + \frac{q^\mu}{2\,M_N}\,G_P(\bar{p},p,\xn) + \frac{Q^{\mu}}{2\,M_N}\,G_T(\ppf,\ppi,\xn) 
+ 2\,M_N\,x_n^{\mu}\, G_X(\ppf,\ppi,\xn)\,\Big)\,\gamma_5\,.
\label{app:Dirac:eqn:finalAxialMatrixElement}
\end{aligned}\end{equation}
%%%
% begin new part
%%%

So far we did not consider possible consequences of the summation over $\vec{n}$ for our decomposition in Eq.~\eqref{app:Dirac:eqn:finalAxialMatrixElement}. The question arises if the four form factors are still minimal if the sum over $\mathbb{Z}^3$ is taken into account. To prove that this is indeed the case, we first define
\begin{equation}
G_A (\bar p, p) = \sum_{\vec n \in \mathbb{Z}^3} 
G_A (\bar p, p, x_n) \,,\quad 
G_P (\bar p, p) = \sum_{\vec n \in \mathbb{Z}^3} 
G_P (\bar p, p, x_n) \,,\quad 
G_T (\bar p, p) = \sum_{\vec n \in \mathbb{Z}^3} 
G_T (\bar p, p, x_n)\,.
\label{def:GAPTloop}
\end{equation}
These three quantities are observables th can be measured on the lattice, in contrast to the form factors with the explicit $\xn$ dependence.  With these quantities we can write Eq.~\eqref{app:Dirac:eqn:finalAxialMatrixElement} more explicitly as
\begin{equation}\begin{aligned}[b]
&\langle N(\bar{p})\vert\,A^{0}_i(0)\,\vert N(p)\rangle = \bar{u}(\bar{p})\,\Big(\gamma^0\,G_A(\bar{p},p) + \frac{\sqrt{M_N^2+\kf[2]} - \sqrt{M_N^2+\ki[2]}}{2\,M_N}\,G_P(\bar{p},p) 
\\ &\hspace{20em}
+ \frac{\sqrt{M_N^2+\kf[2]} +\sqrt{M_N^2+\ki[2]}}{2\,M_N}\,G_T(\ppf,\ppi) \Big)\,\gamma_5\,\frac{\tau_i}{2}\,u(p)\,,\\
%%%
&\langle N(\bar{p})\vert\,A^{j}_i(0)\,\vert N(p)\rangle = \bar{u}(\bar{p})\,\Big(\gamma^{j}\,G_A(\bar{p},p) + \frac{\kf_j-\ki_j}{2\,M_N}\,G_P(\bar{p},p) 
\\ &\hspace{20em}
+ \frac{\kf_j+\ki_j}{2\,M_N}\,G_T(\ppf,\ppi) 
+ 2\,M_N\,\sum_{\vec{n}\in\mathbb{Z}^3}x_n^{j}\, G_X(\ppf,\ppi,\xn)\,\Big)\,\gamma_5\,\frac{\tau_i}{2}\,u(p)\,,
\end{aligned}\end{equation}
where $j\in\{1,2,3\}$. Note that $G_X(\ppf,\ppi,\xn)$ does not contribute to the decomposition of the first matrix element, since $x_n = (0,\kx)$. In the second matrix element for $\kf=\vec{0}$ or $\ki = \vec{0}$, the induced pseudotensor form factor $G_T(\ppf,\ppi)$ could be absorbed into the induced pseudoscalar factor $G_P(\ppf,\ppi)$. We would then have $\ki_j/(2\,M_N)\,(G_T(\ppf,\ppi) - G_P(\ppf,\ppi))$ as an \enquote{effective} form factor. Since $\lim_{L\to\infty}\,G_T(\ppf,\ppi) = 0$, the infinite-volume limit of the induced pseudoscalar form factor $\lim_{L\to \infty} G_P(\ppf,\ppi) = G_P(q^2)$, would not be changed. However, the same is not true in the first matrix element, because for the case of $\kf = (0,0,0)$, the coefficients of $G_P(\ppf,\ppi)$ and $G_T(\ppf,\ppi)$ collapse as
\begin{equation}
\frac{M_N - \sqrt{M_N^2+\ki[2]}}{2\,M_N}\,G_P(\bar{p},p) 
+ \frac{M_N + \sqrt{M_N^2+\ki[2]}}{2\,M_N}\,G_T(\ppf,\ppi) \,.
\end{equation}
Consequently, the difference $G_T(\ppf,\ppi)-G_P(\ppf,\ppi)$ cannot be factorized. This would result in a reduction of the number of the form factors. Therefore, we keep our general decomposition, which comes with a possible advantage. If we do not assume $\kf=\vec{0}$ or $\ki=\vec{0}$, our decomposition of the matrix element and the corresponding results can be applied to lattice results of the axial-vector form factors evaluated in moving frames, where $\kf \neq \vec{0}$ and $\ki\neq\vec{0}$. While we are not aware that such data exist for the axial-vector matrix element, other form factors, e.g. the  scalar form factor of the pion, see Refs.~\cite{Ottnad:2025hkw,vonHippel:2025uhr}, were determined in moving frames to increase the momentum resolution. Therefore, it is advantageous to keep our general decomposition for future lattice data and studies. Furthermore, our decomposition also indicates that in a moving frame, the form factors $G_P(\ppf,\ppi)$ and $G_T(\ppf,\ppi)$ are more distinguishable than in case of $\kf $ or $\ki$ equal to $\vec{0}$. 
\\\par 

We next turn to the fourth structure in Eq.~\eqref{app:Dirac:eqn:finalAxialMatrixElement}, the sum involving the $G_X(\ppf,\ppi,\xn)$ form factor. This form factor could only be incorporated into the remaining form factors if a decomposition of the type 
\begin{equation}
    \bar{u}(\bar{p})\,\sum_{\vec{n}\in\mathbb{Z}^3}\,\vec{x}_n^{\mu}\,G_X(\ppf,\ppi,\xn)\,\gamma_5\,u(p) = \bar{u}(\ppf)\Big(\gamma^{i}\,a(\ppf,\ppi) + \kf_{j}\,b(\ppf,\ppi) + \ki_{j}\,c(\ppf,\ppi)\Big)\,\gamma_5\,u(p)
    \label{appendix:DecompFF:eq-possibleDecompGX}
\end{equation}
would exist. Although we cannot prove the non-existence of such a decomposition, we doubt that it exists for general $G_{X}(\ppf,\ppi,\xn)$. But for the sake of completeness, let us pretend that such a decomposition exists. Then it would only change the form factors of the matrix element $\langle N(\bar{p})\vert\,A^{j}_i(0)\,\vert N(p)\rangle$, while the form factors of the matrix element $\langle N(\bar{p})\vert\,A^{0}_i(0)\,\vert N(p)\rangle$ would remain the same. Altogether, we could end up with up to six different form factors in the general case. This shows that our four form factors are indeed a minimal basis for the decomposition of the axial-vector current between two nucleon states in a finite box.  \\\par 

%%%
% end new part
%%%
In the next step, we want to derive the symmetry properties of the form factors presented in Eq. \eqref{eqn:SymRelFF}. First, we apply the hermitian conjugate to both sides of Eq. \eqref{app:Dirac:eqn:finalAxialMatrixElement} and find
\begin{equation}\begin{aligned}[b]
&\langle N(\ppi)\vert\,A^{\mu}_i(0)\,\vert N(\ppf)\rangle = \bar{u}(\ppi)\,\sum_{\vec{n}\in\mathbb{Z}^3}\,\tilde{F}^\mu_A(\bar{p},p,\xn)\,\frac{\tau_i}{2}\,u(\ppf)\,,\\
&\tilde{F}^\mu_A(\bar{p},p,\xn) = \Big(\gamma^\mu\,G^{\ast}_A(\bar{p},p,\xn) - \frac{(\ppf-\ppi)^\mu}{2\,M_N}\,G_P^{\ast}(\bar{p},p,\xn) - \frac{Q^{\mu}}{2\,M_N}\,G_T^{\ast}(\ppf,\ppi,\xn) - 2\,M_N\,x_n^{\mu}\, G_X^{\ast}(\ppf,\ppi,\xn)\,\Big)\,\gamma_5\,,
\label{app:Dirac:eqn:Hermitian}
\end{aligned}\end{equation}
where we used $(A^{\mu}_i)^{\dagger} = A^{\mu}_i$. Naturally, we can decompose the matrix element in Eq. \eqref{app:Dirac:eqn:Hermitian} the same way as in Eq. \eqref{app:Dirac:eqn:finalAxialMatrixElement}. This is done by simply switching $\ppf\leftrightarrow\ppi$ in the right-hand side of Eq. \eqref{app:Dirac:eqn:finalAxialMatrixElement}. This implies $\tilde{F}^{\mu}_A(\ppf,\ppi,x_n) = F^{\mu}_A(\ppi,\ppf,x_n)$ and we find
\begin{equation}\begin{aligned}[b]
& G_A(\bar{p},p, x_n) = +\,G_A^{\ast}(p,\bar{p}, x_n ) \,, \qquad
    G_P(\bar{p},p, x_n) = +\,G_P^{\ast}(p,\bar{p}, x_n ) \,,
    \qquad \text{ but } \quad
\\
&  G_T(\bar{p},p, x_n) = -\,G_T^{\ast}(p,\bar{p} , x_n) \,, \qquad
    G_X(\bar{p},p , x_n) = -\,G_X^{\ast}(p,\bar{p}, x_n ) \,.
\label{chap:AxialFormFactor:sec:Decomp:eqn-transformationFromHermiticy}
\end{aligned}\end{equation}
In the next step, we transform the matrix element under time reversal $T$ and parity $P$. Additionally, we rotate by $\pi$ around the second axis in the isospin space and denote this symmetry transformation with $R$. 
First, we collect the transformation properties of the axial current,
\begin{equation}\begin{aligned}[b]
&A^{\mu}_i(x) \quad \xlongrightarrow{T}\quad
\begin{dcases}
    (-1)^{\mu}\,A^{\mu}_i(-t,x)\,,\quad &\text{ for } i\in\{1,3\}\,, \\
    -\,(-1)^{\mu}\,A^{\mu}_i(-t,x)\,,\quad &\text{ for } i = 2
\end{dcases}\,,\\[.5em]
&A^{\mu}_i(x) \quad \xlongrightarrow{P}\quad -\,(-1)^{\mu}\,A_i^{\mu}(t,\,-\vec{x})\,,\\[.5em]
%%%
&A^{\mu}_i(x) \quad \xlongrightarrow{R} \quad  
\begin{dcases}
-\,A_i^{\mu}(x)\,,\quad &\text{ for } i \in\{1,3\}\,,\\    
\quad  A_2^{\mu}(x) &\text{ for } i =2     
\end{dcases}\,,
\end{aligned}\end{equation}
with 
\begin{equation}
(-1)^{\mu} = \begin{dcases}
1\,, &\text{ for } \mu = 0\,,\\
-1\,, &\text{ for } \mu\in\{1,2,3\}
\end{dcases}
\end{equation}
and thus we find that $A_i^{\mu}(x) \, \xlongrightarrow{P\,T\,R} \,  A_i^{\mu}(x)$. For the spinors, we recall the transformation properties \cite{Peskin:1995ev}
\begin{equation}\begin{aligned}[b]
%%%
&u(\ppi) \quad \xlongrightarrow{T}\quad \gamma^1\,\gamma^3\,u(-\ppi)\,,&\qquad
&\bar{u}(\ppf) \quad \xlongrightarrow{T}\quad -\,\bar{u}(\ppf)\,\gamma^1\,\gamma^3\,,&&\\
%%%
&u(\ppi) \quad \xlongrightarrow{R} \quad \ii\,\tau_2\,u(\ppi)\,,&\qquad  
&\bar{u}(\ppf) \quad \xlongrightarrow{R} \quad -\,\ii\,\tau_2\,u(\ppi)\,,&&\\
%%%
&u(\ppi)\quad \xlongrightarrow{P} \quad \gamma^0\,u(-\ppi)\,,&\qquad
&\bar{u}(\ppf) \quad  \xlongrightarrow{P}\quad \bar{u}(-\ppf)\,\gamma^0\,.&&
\end{aligned}\end{equation}
We suppressed the transformation of the spins, as these are irrelevant to our current discussion. 
Putting everything together, we find 
\begin{equation}\begin{aligned}[b]
&\eqref{app:Dirac:eqn:finalAxialMatrixElement}\,\xlongrightarrow{P\,T\,R}\; \langle N(\bar{p})\vert\,A^{\mu}_i(0)\,\vert N(p)\rangle = \bar{u}(\bar{p})\,\sum_{\vec{n}\in\mathbb{Z}^3}\,\tilde{F}^\mu_A(\bar{p},p,\xn)\,\frac{\tau_i}{2}\,u(p) \ , \\
&\tilde{F}^\mu_A(\bar{p},p,\xn) = \Big(\gamma^\mu\,G_A^{\ast}(\bar{p},p,\xn) + \frac{q^\mu}{2\,M_N}\,G_P^{\ast}(\bar{p},p,\xn) + \frac{Q^{\mu}}{2\,M_N}\,G_T^{\ast}(\ppf,\ppi,\xn) 
+ 2\,M_N\,x_n^{\mu}\, G_X^{\ast}(\ppf,\ppi,\xn)\,\Big)\,\gamma_5\,.
\label{app:Dirac:eqn:PTRTransformedAxialMatrixElement}
\end{aligned}\end{equation}
The complex conjugates of the form factors appear because $T$ is an anti-unitary operator. The strong interaction is invariant under $CPT$, where $C$ is the charge-conjugation operator. Thus, we find that $PTR=G$, where $G$ is the $G$-parity operator. In the last step, we apply Eq. \eqref{chap:AxialFormFactor:sec:Decomp:eqn-transformationFromHermiticy} to Eq. \eqref{app:Dirac:eqn:PTRTransformedAxialMatrixElement} and therefore we find that $G$-parity implies that
\begin{equation}\begin{aligned}[b]
&G_A(\bar{p},p, x_n) = +\,G_A(p,\bar{p}, x_n ) \,, \qquad
    G_P(\bar{p},p, x_n) = +\,G_P(p,\bar{p}, x_n ) \,,
    \qquad \text{ but } \quad
\\
&G_T(\bar{p},p, x_n) = -\,G_T(p,\bar{p} , x_n) \,, \qquad
    G_X(\bar{p},p , x_n) = -\,G_X(p,\bar{p}, x_n ) \,.
\end{aligned}\end{equation}

\section{Axial amplitudes}\label{app:AxialAmplitudes}
The one-loop integrals in $d$-dimension from Eq. \eqref{def-Js} read
\begin{equation}
J^{\mu}_{\ldots}(\bar{p},p,x_n) = \ii\,\mu^{4-d}\,\IntdDPi{l}\,e^{\ii\,\kl\cdot\kx}\,\Big[ K^\nu_{\ldots}(\ppf,\ppi;l)\big(\delta^{\,\mu}_{\nu}-q_{\nu}\,q^{\mu}/(q^2-m_\pi^2)\big) + L^{\mu}_{\ldots}(\ppf,\ppi;l)\Big]\,.
\end{equation}
We list here all non-vanishing functions:
\begin{align}
&L^{\mu}_{\pi} = \frac{1}{\sqrt{2}}\,\frac{\gamma_5\,\gamma^{\nu}}{l^2-m_\pi^2}\bigg[\delta^{\,\mu}_{\nu} - \frac{q_{\nu}\,q^{\mu}}{3\,(t-m_\pi^2)}\bigg]\,,\notag \\
%%%
&K^{\mu}_{N\pi} = 
\frac{1}{\sqrt{2}}\,\frac{\gamma_5\,\slashed{l}}{l^2-m_\pi^2}\,S_{N}(\ppf-l)\,\bigg[ 2\,g_S\,l^{\mu} -2\,g_T\,\ii\,\sigma^{\mu\nu}\,l_{\nu} 
+\frac{g_V}{2}\,\Big[
\gamma^{\mu}\big(\ppf\cdot l - l^2+ l\cdot \ppi\big) + (\ppf-l+\ppi)^{\mu}\,\slashed{l}
\Big]\bigg]\,,\notag \\
%%%
&L^{\mu}_{N\pi} = \frac{1}{\sqrt{2}}\,\frac{\gamma_5\,
\slashed{l}}{l^2-m_\pi^2}\,S_N(\ppf-l)\,\bigg[\gamma^{\mu}
-\,\frac{\big(\slashed{q}+\slashed{l}\big)\,q^{\mu}}{2\,\big(t-m_\pi^2\big)} + \frac{8\,b_{\chi}\,B_0\,m\,q^{\mu}}{t-m_\pi^2}
+4\,g_F \,\ii\,\sigma^{\mu\nu}\,q_{\nu}\bigg]\,,\notag \\
%%%
&K^{\mu}_{\pi N} = \frac{1}{\sqrt{2}}\,\bigg[ 2\,g_S\,l^{\mu} + 2\,g_T\,\ii\,\sigma^{\mu\nu}\,l_{\nu}
+\frac{g_V}{2}\,\Big[\gamma^{\mu}\,\big(\ppf\cdot l - l^2 + l\cdot\ppi\big) + \big(\ppf-l+\ppi\big)^{\mu}\,\slashed{l}\Big]
\bigg]\,S_{N}(\ppi-l)\,\frac{\gamma_5\,\slashed{l}}{l^2-m_\pi^2}\,,\notag \\
%%%
&L^{\mu}_{\pi N} = \frac{1}{\sqrt{2}}\,\bigg[
\gamma^{\mu} -\frac{\big(\slashed{q} - \slashed{l}\big)\,q^{\mu}}{2\,\big(t-m_\pi^2\big)} -\frac{8\,b_{\chi}\,B_0\,m\,q^{\mu}}{t-m_\pi^2} + 4\,g_F \,\ii\,\sigma^{\mu\nu}\,q_{\nu}\bigg]\, S_{N}(\ppi-l)\,\frac{\gamma_5\,\slashed{l}}{l^2-m_\pi^2}\,,\notag \\
%%%
&K^{\mu}_{\Delta\pi} =  \frac{1}{\sqrt{2}}\,l_{\nu}\,\Big[ \big(f_A^{(-)}-5\,f_A^{(+)}\big)\,S_{\Delta}^{\nu\mu}(\ppf-l)\,\slashed{l} - \big(f_A^{(-)} + 5\,f_A^{(+)}\big)\,  S_{\Delta}^{\nu\rho}(\ppf-l)\,l_{\rho}\,\gamma^{\mu}\Big]\,\frac{\gamma_5}{l^2-m_\pi^2}\,,\notag \\
%%%
&L^{\mu}_{\Delta\pi} = 2\,\sqrt{2}\,f_M\,\,l_{\nu}\,\big( S_{\Delta}^{\nu\mu}(\ppf-l)\,\slashed{q}  - S_{\Delta}^{\nu\rho}(\ppf-l)\,q_{\rho}\,\gamma^{\mu} \big)\,\frac{\gamma_5}{l^2-m_\pi^2}\,,\notag \\
%%%
&K^{\mu}_{\pi\Delta} = \frac{1}{\sqrt{2}}\,\frac{\gamma_5}{l^2-m_\pi^2}\,\Big[
\big(5\,f_A^{(+)}-f_A^{(-)}\big)\,\slashed{l}\,S_{\Delta}^{\mu\nu}(\ppi-l) + \big(5\,f_A^{(+)} + f_A^{(-)}\big)\,\gamma^{\mu}\,l_{\rho}\,S_{\Delta}^{\rho\nu}(\ppi-l)\Big]\,l_{\nu}\,,\notag \\
%%%
&L^{\mu}_{\pi\Delta} = 2\,\sqrt{2}\,f_M\,\frac{\gamma_5}{l^2-m_\pi^2}\,\big(\slashed{q}\,S_{\Delta}^{\mu\nu}(\ppi-l)  - \gamma^{\mu}\,q_{\rho}\,S_{\Delta}^{\rho\nu}(\ppi-l) \big) \,l_{\nu}\,,\notag \\
%%%
&K^{\mu}_{N\pi N} =  \frac{1}{\sqrt{2}}\,\gamma_5\,\slashed{l}\,S_N(\ppf-l)\,\frac{\gamma^{\mu}\,\gamma_5}{l^2-m_\pi^2}\,S_N(\ppi-l)\,\slashed{l}\,\gamma_5\,,\notag \\
%%%
&K^{\mu}_{\Delta\pi N} = 
\frac{f_S}{\sqrt{2}}\,l_{\nu}\,S^{\nu\mu}_{\Delta}(\ppf-l)\,
S_N(\ppi-l)\,\frac{\slashed{l}\,\gamma_5}{l^2-m_\pi^2}\,,
%%%
\qquad
K^{\mu}_{N\pi\Delta}= 
\frac{f_S}{\sqrt{2}}\,\frac{\slashed{l}\,\gamma_5}{l^2-m_\pi^2}\,S_N(\ppf-l)\,S^{\mu\nu}_{\Delta}(\ppi-l)\,l_{\nu}
\,, \notag \\
%%%
&L^{\mu}_{\Delta\pi N} = \sqrt{2}\,f_E\,
l_{\nu}\,\big(S_{\Delta}^{\nu\mu}(\ppf-l)
\,\slashed{q} - S^{\nu\rho}_{\Delta}(\ppf-l)\,q_{\rho}
\,\gamma^{\mu}\big)\,S_{N}(\ppi-l)\,\frac{\slashed{l}\,\gamma_5}{l^2-m_\pi^2}\,, \notag \\
%%%
&L^{\mu}_{N\pi\Delta} = \sqrt{2}\,f_E\,
\frac{\slashed{l}\,\gamma_5}{l^2-m_\pi^2}\,S_{N}(\ppf-l)\,\big(\gamma^{\mu}\,q_{\nu}
\,S_{\Delta}^{\nu\rho}(\ppi-l)-\slashed{q}
\,S_{\Delta}^{\mu\rho}(\ppi-l)\big)\,l_{\rho}\,,\notag \\
%%%
&K^{\mu}_{\Delta\pi\Delta} =  \frac{1}{\sqrt{2}}\,l_{\nu}\,S_{\Delta}^{\nu\rho}(\ppf-l)\,g_{\rho\sigma}\,
\frac{\gamma_5\,\gamma^{\mu}}{l^2-m_\pi^2}\,S_{\Delta}^{\sigma\lambda}(\ppi-l)\,l_{\lambda}\,,
\label{eqn:defKandL}
\end{align}
with the baryon propagators \cite{Lutz:2001yb,Semke:2005sn,Haberzettl:1998rw}
\begin{align}
&S_{N}(k) = \frac{1}{\slashed{k}-M_N }\,,
 \nonumber\\
&S_{\Delta}^{\mu\nu}(k) =\frac{-\,1}{\slashed{k}-M_{\Delta}}
\bigg(g^{\mu\nu}-\frac{\gamma^{\mu}
\gamma^{\nu}}{d-1}+\frac{k^{\mu}\gamma^{\nu}-k^{\nu}\gamma^{\mu}}{(d-1)\,M_{\Delta}}-
\frac{(d-2)\,k^{\mu}k^{\nu}}{(d-1)\,M_{\Delta}^2}\bigg).
\end{align}
The loop functions proportional to the two-body LECs $g_S,g_V, g_T, f_\pm,\,f_M, g_F$, and the subleading axial $\Delta N$ transition LEC $f_E$ are taking from our previous works \cite{Hermsen:2024eth,Lutz:2020dfi}. New are the contributions proportional to the LEC $b_{\chi}$, where $b_{\chi}$ is subleading order \cite{Lutz:2020dfi}. As these bubble terms do not contribute in the infinite volume, we omitted them in our previous works. Due to the need to introduce the $L^{\mu}_{N\pi}$ and $L^{\mu}_{\pi N}$ functions,  we rewrote the $K^{\mu}$ functions in comparison to our previous work, so that now $J^{\mu}_{\ldots}$ are the one-loop amplitudes deduced from the chiral Lagrangian density and not only their on-shell projections. 
\clearpage

\section{Finite-box basis elements }\label{app:basisElements}

We next turn to the tadpole results. Our basis elements are given by 
\begin{align}
&\bar I^{(\bar a , \, h,  \,a)}_{\pi}(\vec{ \bar p}, \,\vec{ p}\,) = \delta_{\bar a,0}\,\delta_{h,0}\,\delta_{a,0}\, \bar I_\pi
+ \,
\sum^{\vec n \neq 0}_{\vec{n}\in\mathbb{Z}^3}\,X_\pi^{(\bar{a} + 2\,h + a)}[\vec{x}_n]\,\big(\vec{\bar p}\cdot \vec{x}_n \big)^{\bar a}\,\big(\vec{x}_n^{\,2}\big)^h\, \big( \vec{p}\cdot \vec{x}_n \big)^a
\ ,\nonumber \\
&\bar  I_\pi = \frac{m_\pi^{2}}{16 \, \pi^2}\,\log\frac{m_\pi^2}{\mu^2} \sim Q^{2}
\,, \nonumber \\ 
&X_\pi^{(n_x)}[\,\vec{x}_n] = \frac{(-1)^{n_x /2}}{4\,\pi^2}\,K_{1 + n_x } \big[\,m^2_\pi,\, \vert \vec{x}_n\vert\, \big] \,, \nonumber \\
&K_{n} [\,m^2, \,x ] = \bigg( \frac{m }{x} \bigg)^n
K_n (\,m \, x\,) \,,
\label{def-tadpole-basis}
\end{align}
where $p^2_0= M_N^2+\vec{ p\,}^2$,
$\bar p^2_0= M_N^2+\vec{\bar  p}\,^2$, and $K_n(m\,x)$ denotes the modified Bessel function. Note that $\bar I^{(\bar a,\,h , \,a)}_{\pi}(\vec{ \bar p},\, \vec{p}\,) = 0$ if $\bar{a}+a$ odd.\\\par 
The basis bubble elements are  given by 
\begin{align}
&\bar I^{(\bar a,\, h , \,a)}_{L\pi}(\vec{ \bar p}, \vec{ p}\,) = \delta_{\bar a,0}\,\delta_{ h,0}\,\delta_{ a,0}\, \bar I_{ L \pi}
+ 
\sum^{\vec n \neq 0}_{\vec{n}\in\mathbb{Z}^3} \,X_{L\pi}^{(\bar{a} + 2\,h +  a)}[\,\vec{\bar{p}}, \, \vec{x}_n\,]\,
 \big(\vec{\bar p}\cdot \vec{x}_n \big)^{\bar{a}}\, \big(\vec{x}_n^{\,2}\big)^h\, \big(\vec{p}\cdot\vec{x}_n \big)^a
 \,, \nonumber \\
&\bar{I}_{L\pi} = \frac{\gamma_N^L-2}{16\,\pi^2} - \frac{1}{16\,\pi^2}\,\int_0^1\,\dd{z} \,\log\,\frac{F_{L\pi}(z)}{M_L^2} \sim Q
\,, \nonumber \\
&X_{L\pi}^{(n_x)}[\,\vec{\bar{p}},\, \vec{x}_n\,] =  \frac{(-1)^{\lfloor (1-n_x ) / 2 \rfloor}}{8\,\pi^2}\,\int_0^1 \dd{z}\,K_{n_x}\big[\, F_{L\pi}(z),\, \vert \vec{x}_n \vert \big]\,
 {\rm cs}_{n_x}\big( z\,\vec{ \bar p}\cdot \vec{x}_n \big)
 \,, \nonumber \\
&F_{L\pi}(z) = z\,M_L^2 + (1-z)\,m_\pi^2- (1-z)\,z\,\bar p^2
\,,\nonumber \\
%%%
% right bubble 
%%%
&\bar I^{(\bar a,\, h , \,a)}_{\pi R}(\vec{ \bar p}, \vec{ p}\,) = \delta_{\bar a,0}\,\delta_{ h,0}\,\delta_{ a,0}\, \bar I_{\pi R}
+ \sum^{\vec n \neq 0}_{\vec{n}\in\mathbb{Z}^3} \,X_{\pi R}^{(\bar{a} + 2\,h + a)}[\,\vec{x}_n,\,\vec{p}\,]\,
 \big(\vec{\bar p}\cdot \vec{x}_n \big)^{\bar a}\,\big(\vec{x}_n^{\,2}\big)^h\, \big( \vec{p}\cdot \vec{x}_n \big)^a
 \,, \nonumber \\
&\bar{I}_{\pi R} = \frac{\gamma_N^R-2}{16\,\pi^2} - \frac{1}{16\,\pi^2}\,\int_0^1\,\dd{z} \,\log\,\frac{F_{\pi R}(z)}{M_R^2} \sim Q
\,,\nonumber \\
&X_{\pi R}^{(n_x)}[\,\vec{x}_n,\,\vec{p}\,] =  \frac{(-1)^{\lfloor (1-n_x) / 2 \rfloor}}{8\,\pi^2}\,\int_0^1 \dd{z}\,K_{n_x} \big[\, F_{\pi R}(z),\, \vert\vec{x}_n\vert \big] \,
 {\rm cs}_{n_x}\big( z\,\vec{ p }\cdot \vec{x}_n  \big)
 \,,\nonumber \\
&F_{\pi R}(z) = z\,M_R^2 + (1-z)\,m_\pi^2- (1-z)\,z\,p^2\,, 
\label{def-bubble-basis}
\end{align}
where we use the short-hand notation ${\text{cs}}_n(x)=\cos (x)$ for $n$ even, but ${\text{cs}}_n(x)=\sin (x)$ for $n$ odd. The brackets $\lfloor x \rfloor := \operatorname{max}\{k\in\mathbb{Z}\,\vert\,k\leq z\}$ denote the floor function. With the expressions of Eq. (\ref{def-bubble-basis}) we separated the ultraviolet-convergent finite-box contributions from the ultraviolet-divergent infinite-volume contributions, $ \bar I_{L\pi}$ and $ \bar I_{\pi R}$. To ensure a consistent power counting for the two domains $m_\pi\sim \Delta$ and $m_\pi \ll \Delta$, the subtaction terms $\gamma^{L/R}_N$ are needed. In Ref. \cite{Hermsen:2024eth} a more detailed discussion is given, we only cite the results:
\begin{equation}
r = \frac{\Delta}{M}\,,\qquad
a = r\,(2+r)\,,\qquad
\gamma^{\Delta}_N = a\,\log\,\frac{1+a}{a}\,,\qquad
\gamma^{N}_{N} = 0\,.
\label{eqn:subtraction-bubble}
\end{equation}
The triangle basis functions take the form
\begin{align}
&\bar I^{(\bar a,\, h , \,a)}_{L^m \pi R^n}(\vec{ \bar p}, \,\vec{p} \,) = \delta_{\bar a,0}\,\delta_{ h,0}\,\delta_{ a,0}\, \bar I_{ L^m \pi R^n}
+ \sum^{\vec n \neq 0}_{\vec{n}\in\mathbb{Z}^3} \,X_{L^m\pi R^n}^{(\bar{a} + 2\,h + a)}[\,\vec{\bar{p}},\,\vec{x}_n,\,\vec{p}\,]\,
 \big(\vec{\bar p}\cdot \vec{x}_n \big)^{\bar a}\,\big(\vec{x}_n^{\,2}\big)^h\, \big( \vec{p}\cdot \vec{x}_n \big)^a
 \,, \nonumber \\ 
&\bar{I}_{L^m\! \pi R^n} = \frac{1}{16\,\pi^2} \int_0^1 \dd{z_L} \int_0^{1-z_L} \dd{z_R}\, \frac{z_L^m\,z_R^n}{ F_{L\pi R}(z_L,z_R) } - \frac{\gamma^{(m,n)}_{L\pi R}}{16\,\pi^2\, M^2} \sim Q^0
\,, \nonumber \\
& X_{L^m \pi R^n}^{(n_x)}[\,\vec{\bar{p}},\,\vec{x}_n,\,\vec{p}\,] =
\frac{(-1)^{\lfloor (1-n_x ) / 2 \rfloor}}{16\,\pi^2}\int_0^1 \dd{z_L} \int_0^{1-z_L} \dd{z_R}
\,K_{-1 + n_x}\big[\, F_{L\pi R}(z_L,z_R),\, \vert\vec{x}_n\vert \big]\,
{\rm cs}_{n_x}\big( (z_L\,\vec{ \bar p} + z_R\,\vec{p}\,)\cdot \vec{x}_n \,\big)\,z_L^m\,z_R^n
\,, \nonumber\\
&F_{L\pi R}(z_L,z_R) = m_\pi^2 
- z_L\,\big( \bar p^2 - M_L^2 + m_\pi^2  \big)  - z_R\,\big( p^2 - M_R^2 + m_\pi^2  \big)
+\,z_L^2\,\bar{p}^2 +  z_L\,z_R\,(\bar{p}^2 +  p^2 - t) + z_R^2\,p^2\,, 
\label{def-triangle-basis}
\end{align}
where we find two options, either choose $m= 0$ or $n= 0$ always.
The first term in Eq. (\ref{def-triangle-basis}) with $ \bar I_{ L^m\! \pi R^n} \sim Q^0 $, is ultraviolet-finite always. Like in Ref. \cite{Hermsen:2024eth}, a finite subtraction is 
implemented to arrive at the scaling $\bar I_{ L^m\! \pi R^n} \sim Q^2 $ in the chiral domain $m_\pi \ll  M_\Delta-M_N $. We recall the required terms \cite{Hermsen:2024eth}:
\begin{align}
&\gamma^{(0,0)}_{\Delta\pi N}=\gamma^{(0,0)}_{N\pi\Delta}=
 \frac{1}{a}\,\log (1+ a) +\log \frac{ 1+a}{a} 
 \,,\nonumber\\
&\gamma^{(1,0)}_{\Delta\pi N}=\gamma^{(0,1)}_{N\pi\Delta}= \frac{1+a}{3\,a}-\frac{1}{3\,a^2}\,
\log (1+ a) - \frac{a}{3}\,\log \frac{ 1+a}{a} 
\,,\nonumber\\
&\gamma^{(2,0)}_{\Delta\pi N}=\gamma^{(0,2)}_{N\pi\Delta}= \frac{-2 + a +a^2-2\, a^3}{10\,a^2}+\frac{1}{5\,a^3}\,
\log (1+ a) + \frac{a^2}{5}\,\log \frac{ 1+a}{a} 
\,,\nonumber\\
&\gamma^{(0,0)}_{\Delta \pi \Delta }=\log \frac{ 1+a}{a}
\,,\nonumber\\
&\gamma^{(1,0)}_{\Delta \pi \Delta } = \gamma^{(0,1)}_{\Delta \pi \Delta }= \frac{1}{2}\,\Big( 1-a\,\log \frac{ 1+a}{a} \Big) 
\,,\nonumber\\
&\gamma^{(2,0)}_{\Delta \pi \Delta }= 
\gamma^{(0,2)}_{\Delta \pi \Delta }= \frac{1}{6}\,\Big( 1-2\, a+2\,a^2\,\log \frac{ 1+a}{a} \Big)\,,
\end{align}
where the definition of $a$ is given in Eq. \eqref{eqn:subtraction-bubble}.

\section{Updated tensor decomposition in the finite-box case}
\label{app:redInBox}

The loop contributions of the form factors 
$G_{A}^{\text{loop}}(\bar p, p, x_n), \, G_P^{\text{loop}}(\bar p, p, x_n), \, G_T^{\text{loop}}(\bar p, p, x_n)$, and $G_X^{\text{loop}}(\bar p, p, x_n)$
can be expressed through the over-complete set of basis functions
\begin{align}
&J^{(\bar{a} \IS h \IS a; \IS n_x)}_{\pi} (\bar p,\,p,\,x_n) = i\,\,\mu^{4-d}\,
 \IntdDPi{l} \, \frac{e^{i\, \vec{l}\,\cdot\, \vec{x}_n}\,(\bar p \cdot l)^{\bar{a}}\,  (l^2)^h\,(l \cdot p)^{a}\,\,(l\cdot x_n)^{n_x}}{l^2-m_\pi^2}
\,, 
\nonumber \\
&J^{(\bar{a} \IS h \IS a; \IS n_x)}_{L \pi}(\bar p,\,p,\,x_n) = -\,i\,\mu^{4-d}\,
\IntdDPi{l} \, \frac{e^{i\, \vec{l}\,\cdot\, \vec{x}_n}\,(\bar p \cdot l)^{\bar{a}}\,  (l^2)^h\,(l \cdot p)^{a}\,\,(l\cdot x_n)^{n_x}}{\big((l-\bar p)^2- M_L^2\big)\big(l^2-m_\pi^2\big)}\,, 
\nonumber \\
&J^{(\bar{a} \IS h \IS a; \IS n_x)}_{\pi R} (\bar p,\,p,\,x_n) = -\,i\,\mu^{4-d}\,
 \IntdDPi{l}  \, \frac{e^{i\, \vec{l}\,\cdot\, \vec{x}_n}\,(\bar p \cdot l)^{\bar{a}}\,  (l^2)^h\,(l \cdot p)^{a}\,\,(l\cdot x_n)^{n_x}}{\big(l^2-m_\pi^2\big)\big((l- p)^2- M_R^2\big)}\,,
\nonumber \\
&J^{(\bar{a} \IS h \IS a; \IS n_x)}_{L \pi R}(\bar p,\,p,\,x_n) = i\,\mu^{4-d}\,
 \IntdDPi{l}  \, \frac{e^{i\, \vec{l}\,\cdot\, \vec{x}_n}\,(\bar p \cdot l)^{\bar{a}}\,  (l^2)^h\,(l \cdot p)^{a}\,\,(l\cdot x_n)^{n_x}}{\big((l- \bar{p})^2- M_L^2\big)\big(l^2-m_\pi^2\big)\big((l- p)^2- M_R^2\big)}\,,
\label{def-overcomplete-basis}
\end{align}
where $L,R$  either refer to a nucleon or a $\Delta$-isobar field. In Ref. \cite{Hermsen:2025vds} we derived a reduction scheme into a minimal set of basis elements for $n_x=0$, because only these elements are needed for $G_{A}^{\text{loop}}(\bar p, p, x_n)$. In this Section we will extend this reduction to the case $n_x\neq 0$. \\\par

After a Feynman parametrization of the over-complete basis elements in Eq. \eqref{def-overcomplete-basis} the generic target function of our reduction scheme is
\begin{equation}\begin{aligned}[b]
&J_{\kkf}^{(\aaf,\pll,\aai;\,n_x)}[\ppf,\,m^2,\,\vec{v},\,\xn,\,\ppi] 
= \mu^{4-d}\,(-1)^{\kkf-1}\,i\,\Gamma(\kkf)\,\IntdDPi{l} \, \frac{e^{\ii(\kl+\vec{v})\cdot\kx}\,\big(\ppf\cdot l\big)^{\aaf}\,\big(l^2\big)^h\,\big(l\cdot\ppi\big)^{\aai}\,\big(l\cdot x_n\big)^{n_x}}{\big(l^2-m^2 + \ii\,\varepsilon\big)^{\kkf}}\,.
\end{aligned}\end{equation}
The energy integration is done in the same way as in our previous work \cite{Hermsen:2025vds}. We use the tensor decomposition of our previous work \cite{Hermsen:2025vds}. We find after some tedious algebra the updated basis identity of our in-box decomposition scheme:
\begin{equation}\begin{aligned}[b]
\label{chap:redOpenMomenta:secll-lpf-lpi-lxn-eqn:definition-TensorDecomp}
&J_{\kkf}^{(\aaf,\pll,\aai;\,n_x)}[\ppf,\,m^2,\,\vec{v},\,\xn,\,\ppi] = \mu^{4-d}\,(-1)^{\kkf-1}\,i\,\Gamma(\kkf)\,\IntdDPi{l}\frac{e^{\ii(\kl+\vec{v})\cdot\kx}\,\big(\ppf\cdot l\big)^{\aaf}\,\big(l^2\big)^h\,\big(l\cdot\ppi\big)^{\aai}\,\big(l\cdot x_n\big)^{n_x}}{\big[l^2-m^2+\ii\,\varepsilon\big]^{\kkf}}\\
&\hspace{1em}
 = \sum_{\bbf,\bar{h},\bbi,k;\,n_y}^{\infty}\,C^{(\aaf,\,\pll,\,\aai;\,n_x)}_{\bbf\,\bar{h}\,\bbi,k;\,n_y}
\,\big(\ppf\cdot\ppi\big)^k\,\big(\kx^2\big)^{\hhi-\bar{h}+n_x-n_y}\,
\ppf[\bbf-k]\,\big(\kf\cdot\kx\big)^{\aaf-\bbf}
 \\ &\hspace{8em}\times
\,V_{\kkf,\bar{h}+\frac{\bbf+\bbi}{2}+n_y}^{(\aaf-\bbf + 2\,(\pll-\bar{h})+n_x-2\,n_y + \aai-\bbi)}[m^2,\,\kx,\,\vec{v}]\,\big(\ki\cdot\kx\big)^{\aai-\bbi}\,\ppi[\bbi-k]
\,\,,\\\\
&C^{(\aaf,\,\pll,\,\aai;\,n_x)}_{\bbf\,\bar{h}\,\bbi,k;\,n_y} = 
\begin{dcases}
&0\,, \text{ if } \quad
	\bbf-k\text{ odd }\quad
 \vert\vert\quad
 \bbi-k\text{ odd }
 \quad\vert\vert \quad
\bbf+\bbi \text{ odd }\\ 
&\hspace{5em} \vert\vert\quad 
n_y > \hhi-\bar{h} + (\aaf-\bbf + n_x +\aai-\bbi)/2\,,\\
&k!\binom{\aaf}{\bbf}\,\binom{\bbf}{k}\,(\bbf-k-1)!!\,\binom{\aai}{\bbi}\,\binom{\bbi}{k}\,(\bbi-k-1)!!\,\binom{\pll}{\bar{h}}
\\ &\qquad 
\times \,\prod_{n=0}^{\bar{h}-1}\big(d+2(n + \aaf + \pll-\bar{h} +\aai)-\bbf-\bbi\big)
\\ &\qquad \times\,
(-1)^{\hhi+\bar{h}+n_y}\,\sum_{i=0}^{n_x}\binom{n_x}{2\,n_y-i}\binom{2\,n_y-i}{i}\binom{\aaf-\bbf +2\,(\hhi-\bar{h})+\aai-\bbi}{i}
\\
&\hspace{2em}
\times\,(2\,(n_y-i)-1)!!\,i!
\,,\text{ else }
\end{dcases}
\end{aligned}\end{equation}
where we recall the generalized basis elements from our previous work \cite{Hermsen:2025vds}
\begin{align}
&V_{\bar{k},k}^{(n_x)}[m^2,\,\vec{x}_n,\,\vec{v}\,] = I_{\bar{k}}^{(k)}[m^2]\,\delta_{n_x,0}\,\delta_{\kx,\vec{0}} + X_{\bar{k},k}^{(n_x)}[m^2,\,\vec{x}_n,\,\vec{v}\,]\,(1-\delta_{\kx,\vec{0}})\,,\notag\\[.75em]
%%%
%definition infinite volume
%%%
&I^{(k)}_{\bar{k}}[m^2] = (-1)^{\bar{k}-1}\,\Gamma(\bar{k})\Bigg[\prod_{n=0}^{k-1}\frac{1}{d+2\,n}\Bigg]\,\int\frac{\dd[d]{l}}{(2\,\pi)^{d}}\frac{\mu^{4-d}\,i\,\big(l^2\big)^{k}}{\big(l^2-m^2+i\,\varepsilon\big)^{\bar{k}}}\,,\notag\\[.75em]
%%%
% definition finite volume effect
%%%
&X^{(n_x)}_{\bar{k},k}[m^2,\,\vec{x}_n,\,\vec{v}\,]
= (-1)^{n_x+k+(1+3\,d_t)/2}\,\frac{\Gamma\big(\bar{k}-k-n_x - d_t/2\big)}{2^{k+n_x+d_t}\,\pi^{d_t/2}}
\,\int\frac{\dd[d_s]{l}}{(2\,\pi)^{d_s}}\,\frac{\mu^{4-d_t-d_s}\,i^{n_x}\,e^{i\,(\vec{l}+\vec{v})\cdot\vec{x}_n}}{\big[\vec{l}^{2}+m^2\big]^{\bar{k}-k-n_x - d_t/2}}
\notag\\ 
&\phantom{X^{(n_x)}_{\bar{k},k}[m^2,\,\vec{x}_n,\,\vec{v}\,]}\,
= (-1)^{n_x+k}\,\frac{i^{n_x}\,e^{i\,\vec{v}\cdot\vec{x}_n}}{2^{\bar{k}+1}\,\pi^{2}}
K_{2-\bar{k}+n_x+k}[m^2, \vert \vec{x}_n\vert\,]\,.
\label{app:eqn-defGenBasis}
\end{align}
We recall that our tensor decomposition and the following reduction scheme hold for any value of $m^2$, where the rewrite into the modified Bessel function is only valid for $\Im(m^2)\neq 0$  and  $\Re(m^2)>0$. In the following, we consider two types of heavy fields $L,R$, and one light field $Q$. In the main part of this work, we set $L,R\in\{N,\Delta\}$ and $Q=\pi$. We continue with a further set of intermediate functions \cite{Hermsen:2025vds},
\begin{align}
&\kernelQ{\kki}{n_x}[\,\vec{x}_n\,] = V_{1,k}^{(n_x)}[m_Q^2,\,\kx,\,\vec{0}\,]\,,  \notag\\
%%%
&
\kernelLQ{n}{k}{n_x}[\,\kf,\,\kx]
= \IntFeyPa{\FPL}\,\FPL[n]\,V^{(n_x)}_{2,k}[F_{LQ}(\FPL),\,\kx,\,\FPL\,\kf]
\,, \notag\\
&\kernelQR{n}{k}{n_x}[\,\kx,\,\ki]  = \IntFeyPa{\FPR}\,\FPR[n]\,V^{(n_x)}_{2,k}[F_{QR}(\FPR),\,\kx,\,\FPR\,\ki]  \,, \notag\\
%%%
&\kernelLQR{m}{n}{k}{n_x}[\,\kf ,\,\kx,\, \ki ]
 = 
\IntFeyPa{\FPL}\,\IntFeyPa[1-\FPL]{\FPR}\,\FPL[m]\,\FPR[n]\,V_{3,k}^{(n_x)}\big[F_{LQR}(\FPL,\,\FPR),\,\kx,\,\FPL\,\kf + \FPR\,\ki\big]\,, \notag\\
%%%
% mod masses
%%%
&  F_{L Q}(z_L) = z_L\,M_L^2 + (1-z_L)\,m_Q^2- (1-z_L)\,z_L\,\bar p^2\,,     \notag\\
& F_{Q R}(z_R) = z_R\,M_R^2 + (1-z_R)\,m_Q^2 - (1-z_R)\,z_R\,p^2\,, \notag\\
&F_{L Q R}(z_L,z_R) = m_Q^2 
- z_L\,\big( \bar p^2 - M_L^2 + m_Q^2  \big)  - z_R\,\big( p^2 - M_R^2 + m_Q^2  \big)
+\,z_L^2\,\bar{p}^2 + z_L\,z_R\,(\bar{p}^2 +  p^2 - t) + z_R^2\,p^2\,.
\label{def-kernel}
\end{align}
Note that $F_{L \pi}(z_L) = F_{L \pi R}(z_L,0)$ and $F_{\pi R}(z_R) = F_{L \pi R}(0,z_R)$ hold. The over-complete set basis elements from Eq.~\eqref{def-overcomplete-basis} is then expressed in the intermediate basis elements
\begin{align}
%%%
% tadpole
%%%
&\genQBox{\aaf}{\hhi}{\aai}{n_x} 
=\hspace{-0.3em}\sum_{\bbf,\bar{h},\bbi,k,n_y=0}^{\infty}\,C^{(\aaf,\,\pll,\,\aai;\,n_x)}_{\bbf\,\bar{h}\,\bbi,k;\,n_y}
\,\big(\ppf\cdot\ppi\big)^k\,\big(\kx^2\big)^{\hhi-\bar{h}+n_x-n_y}\,
\notag \\ &\hspace{7em}\times\, 
\ppf[\bbf-k]\,\big(\kf\cdot\kx\big)^{\aaf-\bbf}
\kernelQ{\bar{h}+\frac{\bbf+\bbi}{2}+n_y}{\aaf-\bbf + 2\,(\pll-\bar{h})+n_x-2\,n_y + \aai-\bbi}
\,\big(\ki\cdot\kx\big)^{\aai-\bbi}\,\ppi[\bbi-k]\,, \notag \\
%%%
% left bubble
%%%
&\genLQBox{\aaf}{\pll}{a}{n_x}=\hspace{-0.3em}\sum_{\vert\mathbf{\ccf}\vert,\,\vert\mathbf{\eef}\vert,\,\vert\mathbf{\cci}\vert,\,\vert\mathbf{n}_z\vert}^{\aaf,\pll,\aai,n_x}\,\sum_{\bbf,\bbi,{\bar{h}},k,n_y=0}^{\infty}\,\binom{\aaf}{\mathbf{\ccf}}\,\binom{\pll}{\mathbf{\eef}}\,\binom{\aai}{\mathbf{\cci}}\,\binom{n_x}{\mathbf{n}_z}\,C^{(\ccf_1 + \eef_2,\, \bar{e}_1,\,\cci_1;\,n_{z1})}_{\bbf\,{\bar{h}}\,\bbi,k;\,n_y}\,(-1)^{n_{z2}}\,2^{\bar{e}_2}
\big(\ppf\cdot\ppi\big)^{k+\ccf_2}\,\big(\kx^2\big)^{\eef_1-\bar{h}+n_{z1}-n_y}
\notag \\ &\hspace{7em} \times
\,\ppf[\bbf-k+2\,\ccf_2+2\,\eef_3]
\,\big(\kf\cdot\kx\big)^{\ccf_1+\eef_2-\bbf + n_{z2}}
\kernelLQ{\ccf_2+\eef_2  +2\,\bar{e}_3+\cci_2 + n_{z2}}{\bar{h}+\frac{\bbf+\bbi}{2}+n_y}{\ccf_1+\eef_2-\bbf+2\,(\bar{e}_1-\bar{h}+n_{z1}-n_y)+\cci_1-\bbi}\,\big(\ki\cdot\kx\big)^{ \cci_1-\bbi}\,\ppi[\bbi-k]\,,\notag \\
%%%
% right bubble 
%%%
&\genQRBox{\aaf}{\pll}{a}{n_x}=\hspace{-0.3em}\sum_{\vert\mathbf{\ccf}\vert,\,\vert\mathbf{\eef}\vert,\,\vert\mathbf{\cci}\vert,\,\vert\mathbf{n}_z\vert}^{\aaf,\pll,\aai,n_x}\,\sum_{\bbf,\bbi,{\bar{h}},k,n_y=0}^{\infty}\,\binom{\aaf}{\mathbf{\ccf}}\,\binom{\pll}{\mathbf{\eef}}\,\binom{\aai}{\mathbf{\cci}}\,\binom{n_x}{\mathbf{n}_z}\,C^{(\ccf_1,\, \bar{e}_1,\,\bar{e}_2+\cci_1;\,n_{z1})}_{\bbf\,{\bar{h}}\,\bbi,k;\,n_y}\,(-1)^{n_{z2}}\,2^{\bar{e}_2}\,
\big(\ppf\cdot\ppi\big)^{k+\ccf_2}\,\big(\kx^2\big)^{\eef_1-\bar{h}+n_{z1}-n_y}
 \notag \\ &\hspace{7em} \times\,
\,\ppf[\bbf-k]\,\big(\kf\cdot\kx\big)^{\ccf_1-\bbf}
\kernelQR{\ccf_2+\eef_2  +2\,\bar{e}_3+\cci_2 + n_{z2}}{\bar{h}+\frac{\bbf+\bbi}{2}+n_y}{\ccf_1-\bbf+2\,(\bar{e}_1-\bar{h}+n_{z1}-n_y)+\bar{e}_2 + \cci_1-\bbi}\,\big(\ki\cdot\kx\big)^{\eef_2 + \cci_1-\bbi+n_{z2}}\,\ppi[\bbi-k+2\,\bar{e}_3+2\,\cci_2]\,, \notag \\
%%%
% triangle
%%%
&\genLQRBox{\aaf}{\hhi}{\aai}{n_x} =\hspace{-0.3em} 
\sum_{\vert\mathbf{\ccf}\vert,\,\vert\mathbf{\eef}\vert,\,\vert\mathbf{\cci}\vert,\,\vert\mathbf{n}_z\vert}^{\aaf,\pll,\aai;\,n_x}\,\sum_{\bbf,\bar{h},\bbi,k,n_y}^{\infty}\,\binom{\aaf}{\mathbf{\ccf}}\,\binom{\pll}{\mathbf{\eef}}\,\binom{\aai}{\mathbf{\cci}}\,\binom{n_x}{\mathbf{n}_z}\,C^{(\ccf_1+\eef_2,\,\eef_1,\,\eef_3+\cci_1;\,n_{z1})}_{\bbf\,\hhi\,\bbi,k;\,n_y}
\,(-1)^{n_{z2} + n_{z3}}
\,2^{\eef_2+\eef_3+\eef_5}
\notag \\ &\hspace{7em} \times \,
\big(\ppf\cdot \ppi\big)^{k + \ccf_3+\eef_5+\cci_2}\,\big(\kx^2\big)^{\eef_1-\bar{h}+n_{z1}-n_y}
\ppf[\bbf -k + 2\,\ccf_2+2\,\eef_4 ]\,\big(\kf\cdot\kx\big)^{\ccf_1 + \eef_2 - \bbf + n_{z2}}
 \notag \\ &\hspace{7em} \times 
\,\kernelLQR{\ccf_2+\eef_2+2\,\eef_4+\eef_5 + \cci_2+n_{z2}}{\ccf_3+\eef_3+\eef_5+2\,\eef_6+\cci_3+n_{z3}}{\bar{h}+\frac{\bbf+\bbi}{2}+n_y}{\ccf_1+\eef_2-\bbf + 2\,(\eef_1-\bar{h})+n_{z1}-2\,n_y + \eef_3+\cci_1-\bbi}\,
\big(\ki\cdot\kx\big)^{\eef_3 + \cci_1 - \bbi + n_{z3}}
\,\ppi[\bbi - k + 2\,\eef_6+2\,\cci_3]\,,
 \label{res-overcomplete-basis}
\end{align}
where we use the multi-indices
\begin{equation}
\mathbf{\bar{c}} = (\bar{c}_1,\,\ldots,\,\bar{c}_i)\,,\qquad
\mathbf{\bar{e}} = (\bar{e}_1,\,\ldots,\,\bar{e}_j)\,,\qquad
\mathbf{c} = (c_1,\,\ldots,\,c_i)\,,\qquad 
\mathbf{n}_z = (n_{z1},\,\ldots,\,n_{zi})\,,
\end{equation}
with $i=2,j=3$ if the integral is a bubble and $i=3,\, j=6$ if it is a triangle, respectively. Furthermore, we used the short-hand notations
\begin{equation}
\sum_{\vert\mathbf{\bar{c}}\vert,\,\vert\mathbf{\bar{e}}\vert,\,\vert\mathbf{c}\vert,\,\vert\mathbf{n}_z\vert}^{\bar{a},h,a,n_x} = 
\sum_{\vert\mathbf{\bar{c}}\vert=\bar{a}}
\sum_{\vert\mathbf{\bar{e}}\vert=h}
\sum_{\vert\mathbf{c}\vert=a}
\sum_{\vert\mathbf{n}_z\vert=n_x}\,, \qquad
\sum_{\vert\alpha\vert=k} = \sum_{\substack{\alpha_1+\ldots + \alpha_n = k \\ \alpha_1,\,\ldots\,,\alpha_n\geq 0} }\,,\qquad 
\binom{k}{\alpha} = \binom{k}{\alpha_1,\,\ldots,\,\alpha_n}\,.
\end{equation}
In our previous work we derived a reduction scheme of the intermediate basis in Eq. \eqref{def-kernel} with $k\to0$ for all elements, additional $n\to0$ for the bubbles and either $n\to0$ or $m\to0$ for the triangle elements. We then defined the unrenormalized in-box basis elements
\begin{align}
&I_{Q}^{(\aaf,\,\hhi,\,\aai)}(\kf,\ki) = \SumAll\,\big(\kf\cdot\kx\big)^{\aaf}\,\big(\kx^2\big)^{\hhi}\,\big(\ki\cdot\kx\big)^{\aai}\,\kernelQ{0}{\aaf+2\,\hhi+\aai}[\,\kx\,]\,,\notag\\
%%%
&I_{LQ}^{(\aaf,\,\hhi,\,\aai)}(\kf,\ki) = \SumAll\,\big(\kf\cdot\kx\big)^{\aaf}\,\big(\kx^2\big)^{\hhi}\,\big(\ki\cdot\kx\big)^{\aai}\,\kernelLQ{0}{0}{\aaf+2\,\hhi+\aai}[\,\kf,\,\kx\,]\,,\notag\\
%%%
&I_{QR}^{(\aaf,\,\hhi,\,\aai)}(\kf,\ki) = \SumAll\,\big(\kf\cdot\kx\big)^{\aaf}\,\big(\kx^2\big)^{\hhi}\,\big(\ki\cdot\kx\big)^{\aai}\,\kernelQR{0}{0}{\aaf+2\,\hhi+\aai}[\,\kx,\,\ki\,] \,,\notag\\
%%%
&I_{L^mQR^n}^{(\aaf,\,\hhi,\,\aai)}(\kf,\ki) = \SumAll\,\big(\kf\cdot\kx\big)^{\aaf}\,\big(\kx^2\big)^{\hhi}\,\big(\ki\cdot\kx\big)^{\aai}\,\kernelLQR{m}{n}{0}{\aaf +2\,\hhi + \aai }[\,\kf ,\,\kx, \, \ki[2] ]\,.
\end{align}

The renormalized results for $Q\to\pi$ can be found in Eqs. \eqref{def-tadpole-basis},
\eqref{def-bubble-basis}, and \eqref{def-triangle-basis}. 

\clearpage
\section{Results for $\bar{J}^P_{\ldots}$ and $\bar{J}^T_{\ldots}$ with an internal $\Delta$-isobar}
\label{res:JPJT}
\begin{align}
%%%
% JP form factor 
%%% 
&\bar{J}^P_{\Delta\pi} + \bar{J}^P_{\pi\Delta} = 
%%% fAP parts
\frac{5}{9}\,f_A^{+}\,M_N\,\bigg\{
20\,\Big[m_\pi^2\,\alpP{1}{2}{0} - 2\,M_N\,r\,\delta\,\alpP{1}{3}{0}\Big]\Big(\renLQ[\Delta\pi]{0}{0}{0} + \renQR[\pi\Delta]{0}{0}{0}\Big)
\notag \\ &\hspace{6em}
- 3\,r\,\bigg[3\,\alpP{1}{0}{1} - \frac{m_\pi^2}{t}\,\alpP{1}{2}{1}\bigg]\,\Big(\renLQ[\Delta\pi]{1}{0}{0} - \renLQ[\Delta\pi]{0}{0}{1} - \renQR[\pi\Delta]{1}{0}{0} + \renQR[\pi\Delta]{0}{0}{1}\Big)
\notag \\ &\hspace{6em}
- \bigg[1-3\,\frac{m_\pi^2}{t}\bigg]\,\alpP{1}{0}{3}\,\Big(\renLQ[\Delta\pi]{0}{1}{0} + \renQR[\pi\Delta]{0}{1}{0}\Big)
+ 3\,\bigg[1+\frac{m_\pi^2}{t}\bigg]\,\alpP{1}{0}{4}\,\Big(\renSLQ[\Delta\pi]{2} + \renSQR[\pi\Delta]{2}\Big)
\bigg\} 
\notag \\
%%% fAM parts
\notag &\hspace{4em}
- \frac{1}{9}\,f_A^{-}\,M_N\,\bigg\{
4\,\Big[- m_\pi^2\,\alpP{2}{2}{0} + 2\,M_N\,r\,\delta\,\alpP{2}{3}{0}\Big]\,\Big(\renLQ[\Delta\pi]{0}{0}{0} + \renQR[\pi\Delta]{0}{0}{0}\Big)
\notag \\ &\hspace{6em}
- 3\,r\,\bigg[3\,\alpP{2}{0}{1} - \frac{m_\pi^2}{t}\,\alpP{2}{2}{1}\bigg]\Big(\renLQ[\Delta\pi]{1}{0}{0} - \renLQ[\Delta\pi]{0}{0}{1} - \renQR[\pi\Delta]{1}{0}{0} + \renQR[\pi\Delta]{0}{0}{1}\Big)
\notag \\ &\hspace{6em}
- \bigg[1-3\,\frac{m_\pi^2}{t}\bigg]\,\alpP{2}{0}{3}\,\Big(\renLQ[\Delta\pi]{0}{1}{0} + \renQR[\pi\Delta]{0}{1}{0}\Big)
+ 3\,\bigg[1+\frac{m_\pi^2}{t}\bigg]\,\alpP{2}{0}{4}\,\Big(\renSLQ[\Delta\pi]{2} + \renSQR[\pi\Delta]{2}\Big)
\bigg\}
%%% fM parts
\notag \\ &\hspace{4em}
- \frac{32}{9}\,f_M\,M_N\,r^2\,\big[t-m_\pi^2\big]\,\alpP{3}{1}{0}\,\Big(\renLQ[\Delta\pi]{0}{0}{0} + \renQR[\pi\Delta]{0}{0}{0} \Big) + \mathcal{O}(Q^4)
\,, \notag \\[.5em]
%%%
% JT form factor
%%%
&\bar{J}^T_{\Delta\pi} + \bar{J}^T_{\pi\Delta} = 
-\,\frac{10}{9}\,f_A^{+}\,M_N\,r^2\,\bigg\{
2\,t\,\alpT{1}{1}{0}\,\Big(\renLQ[\Delta\pi]{0}{0}{0} - \renQR[\pi\Delta]{0}{0}{0}\Big)
-\,9\,\alpT{1}{0}{1}\,\Big(\renLQ[\Delta\pi]{1}{0}{0} - \renLQ[\Delta\pi]{0}{0}{1} + \renQR[\pi\Delta]{1}{0}{0} - \renQR[\pi\Delta]{0}{0}{1}\Big)
\bigg\} 
\notag \\ &\hspace{4em}
%%% fAM parts
+ \frac{2}{9}\,f_A^{-}\,M_N\,r^2\,\bigg\{
2\,t\,\alpT{2}{1}{0}\,\Big(\renLQ[\Delta\pi]{0}{0}{0} - \renQR[\pi\Delta]{0}{0}{0}\Big)
- 3\,\alpT{2}{0}{1}\,\Big(\renLQ[\Delta\pi]{1}{0}{0} - \renLQ[\Delta\pi]{0}{0}{1} + \renQR[\pi\Delta]{1}{0}{0} - \renQR[\pi\Delta]{0}{0}{1}\Big)
\bigg\} 
\notag \\ &\hspace{4em}
%%% fAM parts
- \frac{32}{9}\,f_M\,M_N\,r^2\,
t\,\alpT{3}{1}{0}\,\Big(\renLQ[\Delta\pi]{0}{0}{0} - \renQR[\pi\Delta]{0}{0}{0}\Big) + \mathcal{O}(Q^5)\,, \notag \\
%%%
% GX term
%%%
&\bar{J}^X_{\Delta\pi} + \bar{J}^X_{\pi \Delta} = 
- \,\frac{2}{3}\,\Big(
5\,f_A^{+}\,r\,\alpX{1}{1}{1} - f_A^{-}\,r\,\alpX{2}{1}{1} + 4\,f_M\,\alpX{3}{1}{1} \Big)\,M_N\,\Big(X_{\Delta\pi}^{(1)} - X_{\pi\Delta}^{(1)}\Big)
\notag \\ &\hspace{6em}
+ \frac{2}{3}\,\Big(5\,f_A^{+}\,\alpX{1}{0}{4} - f_A^{-}\,\alpX{2}{0}{4}\Big)\,\big(\vec{q}\cdot \vec{x}_n\big)\,\frac{M_N}{t}\,\Big(X^{(2)}_{\Delta\pi} + X^{(2)}_{\Delta\pi}\Big) + \mathcal{O}(Q^4)\,.
\label{res:DpiBub}
\end{align}
After reduction and before renormalization, we encounter the terms
\begin{equation}
\frac{4}{9}\,M_N\,r\,\Big(25\,f_A^{+}\,\alpP{0}{1}{0} + f_A^{-}\,\alpP{0}{2}{0}\Big)\,\basisQ[\pi]{0}{0}{0}
-\,\frac{4}{9}\,M_N^3\,r^2\,\Big(25\,f_A^{+}\,\alpP{1}{0}{0} + f_A^{-}\,\alpP{2}{0}{0}\Big)\,\Big(\basisLQ[\Delta\pi]{0}{0}{0} + \basisQR[\pi\Delta]{0}{0}{0} \Big)
\end{equation}
in the $J^P_{\Delta\pi} + J^P_{\pi\Delta}$ contribution. These terms count as order $Q^2$ and are therefore power-counting violating, as 
$J^P_{\Delta\pi} + J^P_{\pi\Delta}  \sim Q^3$. Thus these terms need to renormalized to zero and we expect them to be absorbed by two-loop contributions \cite{Lutz:2020dfi}. Similar terms appear in the contributions $J^P_{\Delta\pi N} + J^{P}_{N\pi\Delta}$ and $J^{P}_{\Delta\pi\Delta}$, where they were also renormalized to zero. They only appear in $J^A_{\ldots}$ and $J^P_{\ldots}$ with at least one internal $\Delta$-isobar. We checked that these contribution also fulfill the Ward identity given in Eq. \eqref{chap:PPFF:eqn-AxialWardbox}.    
{\allowdisplaybreaks
\enlargethispage{2\baselineskip}
We turn to the $\Delta\pi N$-triangles, for which
\begin{align}
%%%
% JP form factor
%%%
&\bar{J}^P_{\Delta\pi N} + \bar{J}^P_{N\pi\Delta} = 
%%% fS term
\frac{1}{9}\,f_S\bigg\{
\Big[r\,t\,\alpP{4}{1}{0} - 4\,m_\pi^2\,\alpP{4}{2}{0} + 8\,M_N\,\delta\,\alpP{4}{3}{0}\Big]\,\Big(\renLQ[\Delta\pi]{0}{0}{0} + \renQR[\pi\Delta]{0}{0}{0}\Big)
\notag \\ &\hspace{6em}
+ 4\,\bigg[2\,\alpP{4}{0}{1} - 3\,\frac{m_\pi^2}{t}\,\alpP{4}{2}{1}\bigg]\,\Big(\renLQ[\Delta\pi]{1}{0}{0} - \renLQ[\Delta\pi]{0}{0}{1} - \renQR[\pi\Delta]{1}{0}{0} + \renQR[\pi\Delta]{0}{0}{1}\Big)
\notag \\ &\hspace{6em}
- \frac{1}{2}\,\bigg[1-3\,\frac{m_\pi^2}{t}\bigg]\,\alpP{4}{0}{3}\,\Big(\renLQ[\Delta\pi]{0}{1}{0} + \renQR[\pi\Delta]{0}{1}{0}\Big)
+\frac{3}{2}\,\bigg[1 + \frac{m_\pi^2}{t}\bigg]\,\alpP{4}{0}{4}\,\Big(\renSLQ[\Delta\pi]{2} + \renSQR[\pi\Delta]{2}\Big)\bigg\}
\notag \\ &\hspace{4em}
%%% Dreieck Basis 
+ \frac{2}{9}\,f_S\,M_N^2\bigg\{
8\,m_\pi^2\,\alpP{5}{2}{0}\,\Big(\renLQR[\Delta-\pi-N]{0}{0}{0}{0}{0}+\renLQR[N-\pi-\Delta]{0}{0}{0}{0}{0}\Big)
\notag \\ &\hspace{6em}
- \bigg[1-3\,\frac{m_\pi^2}{t}\bigg]\,\alpP{5}{0}{3}\,\Big(
\renLQR[\Delta-\pi-N]{0}{0}{0}{1}{0}
+\renLQR[N-\pi-\Delta]{0}{0}{0}{1}{0}
\Big)
+ 3\,\bigg[1 + \frac{m_\pi^2}{t}\bigg]\,\alpP{5}{0}{4}\,\Big(
\renSLQR[\Delta-\pi-N]{0}{0}{2} + \renSLQR[N-\pi-\Delta]{0}{0}{2}
\Big)
\notag \\ &\hspace{6em}
+ 2\,\bigg[1 + 3\,\frac{m_\pi^2}{t}\bigg]\,\alpP{6}{0}{1}\,\Big(
\renLQR[\Delta-\pi-N]{1}{0}{1}{0}{0}
-\renLQR[\Delta-\pi-N]{1}{0}{0}{0}{1}
-\renLQR[N-\pi-\Delta]{0}{1}{1}{0}{0}
+\renLQR[N-\pi-\Delta]{0}{1}{0}{0}{1}
\Big)
- 2\,\Big[t - 3\,m_\pi^2\Big]\,\alpP{7}{1}{0}\,\Big(
\renLQR[\Delta-\pi-N]{2}{0}{0}{0}{0}
+\renLQR[N-\pi-\Delta]{0}{2}{0}{0}{0}
\Big)
\bigg\}
%%%
% fE Terme
%%%
\notag \\ &\hspace{4em}
+\frac{4}{9}\,f_E\,M_N\,r\,\big[t-m_\pi^2\big]\,\alpP{8}{1}{0}\,\Big(\renLQ[\Delta\pi]{0}{0}{0} + \renQR[\pi\Delta]{0}{0}{0}\Big) + \mathcal{O}\big(Q^4\big)
\notag\,, \\[.25em]
%%%
% JT Form faktor
%%%
& \bar{J}^T_{\Delta\pi N} +  \bar{J}^T_{N\pi\Delta}
= \frac{2}{9}\,f_S\,\bigg\{2\,r\,t\,\alpT{4}{1}{0}\,\Big(\renLQ[\Delta\pi]{0}{0}{0} - \renQR[\pi\Delta]{0}{0}{0}\Big)  - 6\,r\,\alpT{4}{0}{1}\,\Big(
\renLQ[\Delta\pi]{1}{0}{0} - \renLQ[\Delta\pi]{0}{0}{1} 
+ \renQR[\pi\Delta]{1}{0}{0} 
- \,\renQR[\pi\Delta]{0}{0}{1}
\Big)
\notag \\ &\hspace{6em}
+4\,t\,m_\pi^2\,\betT{4}{2}{0}\,\Big(\renLQR[\Delta-\pi-N]{0}{0}{0}{0}{0} - \renLQR[N-\pi-\Delta]{0}{0}{0}{0}{0}\Big)
- 12\,m_\pi^2\,\alpT{5}{2}{1}\,\Big(\renLQR[\Delta-\pi-N]{0}{0}{1}{0}{0} - \renLQR[\Delta-\pi-N]{0}{0}{0}{0}{1} 
+ \,\renLQR[N-\pi-\Delta]{0}{0}{1}{0}{0} 
- \renLQR[N-\pi-\Delta]{0}{0}{0}{0}{1}\Big)
\notag \\ &\hspace{6em}
+ t\,\alpT{5}{1}{3}\,\Big(\renLQR[\Delta-\pi-N]{0}{0}{0}{1}{0} - \renLQR[N-\pi-\Delta]{0}{0}{0}{1}{0}\Big)
+ 6\,\alpT{5}{0}{4}\,\Big(\renLQR[\Delta-\pi-N]{0}{0}{2}{0}{0} - \renLQR[N-\pi-\Delta]{0}{0}{0}{0}{2}\Big)
- 6\,\alpT{5}{0}{5}\,\Big(\renLQR[\Delta-\pi-N]{0}{0}{1}{0}{1} - \renLQR[N-\pi-\Delta]{0}{0}{1}{0}{1}\Big)
\notag \\ &\hspace{6em}
+18\,r\,\alpT{5}{0}{6}\,\Big(\renLQR[N-\pi-\Delta]{0}{0}{2}{0}{0} - \renLQR[\Delta-\pi-N]{0}{0}{0}{0}{2}\Big)
- 12\,t\,m_\pi^2\,\betT{6}{2}{0}\,\Big(\renLQR[\Delta-\pi-N]{1}{0}{0}{0}{0} 
- \, \renLQR[N-\pi-\Delta]{0}{1}{0}{0}{0}\Big)
+7\,t\,\alpT{6}{1}{1}\,\Big(\renLQR[\Delta-\pi-N]{1}{0}{1}{0}{0} - \renLQR[N-\pi-\Delta]{0}{1}{0}{0}{1}\Big)
\notag \\ &\hspace{6em}
+ t\,\alpT{6}{1}{2}\,\Big(\renLQR[N-\pi-\Delta]{0}{1}{1}{0}{0} - \renLQR[\Delta-\pi-N]{1}{0}{0}{0}{1}\Big)
+2\,t^2\,\betT{7}{1}{0}\,\Big(\renLQR[\Delta-\pi-N]{2}{0}{0}{0}{0} - \renLQR[N-\pi-\Delta]{0}{2}{0}{0}{0}\Big)\bigg\}
\notag \\ &\hspace{4em}
+ \frac{8}{9}\,f_E\,M_N\bigg\{
2\,r\,t\,\alpT{8}{1}{0}\,\Big(\renLQ[\Delta\pi]{0}{0}{0} - \renQR[\pi\Delta]{0}{0}{0}\Big) + 4\,t\,m_\pi^2\,\betT{9}{2}{0}\,\Big(\renLQR[\Delta-\pi-N]{0}{0}{0}{0}{0} - \renLQR[N-\pi-\Delta]{0}{0}{0}{0}{0}\Big)
+t\,\alpT{9}{1}{3}\,\Big(\renLQR[\Delta-\pi-N]{0}{0}{0}{1}{0} - \renLQR[N-\pi-\Delta]{0}{0}{0}{1}{0}\Big)
\notag \\ &\hspace{6em}
+3\,\alpT{9}{0}{4}\Big(\renLQR[\Delta-\pi-N]{0}{0}{2}{0}{0} -2\,\renLQR[\Delta-\pi-N]{0}{0}{1}{0}{1} + \renLQR[\Delta-\pi-N]{0}{0}{0}{0}{2}
-\renLQR[N-\pi-\Delta]{0}{0}{2}{0}{0}
+2\,\renLQR[N-\pi-\Delta]{0}{0}{1}{0}{1}
-\renLQR[N-\pi-\Delta]{0}{0}{0}{0}{2}
\Big)
\notag \\ &\hspace{6em}
+4\,t\,\alpT{10}{1}{1}\,\Big(\renLQR[\Delta-\pi-N]{1}{0}{1}{0}{0}
-\renLQR[\Delta-\pi-N]{1}{0}{0}{0}{1} + \renLQR[N-\pi-\Delta]{0}{1}{1}{0}{0} - \renLQR[N-\pi-\Delta]{0}{1}{0}{0}{1}\Big)
+ 2\,t^2\,\betT{11}{1}{0}\,\Big(\renLQR[\Delta-\pi-N]{2}{0}{0}{0}{0} - \renLQR[\Delta-\pi-N]{0}{2}{0}{0}{0}\Big)
\bigg\} + \mathcal{O}\big(Q^5\big)\,,\notag \\
%%%
% JX form factor
%%%
& \bar{J}^X_{\Delta\pi N} +  \bar{J}^{X}_{N\pi\Delta}
= \frac{2}{3}\,\Big(f_S\,\alpX{4}{0}{1} + 3\,f_E\,M_N\,r\,\alpX{8}{0}{1}\Big)\Big(X_{\Delta\pi}^{(1)} - X_{\pi\Delta}^{(1)}\Big)
+\frac{1}{3}\,f_S\,\frac{\vec{q}\cdot\vec{x}_n}{t}\,\bigg\{\alpX{4}{0}{4}\,\Big(X_{\Delta\pi}^{(2)} + X_{\pi\Delta}^{(2)}\Big)
\notag \\ &\hspace{6em}
+ 4\,M_N^2\,\alpX{5}{0}{4}\,\Big(X_{\Delta^0\pi N^0}^{(2)} + X_{N^0\pi \Delta^0}^{(2)}\Big)\bigg\}
+ \frac{4}{3}\,f_S\,M_N^2\,\alpX{6}{0}{1}\,\Big(X_{\Delta^1\pi N^{0}}^{(1)} - X_{N^0\pi\Delta^1}^{(1)}\Big) + \mathcal{O}\big(Q^4\big)\, .
\label{res:DpiNTri}
\end{align}
Finally, for the $\Delta\pi\Delta$-triangles, we have
\begin{align}
%%%
% JP Form factor
%%%
&\bar{J}^P_{\Delta \pi \Delta} = 
- \, \frac{1}{27}\,\bigg\{
\Big[r\,t\,\alpP{9}{1}{0} + 34\,r\,m_\pi^2\,\alpP{9}{2}{0} - 20\,M_N\,\delta\,\alpP{9}{3}{0}\Big]\,\Big(\renLQ[\Delta\pi]{0}{0}{0} + \renQR[\pi\Delta]{0}{0}{0}\Big)
\notag \\ &\hspace{6em}
- \bigg[1 + 3\,\frac{m_\pi^2}{t}\bigg]\,\alpP{9}{0}{1}\,\Big(
\renLQ[\Delta\pi]{1}{0}{0}
-\renLQ[\Delta\pi]{0}{0}{1}
-\renQR[\pi\Delta]{1}{0}{0}
+\renQR[\pi\Delta]{0}{0}{1}
\Big)
\notag \\ &\hspace{6em}
- \bigg[1 - 3\,\frac{m_\pi^2}{t}\bigg]\,\alpP{9}{0}{3}\,\Big(\renLQ[\Delta\pi]{0}{1}{0} + \renQR[\pi\Delta]{0}{1}{0}\Big)
+3\,\bigg[1 + \frac{m_\pi^2}{t}\bigg]\,\alpP{9}{0}{4}\,\Big(\renSLQ[\Delta\pi]{2} + \renSQR[\pi\Delta]{2} \Big)
\bigg\}
\notag \\ &\hspace{4em}
+ \frac{4}{27}\,M_N^2\bigg\{
\Big[4\,r^2\,t\,\alpP{10}{1}{0} + 10\,m_\pi^2\,\alpP{10}{2}{0} - 20\,M_N\,r\,\delta\,\alpP{10}{3}{0}\Big]\,\renLQR[\Delta-\pi-\Delta]{0}{0}{0}{0}{0}
+\,\bigg[1-3\,\frac{m_\pi^2}{t}\bigg]\,\alpP{10}{0}{3}\,\renLQR[\Delta-\pi-\Delta]{0}{0}{0}{1}{0}
\notag \\ &\hspace{6em}
- 3\,\bigg[1 + \frac{m_\pi^2}{t}\bigg]\,\alpP{10}{0}{4}\,\renSLQR[\Delta-\pi-\Delta]{0}{0}{2}
+ 3\,\Big[2\,r\,t\,\alpP{11}{1}{0} - r\,m_\pi^2\,\alpP{11}{2}{0}\Big]\,\Big(\renLQR[\Delta-\pi-\Delta]{1}{0}{0}{0}{0}+\renLQR[\Delta-\pi-\Delta]{0}{1}{0}{0}{0}\Big)
\notag \\ &\hspace{6em}
- \bigg[1 + 3\,\frac{m_\pi^2}{t}\bigg]\,\alpP{11}{0}{1}\,\Big(
\renLQR[\Delta-\pi-\Delta]{1}{0}{1}{0}{0}
-\renLQR[\Delta-\pi-\Delta]{1}{0}{0}{0}{1}
-\renLQR[\Delta-\pi-\Delta]{0}{1}{1}{0}{0}
+\renLQR[\Delta-\pi-\Delta]{0}{1}{0}{0}{1}
\Big)
\notag \\ &\hspace{6em}
+ \big[t - 3\,m_\pi^2\big]\,\alpP{12}{1}{0}\,\Big(\renLQR[\Delta-\pi-\Delta]{2}{0}{0}{0}{0}+\renLQR[\Delta-\pi-\Delta]{0}{2}{0}{0}{0}\Big) 
\bigg\}+ \mathcal{O}(Q^4)
\,,\notag \\ 
%%%
% JT Form factor
%%%
&\bar{J}^T_{\Delta \pi \Delta}
= -\,\frac{1}{27}\,\bigg\{r^2\,t\,\alpT{12}{1}{0}\,\Big(\renLQ[\Delta\pi]{0}{0}{0} - \renQR[\pi\Delta]{0}{0}{0}\Big) 
 +12\,r\,\alpT{12}{0}{1}\,\Big(\renLQ[\Delta\pi]{1}{0}{0} - \renLQ[\Delta\pi]{0}{0}{1} + \renQR[\pi\Delta]{1}{0}{0} - \renQR[\pi\Delta]{0}{0}{1} \Big)\bigg\}
\notag\\ &\hspace{4em}
+ \frac{2}{9}\,\bigg\{\Big(4\,M_N^2\,r^2\,\alpT{13}{0}{1} - 11\,r^2\,t\,\alpT{13}{1}{1} - 4\,m_\pi^2\,\alpT{13}{2}{1} + 8\,M_N\,r\,\delta\,\alpT{13}{3}{1}\Big)\Big(\renLQR[\Delta-\pi-\Delta]{0}{0}{1}{0}{0} - \renLQR[\Delta-\pi-\Delta]{0}{0}{0}{0}{1}\Big)
\notag\\ &\hspace{6em}
- 2\,\alpT{13}{0}{4}\,\Big(\renLQR[\Delta-\pi-\Delta]{0}{0}{2}{0}{0} - \renLQR[\Delta-\pi-\Delta]{0}{0}{0}{0}{2}\Big)
\bigg\}
+\frac{1}{9}\,\bigg\{\Big(4\,M_N^2\,r^2\,t\,\alpT{14}{1}{0} - 11\,r^2\,t^2\,\betT{14}{1}{0} - 4\,t\,m_\pi^2\,\betT{14}{2}{0}
\notag\\ &\hspace{6em}
+  8\,M_N\,r\,t\,\delta\,\betT{14}{3}{0}\Big)\Big(\renLQR[\Delta-\pi-\Delta]{1}{0}{0}{0}{0} - \renLQR[\Delta-\pi-\Delta]{0}{1}{0}{0}{0}\Big) -2\,t\,\alpT{14}{1}{1}\,\Big(\renLQR[\Delta-\pi-\Delta]{1}{0}{1}{0}{0}-\renLQR[\Delta-\pi-\Delta]{0}{1}{0}{0}{1}\Big)
\notag\\ &\hspace{6em}
+ 2\,t\,\alpT{14}{1}{2}\,\Big(\renLQR[\Delta-\pi-\Delta]{0}{1}{1}{0}{0} - \renLQR[\Delta-\pi-\Delta]{1}{0}{0}{0}{1}\Big)\bigg\}
- \frac{8}{27}\,r\,t^2\,\betT{15}{1}{0}\,\Big(\renLQR[\Delta-\pi-\Delta]{2}{0}{0}{0}{0} - \renLQR[\Delta-\pi-\Delta]{0}{2}{0}{0}{0}\Big) + \mathcal{O}\big(Q^5\big)\,,\notag\\
%%%
%
%%%
&\bar{J}^X_{\Delta \pi \Delta} = 
\frac{1}{9}\,\alpX{9}{1}{1} \,\Big(X_{\Delta\pi}^{(1)} - X_{\pi\Delta}^{(1)}\Big) 
- \frac{2}{9}\,\frac{\vec{q}\cdot\vec{x}_n}{t}\,\bigg\{\alpX{9}{0}{4}\,\Big(X_{\Delta\pi}^{(2)} + X_{\pi\Delta}^{(2)}\Big) + 4\,M_N^2\,\alpX{10}{0}{4}\,X_{\Delta^0\pi\Delta^0}^{(2)}\bigg\} 
\notag \\ &\hspace{6em}
- \frac{4}{9}\,M_N^2\,\alpX{11}{1}{1}\,\Big(X_{\Delta^1\pi\Delta^0}^{(1)} - X_{\Delta^0\pi\Delta^1}^{(1)}\Big) + \mathcal{O}(Q^4)\,.
\label{res:DpiDTri}
\end{align}

}
\clearpage
\begin{figure*}
    \centering
\begin{subfigure}{0.32\textwidth}
    	\centering
    	 \includegraphics[height=0.17\textheight]{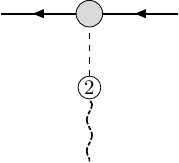}
    \end{subfigure}   
\begin{subfigure}{0.32\textwidth}
        \centering
        \includegraphics[height=0.17\textheight]{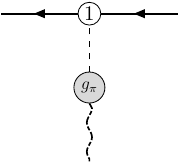}
    \end{subfigure}     
   \begin{subfigure}{0.32\textwidth}
        \centering
        \includegraphics[height=0.17\textheight]{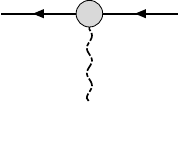}
    \end{subfigure} 
\caption{\small Different types of loop contributions, which enter the matrix element of the pseudoscalar density between two nucleons. The  dashed, dotted, and solid lines represent the pseudoscalar density, pion, and nucleon, respectively. The gray blob stands for one-loop contributions. The chiral order of the vertices is indicated. }
\label{chap:PSFF:figLoopContri}
\end{figure*}

\section{Explicit results for the pseudoscalar form factor of the nucleon}\label{app:resultsPseudoscalarFF}
In this Appendix, we will summarize our results for the pseudoscalar form factor. We first recall the decomposition of the QCD pseudoscalar density from Eq.~\eqref{chap:PseudoFF-startDecomp},
\begin{equation}\begin{aligned}[b]
&\langle N(\ppf)\,\vert\,m\,P_i(0)\,\vert\,N(\ppi)\rangle = \bar{u}(\ppf)\,F_{PS}(\ppf,\ppi)\,\tau_{i}\,u(\ppi)\,,\\
&F_{PS}(\ppf,\ppi) = \frac{i \,m_\pi^2\,M_N}{m_\pi^2-t}\,G_{PS}(\ppf,\ppi)\,\gamma_{5}\,.
\end{aligned}\end{equation}
We continue with the diagrams that enter the amplitude $F_{PS}(\ppf,\ppi)$. Most of the contributions to the pseudoscalar form factor are generated by the pseudoscalar density coupling to a pion, which couples to the nucleon line. The baryonic Lagrangian density from Ref.~\cite{Lutz:2020dfi} then implies one-loop corrections of the nucleon-pion-nucleon coupling. We display these types of diagrams schematically in the first panel of Fig.~\ref{chap:PSFF:figLoopContri}. The next types of contributions, shown in the second panel of Fig.~\ref{chap:PSFF:figLoopContri}, are the one-loop corrections of the mesonic Lagrangian density, discussed e.g. in Ref.~\cite{felix:PhD}, to the pseudoscalar-density-pion coupling.  These corrections are encoded in $g_\pi$, which is the coupling constant of the pseudoscalar density to the pion,
\begin{equation}
\langle 0 \,\vert\,m\,P_{i}(0)\,\vert\,\Phi_{\aai}(\ppi)\rangle = \delta_{i, \aai}\,g_{\pi}\,.
\label{chap:PseudoFF-defgpi}
\end{equation}
Note that compared to e.g. Refs. \cite{Fuchs:2003vw,Bijnens:2014yya,Gasser:1983yg}, we included the quark mass $m$ in the definition of $g_\pi$. Altogether, we then find
\begin{align}
&F_{PS}(\ppf,\ppi) = F_{PS}^{\text{tree}}(\ppf,\ppi) + F_{PS}^{\text{loop}}(\ppf,\ppi)\,,\notag \\
&\frac{1}{\sqrt{2}}\,F_{PS}^{\text{tree}}(\ppf,\ppi) = \frac{\ii\,g_A\,\gamma_5\,\slashed{\qqi}}{2\,\sqrt{2}}\,\frac{m_\pi^2}{t - m_\pi^2}\Big[Z_N + \sqrt{Z_\pi} + \frac{g_\pi}{m_\pi^2\,f} - 2\Big]
\notag \\ &\hspace{2.5em}
-\frac{\ii}{\sqrt{2}}\,\big(2\,B_0\,m\big)\,g_{\chi}^{-}\,M\,\gamma_5 
-\frac{\ii}{\sqrt{2}}\,(2\,B_0\,m)\,M\,\big(4\,g_{\chi}^{+} + g_{\chi}^{-}\big)\,\gamma_5\,\frac{m_\pi^2}{t-m_\pi^2}\,,
\notag \\
%%% loops
&\frac{1}{\sqrt{2}}\,F^{\text{loop}}_{PS}(\bar{p},p)=\frac{g_A}{f^2}\,\Big\{J_{\pi}(\bar{p},p) + J_{ N \pi }(\bar{p},p) +  J_{\pi N}(\bar{p},p)\Big\}
+ \frac{f_S}{3\,f^2}\,  \Big\{J_{ \Delta \pi }(\bar{p},p) +  J_{\pi \Delta}(\bar{p},p)\Big\} 
\notag \\ &\hspace{2.5em}
+ \frac{g_A^3}{4\,f^2}\,J_{N\pi N}(\bar{p},p) 
+ \frac{2\,g_A\,f_S}{3\,f^2}\,\Big\{J_{\Delta\pi N }(\bar{p},p)+ J_{N \pi \Delta }(\bar{p},p)\Big\}
+ \frac{5\,h_A\,f_S^2}{9\,f^2}\,J_{\Delta \pi \Delta}(\bar{p},p) + {\mathcal O} \big( Q^4\big)\,.
\label{chap:WardId:loopfunctions}
\end{align}
First, we will specify the known chiral expansion of the mesonic quantities in a finite box up to chiral order four:
\begin{align}
&m_\pi^2= 2\,B_0\,m + \frac{m_\pi^2}{2\,f^2} \,\renQ[\pi]{0}{0}{0}
 + \frac{2\,m_\pi^4}{f^2}\,l_3\,,\quad 
&&\sqrt{Z_\pi} = 1 + \frac{1}{3\,f^2}\,\renQ[\pi]{0}{0}{0} - \frac{m_\pi^2}{f^2}\,l_4 \,,\notag \\
%%%
&g_{\pi}
= m_\pi^2\,\bigg[f  -  \frac{1}{f}\,\renQ[\pi]{0}{0}{0} + \frac{m_\pi^2\,l_4}{f}\bigg]\,,\quad
&&f_\pi = f\, - \frac{1}{f} \,\renQ[\pi]{0}{0}{0} +\frac{m_\pi^2}{f}\, l_4\,.
\label{app:PseudoScalar:mesonicObservables}
\end{align}
Noten that a expansion of $m_\pi^2$ recovers the strict ChPT result from Ref.~\cite{Gasser:1983yg} upon neglecting higher-order terms in the quark masses and products of the LECs $l_3$ and $l_4$.
Furthermore, with the results from Eq.~\eqref{app:PseudoScalar:mesonicObservables}, it is straightforward to verify that  $m_\pi^2\,f_\pi = g_{\pi}$, which is required by the axial Ward identity in the mesonic case \cite{Gasser:1983yg}. \\\par

We continue with the determination of the pseudoscalar form factor from the amplitude in Eq.~\eqref{chap:WardId:loopfunctions}. Similarly as for the axial-vector current it is convenient to extract the on-shell pseudoscalar form factor from the amplitude via a trace in Dirac space,
\begin{equation}
G_{PS}(\ppf,\ppi) = 2\,\ii\,M_N\,\frac{t - m_\pi^2}{t\,m_\pi^2}\,\tr\bigg[\gamma_{5}\,\frac{\slashed{\bar{p}}+M_N}{2\,M_N}\,F_{PS}(\ppf,\ppi)\,\frac{\slashed{p}+M_N}{2\,M_N}\bigg]\,.
\label{chap:axialWardBox:eq_projectorPseudoscalar}
\end{equation}
We point out that the projector does not imply that the matrix element vanishes at $t=m_\pi^2$. This is because the amplitude $F_{PS}(\ppf,\ppi)$ contains a physical singularity at $t = m_\pi^2$, the so-called pion pole. Furthermore, we note that the projector introduces a superficial singularity at $t=0$. This is not surprising, since we observed this before in the projectors of the axial-vector matrix element given in Eq.~\eqref{res:projection}. But as we will discuss shortly, our final renormalized results for the pseudoscalar form factor will be regular in $t$. Additionally, we want to stress that the projector in Eq. \eqref{chap:axialWardBox:eq_projectorPseudoscalar} is the same in the free space and in the finite volume, since we do not have additional in-box form factors. \\\par

We continue with the tree-level contributions to the pseudoscalar form factor, which are straightforwardly determined by identifying $F_{PS}(\ppf,\ppi) \to F_{PS}^{\text{tree}}(\ppf,\ppi)$ in Eq.~\eqref{chap:axialWardBox:eq_projectorPseudoscalar}. This leads to 
\begin{equation}
G_{PS}^{\text{tree}} = g_A\,\Big(Z_N+\sqrt{Z_\pi} + \frac{f_{\pi}}{f} -2 \Big) + 4\,g_{\chi}^{+}\,m_\pi^2 + g_{\chi}^{-}\,t\,.
\label{chap:axialWardBox:eq_treePS}
\end{equation}

Next we turn to the one-loop contributions to the form factor. First, we specify the $d$-dimensional in-box one-loop functions in Eq.~\eqref{chap:WardId:loopfunctions} as  
\begin{equation}
J_{\ldots}(\ppf,\ppi) = \mu^{4-d}\,\SumAll\,\IntdDPi{l}\,e^{\ii\,\kl\cdot\kx}\,\frac{m_\pi^2}{t-m_\pi^2}\, K_{\ldots}(\ppf,\ppi; l)\,,
\end{equation}
with $\ldots\in\{\pi,L\pi, \pi R, L\pi R\}$ and $L,R\in\{N,\Delta\}$. The functions $K_{\ldots}(\ppf,\ppi; l)$ read
\begin{align}
&K_{\pi} = \frac{\gamma_5\,\slashed{q}}{6\,\sqrt{2}}\,
\,\frac{1}{l^2-m_{\pi}^2}\,,\notag \\
%%%
%N pi
%%%
&K_{N\pi} = \frac{1}{\sqrt{2}}\,\frac{\gamma_5\,\slashed{l}}{l^2-m_\pi^2}\,S_N(\ppf-l)\,\bigg[ \frac{1}{4}\,\big(\slashed{q}+\slashed{l}\big) 
-4\,b_{\chi}\,B_0\,m\,\frac{t}{m_\pi^2}
+ g_S\,\big(l\cdot\qqi\big)
\notag \\ &\hspace{15.0em}
+ g_T\,l_{\nu}\,\ii\,\sigma^{\nu\rho}\,\qqi_{\rho}
+ \frac{g_V}{4}\,\big[
\slashed{l}\,\big(\ppf[2] - \ppi[2]  - l\cdot\qqi\big)
+\slashed{\qqi}\,\big(\ppf\cdot l - l^2 + l\cdot\ppi\big)
\big]\bigg]\,,\notag \\
%%%
% pi N
%%%
&K_{\pi N} = \frac{1}{\sqrt{2}}\,\bigg[\frac{1}{4}\,\big(\slashed{l} - \slashed{q}\big)
-4\,b_{\chi}\,B_0\,m\,\frac{t}{m_\pi^2}
-\,g_S\,\big(l\cdot\qqi\big)
+g_T\,l_{\nu}\,\ii\,\sigma^{\nu\rho}\,\qqi_{\rho}
\notag \\ &\hspace{15.0em}
-\,\frac{g_V}{4}\,\big[
\slashed{l}\,\big(\ppf[2] - \ppi[2]  - l\cdot\qqi\big)
+\slashed{\qqi}\,\big(\ppf\cdot l -l^2 + l\cdot\ppi\big) \big]
\bigg]
\,S_N(\ppi-l)\,\frac{\slashed{l}\,\gamma_5}{l^2-m_\pi^2}\,,\notag \\
%%%
%Delta pi
%%%
&K_{\Delta\pi} = \frac{1}{2\,\sqrt{2}}\,
\,l_{\nu}\,\Big[ \big(f_A^{(-)}-5\,f_A^{(+)}\big)\,S_{\Delta}^{\nu\rho}(\ppf-l)\,\qqi_{\rho}\,\slashed{l} 
- \big(f_A^{(-)} + 5\,f_A^{(+)}\big)\,  S_{\Delta}^{\nu\rho}(\ppf-l)\,l_{\rho}\,\slashed{\qqi}\Big]\,\frac{\gamma_5 }{l^2-m_\pi^2}\,,\notag \\
%%%
% pi Delta
%%%
&K_{\pi\Delta} = \frac{1}{2\,\sqrt{2}}\,
\frac{\gamma_5}{l^2-m_\pi^2}\,\Big[
\big(5\,f_A^{(+)}-f_A^{(-)}\big)\,\slashed{l}\,\qqi_{\nu}\,S_{\Delta}^{\nu\rho}(\ppi-l)
+\big(5\,f_A^{(+)} + f_A^{(-)}\big)\,\slashed{\qqi}\,l_{\nu}\,S_{\Delta}^{\nu\rho}(\ppi-l)\Big]\,l_{\rho}\,,\notag \\
%%%
%
%%%
&K_{N \pi N} =  -\,\frac{1}{2\,\sqrt{2}}\,
\slashed{l}\,\gamma_5\,S_{N}(\ppf-l)\,\frac{\slashed{q}\,\gamma_5}{l^2-m_\pi^2}\,S_{N}(\ppi-l)\,\slashed{l}\,\gamma_5\,,\notag \\
%%%
% Delta pi N
%%%
&K_{\Delta\pi N}= 
\frac{f_S}{2\,\sqrt{2}}\,
l_{\nu}\,S^{\nu\rho}_{\Delta}(\ppf-l)\,\qqi_{\rho}\,
S_N(\ppi-l)\,\frac{\slashed{l}\,\gamma_5}{l^2-m_{\pi}^2}\,,\qquad
%%%
% N pi Delta
%%%
K_{N \pi \Delta} = 
\frac{f_S}{2\,\sqrt{2}}\,\frac{\slashed{l}\,\gamma_5}{l^2-m_\pi^2}\,S_N(\ppf-l)\,
\qqi_{\nu}\,S^{\nu\rho}_{\Delta}(\ppi-l)\,l_{\rho}\,,\notag \\
%%%
%
%%%
&K_{\Delta\pi\Delta} =  \frac{1}{2\,\sqrt{2}}\,
l_{\nu}\,S_{\Delta}^{\nu\rho}(\ppf-l)\,g_{\rho\sigma}\,\frac{\gamma_5\,
\slashed{\qqi}}{l^2-m_\pi^2}\,S_{\Delta}^{\sigma\lambda}(\ppi-l)\,l_{\lambda}\,.
\end{align}
Because the loop functions in Eq. \eqref{chap:WardId:loopfunctions} do not have an explicit $\xn$ dependence, we can, in contrast to the axial-vector form factors, simply determine the loop contributions of the pseudoscalar form factor $J^{PS}_{\ldots}$ from Eq.~\eqref{def:GPSloopToJPS} by identifying $F_{PS}(\ppf,\ppi) \to J_{\ldots}(\ppf,\ppi)$ in Eq.~\eqref{chap:axialWardBox:eq_projectorPseudoscalar}. Then the reduction of projected loop ,contributions into our basis elements from Appendix~\ref{app:basisElements}, and the needed chiral extrapolation and renormalization is the same as in the axial-vector matrix element, which we discussed in Sec.~\ref{sec:Loop}. Thus, we continue with specifying the renormalized tadpole and bubble loop contributions: 
\begin{align}
&\bar{J}^{PS}_{\pi} = -\,\frac{1}{3}\,\renQ[\pi]{0}{0}{0}\,,\notag \\
%%%
% N pi+pi N bubble
%%%
&\bar{J}^{PS}_{N\pi} + \bar{J}^{PS}_{\pi N} = - \, m_\pi^2\,\Big(\renLQ[N\pi]{0}{0}{0} + \renQR[\pi N]{0}{0}{0}\Big) + \frac{4}{3}\,\big(
g_S-2\,g_T\big)\,M_N\,m_\pi^2\,\Big(\renLQ[N\pi]{0}{0}{0} + \renQR[\pi N]{0}{0}{0}\Big) 
\notag \\ &\hspace{2em}
+\frac{4}{3}\,\big(g_S + g_T\big)\,M_N\,\Big(\renLQ[N\pi]{0}{1}{0} + \renQR[\pi N]{0}{1}{0}\Big)
+4\,\big(g_S + g_T\big)\,M_N\,\Big(\renSLQ[N\pi]{2} + \renSQR[\pi N]{2}\Big)
\notag \\ &\hspace{2em}
- \frac{16\,b_{\chi}\,B_0\,m\,M_N}{m_\pi^2}\,\Big(\renLQ[N\pi]{1}{0}{0} - \renLQ[N\pi]{0}{0}{1} - \renQR[\pi N]{1}{0}{0} + \renQR[\pi N]{0}{0}{1}\Big) + \mathcal{O}(Q^4)\,, \notag \\
%%%
% Delta pi + pi Delta bubble
%%%
&\bar{J}^{PS}_{\Delta\pi} + \bar{J}^{PS}_{\pi \Delta} = -\,\frac{10}{9}\,f_A^{+} \,M_N\,\bigg\{ \Big[r^2\,t\,\alpPS{1}{1}{0} + 10\,m_{\pi}^2 \,\alpPS{1}{2}{0}
- 20\,M_N\,r\,\delta\,\alpPS{1}{3}{0}\Big]\,\Big(\renLQ[\Delta\pi]{0}{0}{0} + \renQR[\pi\Delta]{0}{0}{0}\Big)
\notag \\ &\hspace{2em}
-3\,r\,\alpPS{1}{0}{1}\,\Big(\renLQ[\Delta\pi]{1}{0}{0} - \renLQ[\Delta\pi]{0}{0}{1} - \renQR[\pi\Delta]{1}{0}{0} + \renQR[\pi\Delta]{0}{0}{1}\Big)
+\alpPS{1}{0}{3}\,\Big(\renLQ[\Delta\pi]{0}{1}{0} + \renQR[\pi\Delta]{0}{1}{0}\Big)
+3\,\alpPS{1}{0}{4}\,\Big(\renSLQ[\Delta\pi]{2} + \renSQR[\pi\Delta]{2}\Big) \bigg\}
\notag \\ &\hspace{1em}
+\,\frac{2}{9}\,f_A^{-}\,M_N\,\bigg\{\Big[r^2\,t\,\alpPS{2}{1}{0} - 2\,m_\pi^2\,\alpPS{2}{2}{0} + 4\,M_N\,r\,\delta\,\alpPS{2}{3}{0}\Big]\Big(\renLQ[\Delta\pi]{0}{0}{0} + \renQR[\pi\Delta]{0}{0}{0}\Big)
\notag \\ &\hspace{2em}
-3\,r\,\alpPS{2}{0}{1}\,\Big(\renLQ[\Delta\pi]{1}{0}{0} - \renLQ[\Delta\pi]{0}{0}{1} - \renQR[\pi\Delta]{1}{0}{0} + \renQR[\pi\Delta]{0}{0}{1}\Big)
+\alpPS{2}{0}{3}\,\Big(\renLQ[\Delta\pi]{0}{1}{0} + \renQR[\pi\Delta]{0}{1}{0}\Big)
+3\,\alpPS{2}{0}{4}\,\Big(\renSLQ[\Delta\pi]{2} + \renSQR[\pi\Delta]{2}\Big)
\bigg\}\,.
\label{app:PseudoScalar:bubbles}
\end{align}

Some remarks are in order. First, the LEC $b_{\chi}$ only contributes in the finite-box case, similar to the induced pseudoscalar $N\pi$-bubbles in Eq. \eqref{res:Npibubbles}. Also, the expression is regular at $t=0$, as it was the case for the axial-vector $N\pi$-bubbles from Eq. \eqref{res:Npibubbles}, but not for the induced pseudoscalar $N\pi$-bubbles. The $\alpPS{a}{b}{c}$ are rational functions in $r=\Delta/M$, given in Eq.~\eqref{app:AlpFactors:PSCase}. They are normalized as $\alpPS{a}{b}{c}\to 1$ when $r\to 0$.\\\par 

We close this section by listing the triangle contributions:
\begin{align}
\allowdisplaybreaks
&\bar{J}^{PS}_{N\pi N} = \frac{1}{3}\,\bigg\{ \renQ[\pi]{0}{0}{0} + m_\pi^2\,\Big(\renLQ[N\pi]{0}{0}{0} + \renQR[\pi N]{0}{0}{0}\Big) 
- 4\,\Big(\renLQ[N\pi]{1}{0}{0} - \renLQ[N\pi]{0}{0}{1} - \renQR[\pi N]{1}{0}{0} + \renQR[\pi N]{0}{0}{1}\Big)\bigg\}
\notag \\ &\hspace{2em}
+ \frac{4}{3}\,M_N^2\,\bigg\{-m_\pi^2\,\renLQR[N-\pi-N]{0}{0}{0}{0}{0} + 2\,\renLQR[N-\pi-N]{0}{0}{0}{1}{0} + 6\,\renSLQR[N-\pi-N]{0}{0}{2}
+4\,\Big(\renLQR[N-\pi-N]{1}{0}{1}{0}{0} - \renLQR[N-\pi-N]{1}{0}{0}{0}{1} 
- \renLQR[N-\pi-N]{0}{1}{1}{0}{0} + \renLQR[N-\pi-N]{0}{1}{0}{0}{1}\Big)
\notag \\ &\hspace{6em}
+2\,t\,\Big(\renLQR[N-\pi-N]{2}{0}{0}{0}{0} + \renLQR[N-\pi-N]{0}{2}{0}{0}{0}\Big)
\bigg\} + \mathcal{O}(Q^4)\,, \notag \\
%%%
% Dpi N - Npi D
%%%
&\bar{J}^{PS}_{\Delta\pi N} + \bar{J}^{PS}_{N\pi \Delta} = -\,\frac{1}{9}\,f_S\,\bigg\{
\Big[4\,r\,t\,\alpPS{4}{1}{0} - 4\,m_\pi^2\,\alpPS{4}{2}{0} 
+ 8\,M_N\,\delta\,\alpPS{4}{3}{0}\Big]\Big(\renLQ[\Delta\pi]{0}{0}{0} + \renQR[\pi\Delta]{0}{0}{0}\Big)
\notag \\ &\hspace{6em}
-4\,\alpPS{4}{0}{1}\,\Big(\renLQ[\Delta\pi]{1}{0}{0} - \renLQ[\Delta\pi]{0}{0}{1} - \renQR[\pi\Delta]{1}{0}{0} + \renQR[\pi\Delta]{0}{0}{1}\Big)
+\alpPS{4}{0}{3}\,\Big(\renLQ[\Delta\pi]{0}{1}{0} + \renQR[\pi\Delta]{0}{1}{0}\Big)
+3\,\alpPS{4}{0}{4}\,\Big(\renSLQ[\Delta\pi]{2} + \renSQR[\pi\Delta]{2}\Big)
\bigg\}
\notag \\ &\hspace{2em}
-\frac{4}{9}\,f_S\,M_N^2\,\bigg\{4\,m_\pi^2\,\alpPS{5}{2}{0}\Big(\renLQR[\Delta-\pi-N]{0}{0}{0}{0}{0} + \renLQR[\Delta-\pi-N]{0}{0}{0}{0}{0}\Big)
+ \alpPS{5}{0}{3}\,\Big(\renLQR[\Delta-\pi-N]{0}{0}{0}{1}{0} + \renLQR[N-\pi-\Delta]{0}{0}{0}{1}{0}\Big)
+ 3\,\alpPS{5}{0}{4}\,\Big(\renSLQR[\Delta-\pi-N]{0}{0}{2} + \renSLQR[N-\pi-\Delta]{0}{0}{2}\Big)
\notag \\ &\hspace{6em}
+4\,\alpPS{6}{0}{1}\,\Big(\renLQR[\Delta-\pi-N]{1}{0}{1}{0}{0} - \renLQR[\Delta-\pi-N]{1}{0}{0}{0}{1} - \renLQR[N-\pi-\Delta]{0}{1}{1}{0}{0} + \renLQR[N-\pi-\Delta]{0}{1}{0}{0}{1} \Big)
+2\,t\,\alpPS{7}{1}{0}\,\Big(\renLQR[\Delta-\pi-N]{2}{0}{0}{0}{0}
+\renLQR[N-\pi-\Delta]{0}{2}{0}{0}{0}\Big)
\bigg\}+ \mathcal{O}(Q^4)\,,\notag \\
%%%
%Delta pi Delta
%%%
&\bar{J}^{PS}_{\Delta\pi\Delta} = \frac{1}{27}\,\bigg\{
\Big[3\,r\,t\,\alpPS{9}{1}{0} + 32\,r\,m_\pi^2\,\alpPS{9}{2}{0} -20\,M_N\,\delta\,\alpPS{9}{3}{0}\Big]\,\Big(\renLQ[\Delta\pi]{0}{0}{0} + \renQR[\pi\Delta]{0}{0}{0}\Big)
\notag \\ &\hspace{6em}
- 4\,\alpPS{9}{0}{1}\,\Big(\renLQ[\Delta\pi]{1}{0}{0} - \renLQ[\Delta\pi]{0}{0}{1} - \renQR[\pi\Delta]{1}{0}{0} + \renQR[\pi\Delta]{0}{0}{1}\Big)
+2\,\alpPS{9}{0}{3}\,\Big(\renLQ[\Delta\pi]{0}{1}{0} + \renQR[\pi\Delta]{0}{1}{0}\Big)
+6\,\alpPS{9}{0}{4}\,\Big(\renSLQ[\Delta\pi]{2} + \renSQR[\pi\Delta]{2}\Big)
\bigg\}
\notag \\ &\hspace{2em}
+\frac{4}{27}\,M_N^2\,\bigg\{\Big[-2\,r^2\,t\,\alpPS{10}{1}{0} - 10\,m_\pi^2\,\alpPS{10}{2}{0} + 20\,M_N\,r\,\delta\,\alpPS{10}{3}{0}\Big]\,\renLQR[\Delta-\pi-\Delta]{0}{0}{0}{0}{0}
+2\,\alpPS{10}{0}{3}\,\renLQR[\Delta-\pi-\Delta]{0}{0}{0}{1}{0} + 6\,\alpPS{10}{0}{4}\,\renSLQR[\Delta-\pi-\Delta]{0}{0}{2}
\notag \\ &\hspace{6em}
-3\,r\,t\,\alpPS{11}{1}{0}\,\Big(\renLQR[\Delta-\pi-\Delta]{1}{0}{0}{0}{0} + \renLQR[\Delta-\pi-\Delta]{0}{1}{0}{0}{0}\Big)
+4\,\alpPS{11}{0}{1}\,\Big(\renLQR[\Delta-\pi-\Delta]{1}{0}{1}{0}{0} - \renLQR[\Delta-\pi-\Delta]{1}{0}{0}{0}{1}
-\renLQR[\Delta-\pi-\Delta]{0}{1}{1}{0}{0} - \renLQR[\Delta-\pi-\Delta]{0}{1}{0}{0}{1}\Big)
\notag \\ &\hspace{6em}
+2\,t\,\alpPS{12}{1}{0}\,\Big(\renLQR[\Delta-\pi-\Delta]{2}{0}{0}{0}{0} + \renLQR[\Delta-\pi-\Delta]{0}{2}{0}{0}{0}\Big)
\bigg\}+ \mathcal{O}(Q^4)\,.
\label{app:PseudoScalar:triangles}
\end{align}

\clearpage
\section{Form factors expressed in terms of correlation functions}\label{app:FFinCorrFunc}
It is straightforward to inverse the system in Eq.~\ref{eqn:systemCorrAndFF}. For a compact display of the solution, we define 
\begin{align}
&c_1 = M_N\,\big(\kq\cdot\kQ\big) + \bar{E}\,\big(\kq\cdot\ki\big) + E\,\big(\kq\cdot\kf\big)\,,\qquad 
&&\hspace{-15em}c_2 = M_N\,\kq[2] - \bar{E}\,\big(\kq\cdot\ki\big) + E\,\big(\kq\cdot\kf\big)\,,\notag \\ 
%%%
&c_3 = 3\,\big(M_N+\bar{E}\big)\big(M_N+E\big) - \kf\cdot\ki\,,\qquad
&&\hspace{-15em}c_4 = M_N\,\big(\kq\cdot\kQ\big) - \bar{E}\,\big(\kQ\cdot\ki\big) + E\,\big(\kQ\cdot\kf\big)\,,\notag \\
&c_5 = \big(M_N+\bar{E}\big)\big(M_N  + E\big) + \kf[2]\,\big(\kq\cdot\ki\big) - \ki[2]\,\big(\kq\cdot\kf\big)\,,&&\notag \\
%%%
&c_6 = \big(E\,\kf-\bar{E}\,\ki\big)\Big(\kf\,\big[E\,c_5 - c_2\,(\kq\cdot\ki)\big] - \ki\,\big[\bar{E}\,c_5 - c_1\,(\kq\cdot\kf)\big] + \kq\big[M_N\,c_5 - c_2\,c_3\big]\Big)\,.
\end{align}
Then we find: 
\begin{align}
&G_A(\ppf,\ppi) = \frac{M_N^2}{c_6}\,\bigg[2\,\big(\bar{E}-E\big)\big((\kf\cdot\ki)^2 - \kf[2]\,\ki[2]\big)\,\sum_{i = 1}^3\,\kq_i\,C_{-}^{(0,i)} + c_2\,\big(\bar{E}\,(\kq\cdot\ki) - E\,(\kq\cdot\kf)\big)\,\sum_{i,j=1}^3\,\delta_{i,j}\,C_{+}^{(i,j)}
\notag \\ &\hspace{15em}
+ \big(\bar{E}\,\ki - E\,\kf\big)\big(\bar{E}\,\ki - E\,\kf - \kq\,M_N\big)\,\sum_{i,j=1}^3\,\kq_{i}\,\kq_{j}\,C_{+}^{(i,j)}
\bigg]\,,\notag \\
%%%
% GP
%%%
&G_P(\ppf,\ppi) = \frac{M_N^3}{c_6}\,\bigg[\frac{2\,\big(-c_2\,c_3\,(\kq\cdot\kQ) + c_4\,c_5\big)}{c_2}\,\sum_{i = 1}^3\,\kq_i\,C_{-}^{(0,i)} 
- \big((\bar{E}+E)\,c_5 - (\kq\cdot\kQ)\,c_1 \big)\,\sum_{i,j=1}^3\,\delta_{i,j}\,C_{+}^{(i,j)} 
\notag \\ &\hspace{15em}
- \frac{c_1\,c_4 - c_2\,c_3\,(\bar{E}+E)}{c_2}\,\sum_{i,j=1}^3\,\kq_{i}\,\kq_{j}\,C_{+}^{(i,j)}
\bigg]\,,\notag \\
%%%
% GT
%%%
&G_T(\ppf,\ppi) = \frac{M_N^3}{c_6}\,\bigg[2\,\big(c_3\,\kq[2] - c_5\big)\,\sum_{i = 1}^3\,\kq_i\,C_{-}^{(0,i)} 
+ \big((\bar{E}-E)\,c_5 - \kq[2]\,c_1 \big)\,\sum_{i,j=1}^3\,\delta_{i,j}\,C_{+}^{(i,j)}
\notag \\ &\hspace{15em}
+ \big(c_1 - c_3\,(\bar{E}-E)\big)\,\sum_{i,j=1}^3\,\kq_{i}\,\kq_{j}\,C_{+}^{(i,j)}
\bigg]\,,\notag \\
%%%
%
%%%
&G_X^{(k)}(\ppf,\ppi) = -\frac{M_N}{2\,c_2}\,\sum_{i = 1}^3\,\kq_{i}\,C_{-}^{(k,i)} 
\quad\text{ for } k\in\{1,2,3\}
\,.
\end{align}

\clearpage
\section{List of $\alpha$ factors}
\label{app:alpFactors}
We give the $\alpha_{a,b}^{P,c}$ factors used in Eqs. 
\eqref{res:DpiBub} - 
\eqref{res:DpiDTri}:
\begin{align}
&\alpha_{1,2}^{P,0} =  \frac{(2+r)^2\,(40+48\,r+42\,r^2+18\,r^3+9\,r^4+2\,r^5)}{160\,(1+r)^2}\,,  &&\notag \\
%%%
&\alpha_{1,3}^{P,0} =  \frac{(2+r)^3\,(20 + 36\,r + 29\,r^2 + 5\,r^3)}{160\,(1+r)^3}\,, &&\hspace{-17em} \alpha_{1,0}^{P,1}  = \frac{(2+r)^2\,(3+3\,r+r^2)}{12\,(1+r)^2}\,,\notag \\
%%%
&\alpha_{1,2}^{P,1} =  \frac{(2+r)^2\,(1-r-r^2)}{4\,(1+r)^2}\,, &&\hspace{-17em}
\alpha_{1,0}^{P,3} = \alpha_{1,0}^{P,4}=  \frac{2+3\,r+3\,r^2+r^3}{2\,(1+r)^2}\,,\notag \\
%%%
&\alpha_{2,2}^{P,0} =  \frac{(2+r)^2\,(8 - 6\,r^2 - 18\,r^3 - 9\,r^4 - 2\,r^5)}{32\,(1+r)^2}\,, &&\hspace{-17em} \alpha_{2,3}^{P,0} =  \frac{(2+r)^3\,(4-5\,r^2-5\,r^3)}{32\,(1+r)^3}\,,  \notag \\
%%%
&\alpha_{2,0}^{P,1} = \frac{(2+r)^2\,(3+3\,r+r^2)}{12\,(1+r)^2}\,,  &&\hspace{-17em} \alpha_{2,2}^{P,1} =  \frac{(2+r)^2\,(1-r-r^2)}{4\,(1+r)^2}\,, \notag \\
%%%
&\alpha_{2,0}^{P,3} = \alpha_{2,0}^{P,4} = \frac{2+3\,r+3\,r^2+r^3}{2\,(1+r)^2}\,,
&&\hspace{-17em} \alpha_{3,1}^{P,0} = \frac{(2+r)^4}{16\,(1+r)^2}\,,
\notag \\
%%%
&\alpha_{4,1}^{P,0} =  \frac{(2+r)^2\,(5-3\,r-5\,r^2)}{20\,(1+r)^2}\,, &&   \notag \\
&\alpha_{4,2}^{P,0} = \frac{80+260\,r+204\,r^2+39\,r^3+21\,r^4+20\,r^5+5\,r^6}{80\,(1+r)^2}\,,  &&\notag \\
%%%
&\alpha_{4,3}^{P,0} = \frac{(2+r)^4\,(1+2\,r+2\,r^2)}{16\,(1+r)^3}\,, && \hspace{-17em}
\alpha_{4,0}^{P,1} = \frac{(2+r)\,(80+109\,r+65\,r^2+15\,r^3)}{160\,(1+r)^2}\,,\notag \\
&\alpha_{4,2}^{P,1} = \frac{(2+r)\,(40+41\,r+5\,r^2-5\,r^3)}{80\,(1+r)^2}\,, && \hspace{-17em}
\alpha_{4,0}^{P,3} = \alpha_{4,0}^{P,4} = \frac{1 - r - r^2}{(1+r)^2}\,,\notag \\
%%%
&\alpha_{5,2}^{P,0} =  \alpha_{5,0}^{P,3} = \alpha_{5,0}^{P,4} =\frac{2+r}{2}\,,  && \hspace{-17em} \alpha_{6,0}^{P,1} = \frac{2+r}{2}\,,\notag \\
&\alpha_{7,1}^{P,0} = \frac{2+r}{2}\,, && \hspace{-17em} \alpha_{8,1}^{P,0} = \frac{(2+r)^2\,(1+3\,r+r^2)}{4\,(1+r)}\,,\notag \\
&\alpha_{9,1}^{P,0} = \frac{(2+r)^3\,(10+141\,r+236\,r^2+101\,r^3+15\,r^4)}{80\,(1+r)^4}\,,  && \notag \\
&\alpha_{9,2}^{P,0} = \frac{(2+r)\, (1360+3208 \, r+3916 \, r^2+3158 \, r^3+1555 \, r^4+427 \, r^5+50 \, r^6)}{2720 \, (1+r)^4}\,, &&\notag \\
&\alpha_{9,3}^{P,0} = \frac{(2+r)^2 \, (20+80 \, r+211 \, r^2+293 \, r^3+218 \, r^4+84 \, r^5+12 \, r^6)}{80\,(1+r)^5}\,, && \notag \\
%%%
&\alpha_{9,0}^{P,1} =\frac{(2+r)^2 \, (10+36 \, r+43 \, r^2+18 \, r^3+7 \, r^4)}{40\,(1+r)^4}\,,&& \notag \\
%%%
&\alpha_{9,0}^{P,3} = \alpha_{9,0}^{P,4} = \frac{2+7\,r+9\,r^2+7\,r^3+2\,r^4}{2\,(1+r)^4}\,,  &&\notag \\
%%%
&\alpha_{10,1}^{P,0} = \frac{(2+r)^4\,(40 + 195\,r + 227\,r^2 + 97\,r^3 + 12\,r^4)}{640\,(1+r)^4}\,, &&\notag \\
%%%
&\alpha_{10,2}^{P,0} = \frac{(2+r)^2 \, (400+1600 \, r+2700 \, r^2+2460 \, r^3+1516 \, r^4+922 \, r^5+571 \, r^6 
+225 \, r^7+34 \, r^8)}{1600\,(1+r)^4}\,,  &&\notag \\
%%%
&\alpha_{10,3}^{P,0} =\frac{(2+r)^3 \, (20+60 \, r+69 \, r^2+31 \, r^3+3 \, r^4)}{160\,(1+r)^3}\,,&& \hspace{-17em}
\alpha_{10,0}^{P,3} = \alpha_{10,0}^{P,4} = \frac{(2+r)^2\,(2+2\,r-r^2)}{8\,(1+r)^2}\,, \notag \\
%%%
&\alpha_{11,1}^{P,0} = \frac{(2+r)^3 (20+51 \, r+58 \, r^2+28 \, r^3+2  \, r^4)}{160\,(1+r)^4}\,,  &&\notag \\
%%%
&\alpha_{11,2}^{P,0} = \frac{(2+r)^3 \, (10+28 \, r+19 \, r^2 - 6 \, r^3-9 \, r^4)}{80\,(1+r)^4}\,,  &&\notag \\
%%%
&\alpha_{11,0}^{P,1} = \frac{(2+r)^2\,(2 + 2\,r - r^2)}{8\,(1+r)^2}\,,&&\hspace{-17em}
\alpha_{12,1}^{P,0} = \frac{(2+r)^2\,(2 + 2\,r - r^2)}{8\,(1+r)^2}\,. 
\end{align} 

We now turn to the $\alpT{a}{b}{c}$ and $\betT{a}{b}{c}$ factors used in Eqs. 
\eqref{res:DpiBub} - 
\eqref{res:DpiDTri}:
\begin{align}
&\alpT{1}{1}{0} = \frac{(2+r)^4\,(1+2\,r)}{16\,(1+r)^2}\,, && \hspace{-8em}
\alpT{1}{0}{1} = \frac{(2+r)^2\,(3+r)}{12\,(1+r)^2}\,,&& \hspace{-8em}
\alpT{2}{1}{0} = \frac{(2+r)^4\,(1+2\,r)}{16\,(1+r)^2}\,,\notag\\
%%%
&\alpT{2}{0}{1} = \frac{(2+r)^2}{4\,(1+r)}\,, && \hspace{-8em}
\alpT{3}{1}{0} = \frac{(2+r)^4}{16\,(1+r)^2}\,, &&\hspace{-8em}
\alpT{4}{1}{0} = \frac{(2+r)^3\,(1+3\,r+r^2)}{8\,(1+r)^2}\,,\notag\\
%%%
&\alpT{4}{0}{1} = \frac{(2+r)^3}{8\,(1+r)^2}\,,&& \hspace{-8em}
\betT{4}{2}{0} = \frac{8 + 2\,r-3\,r^2-r^3}{8\,(1+r)^2}\,, && \hspace{-8em}
\alpT{5}{2}{1} = \frac{8 + 10\,r + 5\,r^2+r^3}{8\,(1+r)^2}\,, \notag\\
%%%
&\alpT{5}{1}{3} = \frac{2-7\,r-6\,r^2-r^3}{2\,(1+r)^2}\,, && \hspace{-8em}
\alpT{5}{0}{4} = \frac{2-r-r^2}{2\,(1+r)^2}\,,&& \hspace{-8em}
\alpT{5}{0}{5} = \frac{2-7\,r - 6\,r^2-r^3}{2\,(1+r)^2}\,, \notag\\
%%%
&\alpT{5}{0}{6} = \frac{6 + 5\,r+r^2}{6\,(1+r)^2}\,, && \hspace{-8em}
\betT{6}{2}{0} = \frac{8+10\,r+5\,r^2+r^3}{8\,(1+r)^2}\,, && \hspace{-8em}
\alpT{6}{1}{1} = \frac{14 - 13\,r -12\,r^2-r^3}{14\,(1+r)^2}\,,\notag\\
%%%
&\alpT{6}{1}{2} = \frac{2-43\,r-36\,r^2-7\,r^3}{2\,(1+r)^2}\,, && \hspace{-8em}
\betT{7}{1}{0} = \frac{2-7\,r-6\,r^2-r^3}{2\,(1+r)^2}\,, && \hspace{-8em}
\alpT{8}{1}{0} = \frac{(2+r)^3\,(2+2\,r+r^2)}{16\,(1+r)^2}\,,\notag\\
%%%
&\betT{9}{2}{0} = \frac{8+14\,r+5\,r^2}{8\,(1+r)}\,,&& \hspace{-8em}
\alpT{9}{1}{3} = \frac{2-r-r^2}{2\,(1+r)}\,,&& \hspace{-8em}
\alpT{9}{0}{4} = \frac{2-r-r^2}{2\,(1+r)}\,,\notag \\
%%%
&\alpT{10}{1}{1} = \frac{2-r-r^2}{2\,(1+r)}\,, && \hspace{-8em}
\betT{11}{1}{0} = \frac{2-r-r^2}{2\,(1+r)}\,, && \hspace{-8em}
\alpT{12}{1}{0} = \frac{(2+r)^3\,(1+3\,r-r^3)}{8\,(1+r)^4}\,,\notag\\
%%%
&\alpT{12}{0}{1} = \frac{(2+r)^2\,(4+9\,r+12\,r^2+5\,r^3)}{16\,(1+r)^4}\,, && && \hspace{-8em}
\alpT{13}{0}{1} = \frac{(2+r)^4}{16\,(1+r)^2}\,, \notag\\
%%%
&\alpT{13}{1}{1} = \frac{(2+r)^3\,(55+137\,r+122\,r^2+37\,r^3)}{440\,(1+r)^4}\,, && && \hspace{-8em}
\alpT{13}{2}{1} = \frac{(2+r)^2\,(4+4\,r+3\,r^2)}{16\,(1+r)^2}\,, \notag\\
%%%
&\alpT{13}{3}{1} = \frac{(2+r)^3\,(2+3\,r+2\,r^2)}{16\,(1+r)^3}\,,&& && \hspace{-8em}
\alpT{13}{0}{4} = \frac{(2+r)^2\,(2+2\,r-r^2)}{8\,(1+r)^2}\,, \notag\\
%%%
&\alpT{14}{1}{0} = \frac{(2+r)^4}{16\,(1+r)^2}\,, && \hspace{-8em}
\betT{14}{1}{0} = \frac{(2+r)^3\,(55+137\,r+122\,r^2+37\,r^3)}{440\,(1+r)^4}\,,&& \notag\\
%%%
&\betT{14}{2}{0} = \frac{(2+r)^2\,(4+4\,r+3\,r^2)}{16\,(1+r)^2}\,,&& && \hspace{-8em}
\betT{14}{3}{0} = \frac{(2+r)^3\,(2+3\,r+2\,r^2)}{16\,(1+r)^3} \,, \notag\\
%%%
&\alpT{14}{1}{1} = \frac{(2+r)^2 \, (6+34 \, r+51 \, r^2+48 \, r^3+17 \, r^4)}{24\,(1+r)^4}\,, && && \notag\\
%%%
&\alpT{14}{1}{2} = \frac{(2+r)^2 \, (6+2 \, r-21 \, r^2-48 \, r^3-23 \, r^4)}{24 \, (1+r)^4}\,, && && \notag\\
%%%
&\betT{15}{1}{0} = \frac{(2+r)^2 \, (4+9 \, r+12 \, r^2+5 \, r^3)}{16 \, (1+r)^4}\,.&& &&
\end{align}
For the $\alpX{a}{b}{c}$ factors, used in Eqs. 
\eqref{res:JXq} and 
\eqref{res:DpiBub} - 
\eqref{res:DpiDTri}, 
we have:
\begin{align}
&\alpX{1}{1}{1} = \frac{(2+r)^2\,(2+r+r^2)}{8\,(1+r)^2}\,,&& \hspace{-8em}
\alpX{1}{0}{4} = \frac{(2+r)\,(1+r+r^2)}{2\,(1+r)^2}\,, && 
\alpX{2}{1}{1} = \frac{(2+r)^3}{8\,(1+r)}\,,\notag \\
%%%
&\alpX{2}{0}{4} = \frac{(2+r)\,(1+r+r^2)}{2\,(1+r)^2}\,, && \hspace{-8em}
\alpX{3}{1}{1} = \frac{(2+r)^2\,(1+r+r^2)}{4\,(1+r)^2}\,,  &&\notag \\
&\alpX{4}{0}{1} = \frac{(2+r)\,(20+14\,r+10\,r^2+5\,r^3)}{40\,(1+r)^2} \,,&& && \alpX{4}{0}{4} = \frac{1-r-r^2}{(1+r)^2} \,, \notag \\
%%%
&\alpX{5}{0}{4} = \frac{2+r}{2}\,, && \hspace{-8em} \alpX{6}{0}{1} = \frac{2+r}{2}\,, &&\alpX{8}{0}{1} = \frac{6+9\,r+5\,r^2+r^3}{6\,(1+r)^2}\,,\notag \\
&\alpX{9}{1}{1} = \frac{(2+r)^2\,(10+36\,r+43\,r^2+18\,r^3+7\,r^4)}{40\,(1+r)^4}\,, &&
&& \alpX{9}{0}{4} = \frac{2 +7\,r + 9\,r^2+7\,r^3+2\,r^4}{2\,(1+r)^4}\,, \notag \\
&\alpX{10}{0}{4} = \frac{(2+r)^2\,(2+2\,r-r^2)}{8\,(1+r)^2}\,, && &&
\alpX{11}{1}{1} = \frac{(2+r)^2\,(2+2\,r-r^2)}{8\,(1+r)^2}\,.
\end{align}
We close the Section by giving the functions $\alpPS{a}{b}{c}$ used in the Eqs.~\eqref{app:PseudoScalar:bubbles} and 
\eqref{app:PseudoScalar:triangles}:
\begin{align}
&\alpPS{1}{1}{0} = \frac{(2+r)^4\,(1+2\,r)}{16\,(1+r)^2}\,, &&\hspace{-6em}
\alpPS{1}{2}{0} = \frac{(2+r)^2\,(20+24\,r+19\,r^2+3\,r^3)}{80\,(1+r)^2}\,, &&\notag \\
%%%
&\alpPS{1}{3}{0} = \frac{(2+r)^3\,(20+36\,r+29\,r^2+5\,r^3)}{160\,(1+r)^3}\,,
&&\hspace{-6em} \alpPS{1}{0}{1} = \frac{(2+r)^2}{4}\,, \notag\\
%%%
&\alpPS{1}{0}{3} = \alpPS{1}{0}{4} = \frac{2+3\,r+3\,r^2+r^3}{2\,(1+r)^2}\,, && \notag\\
%%%
&\alpPS{2}{1}{0} = \frac{(2+r)^4\,(1+2\,r)}{16\,(1+r)^2}\,, 
&&\hspace{-6em} \alpPS{2}{2}{0} = \frac{(2+r)^2\,(4-r^2-3\,r^3)}{16\,(1+r)^2}\,, \notag\\
%%%
&\alpPS{2}{3}{0} = \frac{(2+r)^3\,(4-5\,r^2-5\,r^3)}{32\,(1+r)^3}\,, 
&&\hspace{-6em} \alpPS{2}{0}{1} = \frac{(2+r)^2}{4}\,, \notag \\
%%%
&\alpPS{2}{0}{3} = \alpPS{2}{0}{4} = \frac{2+3\,r+3\,r^2+r^3}{2\,(1+r)^2}\,, &&\notag\\
%%%
&\alpPS{4}{1}{0} = \frac{(2+r)^2\,(20+11\,r-5\,r^2-5\,r^3)}{80\,(1+r)^2}\,,
&&\hspace{-6em}\alpPS{4}{2}{0} = \frac{16 +64\,r+64\,r^2+22\,r^3+3\,r^4}{16\,(1+r)^2}\,, \notag \\
%%%
&\alpPS{4}{3}{0} = \frac{(2+r)^4\,(1+2\,r+2\,r^2)}{16\,(1+r)^3}\,,
%%% 
&&\hspace{-6em}\alpPS{4}{0}{1} = \frac{40 + 34\,r - 43\,r^2-55\,r^3-15\,r^4}{40\,(1+r)^2}\,, \notag \\ &\alpPS{4}{0}{3} = \alpPS{4}{0}{4} = \frac{1-r-r^2}{(1+r)^2}\,,\notag \\
%%%
& \alpPS{5}{2}{0} = \alpPS{5}{0}{3} = \alpPS{5}{0}{4} = \frac{2+r}{2}\,,
&&\hspace{-6em}\alpPS{6}{0}{1} = \frac{2+r}{2}\,, 
\qquad \alpPS{7}{1}{0} = \frac{2+r}{2}\,,\notag\\
%%%
&\alpPS{9}{1}{0} = \frac{(2+r)^3\,(30+263\,r+523\,r^2+328\,r^3+65\,r^4)}{240\,(1+r)^4} \,, && \hspace{-6em}
\alpPS{9}{2}{0} = \frac{(2+r)\,(64+132\,r+113\,r^2+49\,r^3+8\,r^4)}{128\,(1+r)^4}\,,  && \notag \\
%%%
&\alpPS{9}{3}{0} = \frac{(2+r)^2\,(20+80\,r+211\,r^2+293\,r^3 + 218\,r^4 + 84\,r^5 + 12\,r^6)}{80\,(1+r)^5}\,, &&\notag\\
%%%
&\alpPS{9}{0}{1} = \frac{(2+r)^2\,(10+36\,r+43\,r^2+18\,r^3+7\,r^4)}{40\,(1+r)^4}\,, 
&&\hspace{-6em} \alpPS{9}{0}{3} = \alpPS{9}{0}{4} = \frac{2 + 7\,r + 9\,r^2+7\,r^3+2\,r^4}{2\,(1+r)^4}\,, \notag\\
%%%
&\alpPS{10}{1}{0} = \frac{(2+r)^4\,(20 + 145\,r+266\,r^2+186\,r^3+46\,r^4)}{320\,(1+r)^4} \,,
&&\hspace{-6em}\alpPS{10}{2}{0} = \frac{(2+r)^2\,(20 + 40\,r+39\,r^2+19\,r^3+2\,r^4)}{80\,(1+r)^2}\,, \notag\\
%%%
&\alpPS{10}{3}{0} = \frac{(2+r)^3\,(20 + 60\,r + 69\,r^2 + 31\,r^3 + 3\,r^4)}{160\,(1+r)^3}\,,
&&\hspace{-6em} \alpPS{10}{0}{3} = \alpPS{10}{0}{4} = \frac{(2+r)^2\,(2+2\,r-r^2)}{8\,(1+r)^2}\,, \notag\\
%%%
&\alpPS{11}{1}{0} = \frac{(2+r)^3\,(10+23\,r+39\,r^2+34\,r^3+11\,r^4)}{80\,(1+r)^4}\,, 
&&\hspace{-6em} \alpPS{11}{0}{1} = \frac{(2+r)^2\,(2+2\,r-r^2)}{8\,(1+r)^2}\,,\notag\\
&\alpPS{12}{1}{0} = \frac{(2+r)^2\,(2+2\,r-r^2)}{8\,(1+r)^2}\,.
\label{app:AlpFactors:PSCase}
\end{align}

\end{widetext}

\bibliography{literature.bib} 

\end{document}